%% file: EOI.tex
\documentclass[12pt]{report}
\usepackage{graphicx}
\usepackage{amssymb}
\usepackage{amsmath}
\usepackage{amsfonts}
\usepackage{array,hhline,dcolumn} 
\usepackage{epstopdf}
\usepackage[dvips]{epsfig}

\begin{document}

\thispagestyle{empty}

\title{\vspace{-1.in} Expression of Interest for \\
A Novel Search for CP Violation in the Neutrino Sector: \\
\begin{figure}[h]
\centering
\vspace{-0.25in}
\includegraphics[width=4.5in]{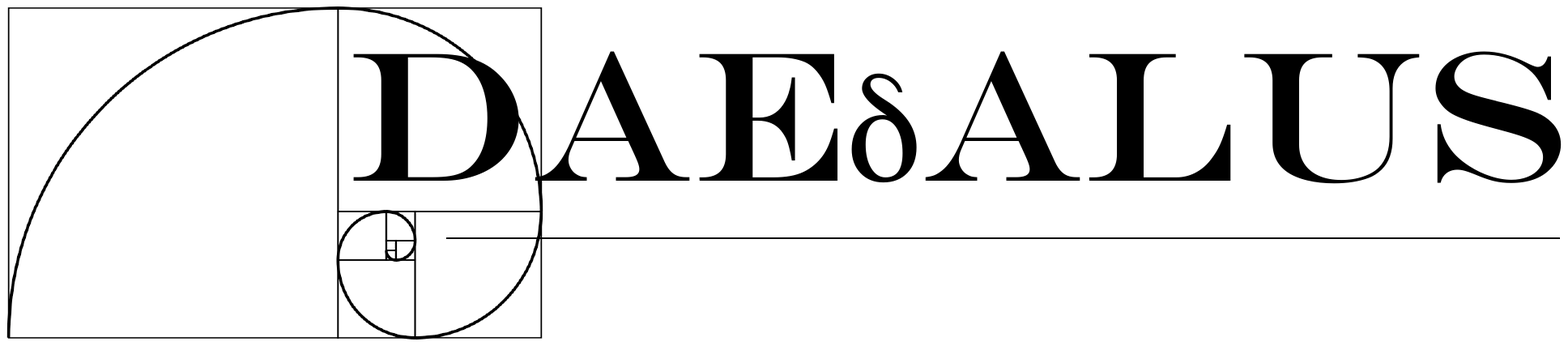} \vspace{-1.5in}
\end{figure}
}

\author{J. Alonso$^{13}$, F.T. Avignone$^{18}$, W.A. Barletta$^{13}$,\\ 
R. Barlow$^5$, H.T. Baumgartner$^{13}$, A. Bernstein$^{11}$, E. Blucher$^4$,\\ 
L. Bugel$^{13}$, L. Calabretta$^9$, L. Camilleri$^6$, R. Carr$^6$,\\ 
J.M. Conrad$^{13,*}$, S.A. Dazeley$^{11}$, Z. Djurcic$^2$, A.~de~Gouv\^ea$^{17}$,\\
P.H. Fisher$^{13}$, C.M. Ignarra$^{13}$, B.J.P. Jones$^{13}$, C.L. Jones$^{13}$,\\G. Karagiorgi$^{13}$, T. Katori$^{13}$, S.E. Kopp$^{20}$, R.C. Lanza$^{13}$,\\ 
W.A. Loinaz$^1$, P. McIntyre$^{19}$, G. McLaughlin$^{16}$, G.B. Mills$^{12}$,\\
J.A. Nolen$^2$, V. Papavassiliou$^{15}$, M. Sanchez$^{2,10}$,K. Scholberg$^7$,\\
W.G. Seligman$^6$, M.H. Shaevitz$^{6,*}$, S. Shalgar$^{17}$, T. Smidt$^{13}$,\\ 
M.J. Syphers$^{14}$, J. Spitz$^{22}$, H.-K. Tanaka$^{13}$, K. Terao$^{13}$,\\ 
C. Tschalaer$^{13}$, M. Vagins$^{3,21}$, R. Van de Water$^{12}$,\\  
M.O. Wascko$^8$, R. Wendell$^7$, L. Winslow$^{13}$\\}

\maketitle

\newpage

~~~

\begin{center}
$^1$Amherst College, Amherst, MA 01002, USA \\
$^2$Argonne National Laboratory, Argonne, IL 60439, USA \\
$^3$University of California, Irvine, CA 92697, USA \\ 
$^4$University of Chicago, Chicago, IL 60637, USA \\
$^5$The Cockcroft Institute for Accelerator Science \& \\
the University of Manchester, Oxford Road, Manchester M13 9PL, UK\\
$^6$Columbia University, New York, NY 10027, USA\\
$^7$Duke University, Durham, NC 27708, USA \\
$^8$Imperial College London. London, SW7 2AZ, UK\\
$^9$Istituto Nazionale di Fisica Nucleare, Laboratori Nazionali del Sud, I-95123, Italy\\
$^{10}$Iowa State University, Ames, IA 50011, USA\\
$^{11}$Lawrence Livermore National Laboratory, Livermore, CA 94551, USA \\
$^{12}$Los Alamos National Laboratory, Los Alamos, NM 87545, USA \\
$^{13}$Massachusetts Institute of Technology, Cambridge, MA 02139, USA\\
$^{14}$Michigan State University, East Lansing, MI 48824, USA\\
$^{15}$New Mexico State University, Las Cruces, NM 88003, USA\\
$^{16}$North Carolina State University, Raleigh, NC 27695, USA\\
$^{17}$Northwestern University, Evanston, IL 60208, USA \\
$^{18}$University of South Carolina, Columbia, SC 29208,USA \\
$^{19}$Texas A\&M University, College Station, TX 77843, USA \\
$^{20}$University of Texas, Austin, TX 78712, USA\\
$^{21}$University of Tokyo, Kashiwa, 277-8583, Japan\\
$^{22}$Yale University, New Haven, CT 06520 USA \\
~~~~\\
~~~~\\
$^*$ Corresponding authors. For further information contact:\\ conrad@mit.edu, shaevitz@nevis.columbia.edu\\

\end{center}

\newpage

\input{theabstract_v2.tex}

\newpage

\tableofcontents

\newpage

\chapter{Executive Summary}
\label{se:intro}
\input{intro_v5.tex}

\newpage

\chapter{The Physics Opportunities of DAE$\delta$ALUS}
\label{se:physics}

\section{$CP$ Violation and Neutrinos}

\input{physintro_v3.tex}

\section{DAE$\delta$ALUS and $CP$ Violation}
\label{sse:CP}

\input{CP_v3.tex}

\section{A Joint Analysis With the Fermilab Beam}
\label{sse:joint}

\input{joint_v5.tex}

\section{Physics with a Near Accelerator}

\label{sse:near}

\input{near_v6.tex}

\newpage

\chapter{Preliminary Design}
\label{se:design}

\input{designintro_v1.tex}

\section{Accelerator Design}
\label{sse:accels}

\input{accel_intro_v2.tex}

\subsection{Compact Cyclotrons}
\label{ssse:compact}

\input{compact_v4.tex}

\subsection{H$_2^+$ Cyclotrons}
\label{ssse:h2plus}

\input{h2plus_v5.tex}

\subsection{Stacked Cyclotrons}
\label{ssse:stacked}

\input{stackedcyclo_v2.tex}

\subsection{Other Types of Accelerators: Superconducting Linacs}
\label{ssse:othertypes}

\input{otheraccel_v1.tex}

\subsection{Neutrino Source Design}
\label{ssse:dump}

\input{nu-source_v1.tex}

\section{Detector Design}
\label{sse:detector}

\input{detectorintro_v1.tex}

\subsection{Gd Doping}
\label{ssse:Gd}

\input{Gd_v2.tex}

\subsection{Photocathode Coverage}
\label{ssse:photocover}

\input{photocover_v1.tex}

\section{A Three-Phase Implementation Plan}
\label{se:schedule}

\input{schedule_v1.tex}

\newpage

\chapter{Impact On Other Analyses}
\label{se:coord}

\input{coord_v1.tex}

\section{Neutrino-electron Events as a Calibration Source}
\label{calib}
\input{calib_v3.tex}

\section{Impact On Other Large Detector Analyses}

\input{backgrounds_v4.tex}

\newpage

\chapter{Conclusion}
\label{se:conclusion}

\input{conclusion_v1.tex}

\newpage

\bibliography{bib_v1}
\bibliographystyle{unsrt}

\end{document}

%% file: theabstract_v2.tex
\abstract{DAE$\delta$ALUS, a 
{\bf D}ecay-{\bf A}t-rest {\bf E}xperiment for \mbox{\boldmath{$\delta$}}$_{CP}$ 
studies {\bf A}t the {\bf L}aboratory for {\bf U}nderground {\bf
S}cience, provides a new approach to the search for $CP$
violation in the neutrino sector.  The design utilizes low-cost,
high-power proton accelerators under development for commercial uses.
These provide neutrino beams with energy up to 52 MeV from pion and
muon decay-at-rest.  The experiment searches for $\bar{\nu}_\mu
\rightarrow \bar{\nu}_e$ at short baselines corresponding to the
atmospheric $\Delta m^2$ region.  The $\bar \nu_e$ will be detected,
via inverse beta decay, in the 300 kton fiducial-volume Gd-doped water
Cherenkov neutrino detector proposed for the Deep Underground Science
and Engineering Laboratory (DUSEL).

  DAE$\delta$ALUS opens new opportunities for DUSEL.  It provides a
  high-statistics, low-background alternative for $CP$ violation
  searches which matches the capability of the conventional long-baseline
  neutrino experiment, LBNE.  Because of the complementary designs, when
  DAE$\delta$ALUS antineutrino data are combined with LBNE neutrino
  data, the sensitivity of the CP-violation search improves beyond
  any present proposals, including the proposal for Project X.  Also,
  the availability of an on-site neutrino beam opens opportunities for
  additional physics, both for the presently planned DUSEL detectors
  and for new experiments at a future 300 ft campus.}

%% file: intro_v5.tex
This Expression of Interest (EOI) describes DAE$\delta$ALUS, a 
{\bf D}ecay-{\bf A}t-rest {\bf E}xperiment for \mbox{\boldmath{$\delta$}}$_{CP}$ 
studies {\bf A}t the {\bf L}aboratory for {\bf U}nderground {\bf
S}cience. 
The primary physics goal of the experiment is to
search for $CP$-violation in the neutrino sector using a novel design
which provides high-statistics and low backgrounds
\cite{Conrad:2009mh}.  DAE$\delta$ALUS searches for CP violation in
$\bar \nu_\mu \rightarrow \bar \nu_e$ oscillations at the atmospheric
mass splitting by comparing absolute neutrino rates in a single
detector that is exposed to neutrino beam sources at three distances.
This method exploits the length-dependence of the $CP$-violating
interference terms in the oscillation formula.  The near-source
measures the initial flux.  The mid-source is at half of
oscillation maximum.  The far source is at oscillation maximum.

As discussed in Chapter ~\ref{se:physics}, DAE$\delta$ALUS has several
advantages over conventional searches.  The experiment can demonstrate
that $\delta_{CP}$, the $CP$-violating parameter in the three-neutrino
mixing matrix, is not 0 or 180$^\circ$, regardless of the mass
hierarchy.  Because this is a short-baseline experiment, it does not
suffer from ``apparent $CP$-violation'' caused by matter effects.
DAE$\delta$ALUS matches the sensitivity of conventional long-baseline
experiments ({\it e.g.}, LBNE \cite{Barger:2007yw}) to $\delta_{CP}$
and the mixing angle $\theta_{13}$.  Because the time-frame for
constructing the accelerators is relatively short, construction need
not begin before present experiments provide a measure of
$\theta_{13}$.  Also because of its complementary design,
DAE$\delta$ALUS substantially improves the sensitivity of the search
when the data are combined with the long-baseline results. Beyond
oscillation physics, DAE$\delta$ALUS enhances the physics
opportunities at the Deep Underground Science and Engineering
Laboratory by providing an on-site high-intensity neutrino source.

The neutrino beams will be produced via high-power proton cyclotron
accelerators that are under development for Accelerator Driven Systems
(ADS) for subcritical thorium-based reactors and for active
interrogation for homeland security.  DAE$\delta$ALUS requires
accelerators that can target protons between 650 and 1500 MeV at 1 MW
or more.  The beams will be delivered in alternating time periods
(presently planned for 100 $\mu$s duration), in
order to allow events from each position to be uniquely identified.
We propose to run the near, mid and far sites each with a 20\% duty
factor.  Multiple accelerators may be required at each site, depending
on the cyclotron design.

\begin{figure}[t]\begin{center}
{\includegraphics[width=4.5in]{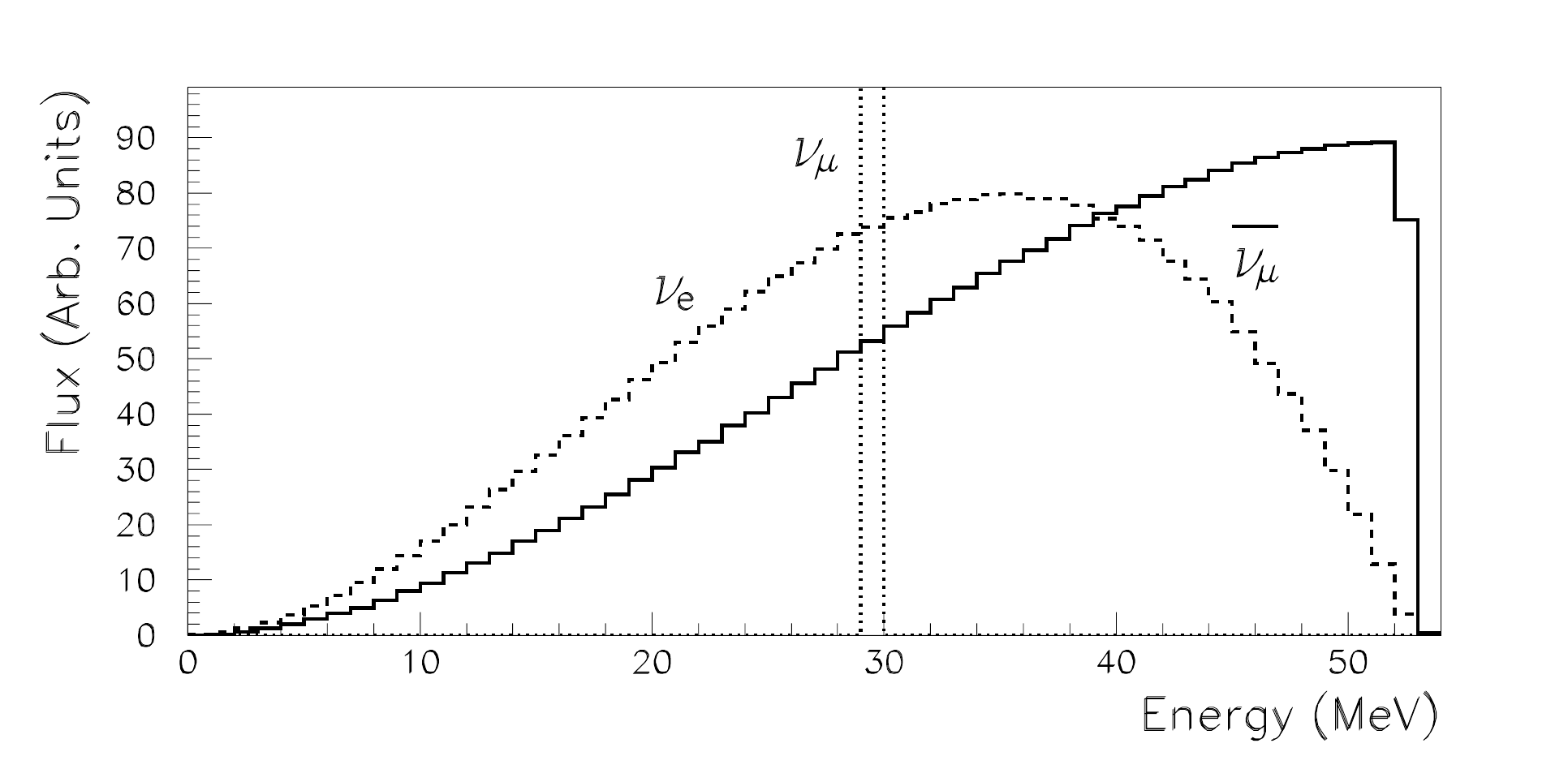}
} \end{center}
\vspace{-0.25in}
\caption{Energy distribution of neutrinos in a DAR beam 
\label{fluxplot} }
\end{figure}

\begin{figure}[t]\begin{center}
{\includegraphics[width=4.5in]{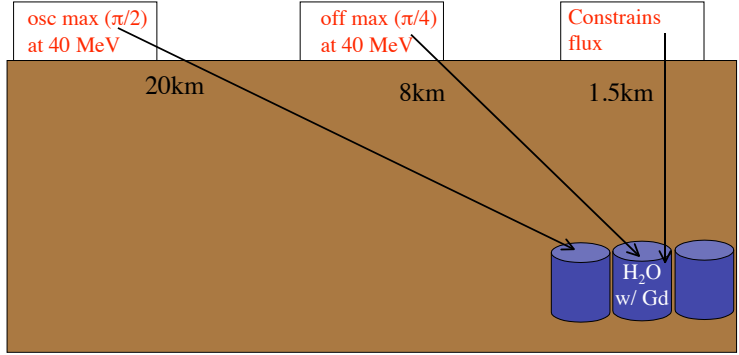}
} \end{center}
\vspace{-0.25in}
\caption{Schematic of the DAE$\delta$ALUS Experiment.  Three 
neutrino source locations are used in conjunction with the 300 kton water 
Cherenkov detector complex at the 4850 level of DUSEL. 
\label{schematic} }
\end{figure}

The preliminary design of the experiment is described in Chapter
~\ref{se:design}.  Cyclotrons deliver protons into a beam stop
creating a high intensity, isotropic decay-at-rest (DAR) neutrino
beam arising from the stopped pion decay chain:
$\pi^{+}\rightarrow\nu_{\mu} \mu^{+}$ followed by $\mu^+ \rightarrow
e^{+}\bar{\nu}_{\mu}\nu_{e}$.  The resulting $\bar \nu_\mu$ flux rises
with energy to the 52.8 MeV endpoint (see Fig.~\ref{fluxplot}), with a
well-known energy dependence.  Because most $\pi^-$ capture before decay,
the $\bar \nu_e$ fraction in the beam is $\sim 4\times 10^{-4}$.  For
a DAR beam design, $CP$-violation in $\bar \nu_\mu \rightarrow \bar
\nu_e$ oscillations can be addressed with neutrino sources located at
1.5 km (near), 8 km (mid) and 20 km (far).  A schematic is shown in
Fig~\ref{schematic}.  The advantage of a DAR beam is that the nature of
the weak interaction alone drives the energy dependence.  The flux from
the three beams will be identical up to the relative normalization.
This experiment will use the 300-kton 3-unit water Cherenkov detector at
the 4850 ft level of DUSEL as the neutrino target.  Water provides a
target of free protons for the inverse beta decay (IBD) interaction:
$\bar \nu_e + p \rightarrow n + e^+$.  IBD interactions are identified
via a coincidence signal: the Cherenkov ring produced by the positron
followed by the capture of the neutron. Doping with gadolinium (Gd) is
proposed in order to enhance the neutron capture signal
\cite{Beacom:2003nk, Watanabe:2008ru, Dazeley:2008xk, Kibayashi:2009ih}.

We propose a two phase search for $CP$-violation, where each phase
represents five years of running.  The purpose of Phase 1, which will
have 1 MW, 2 MW and 3 MW accelerators at the near, mid and far
locations is to explore the oscillation space with $\sim$3$\sigma$.
Once the region of interest is localized, additional neutrino sources
can be added at appropriate locations for Phase 2 to give the best
possible measurement.  For simplicity, in this EOI, we will generally
show results for the Phase 2 configuration of (1~MW, 2~MW, 7~MW), but
it should be understood that this is just one example.

The analysis, described in Sec.~\ref{sse:CP}, follows three steps.
First, the absolute normalization of the flux from the near
accelerator is measured using neutrino-electron scatters in the
detector, for which the cross section is known to 1\%.  The relative
flux normalization between the sources is then determined using the
comparative rates of $\nu_e$-oxygen interactions in the the
detector.  Once the normalizations of the accelerators are known, then
the IBD data can be fit to extract the $CP$-violating parameter
$\delta_{CP}$.  The $\chi^{2}$ statistic of the fit accounts for
systematic uncertainties through parameters that vary along with the
oscillation parameters, but which are constrained by pull terms.  The
oscillation probability also depends on the mixing angle
$\theta_{13}$.  Therefore, the resulting measurement must be described in
a $\sin^2 2\theta_{13}$--$\delta_{CP}$ plane.

\begin{figure}[p]\begin{center}
{\includegraphics[width=4.5in]{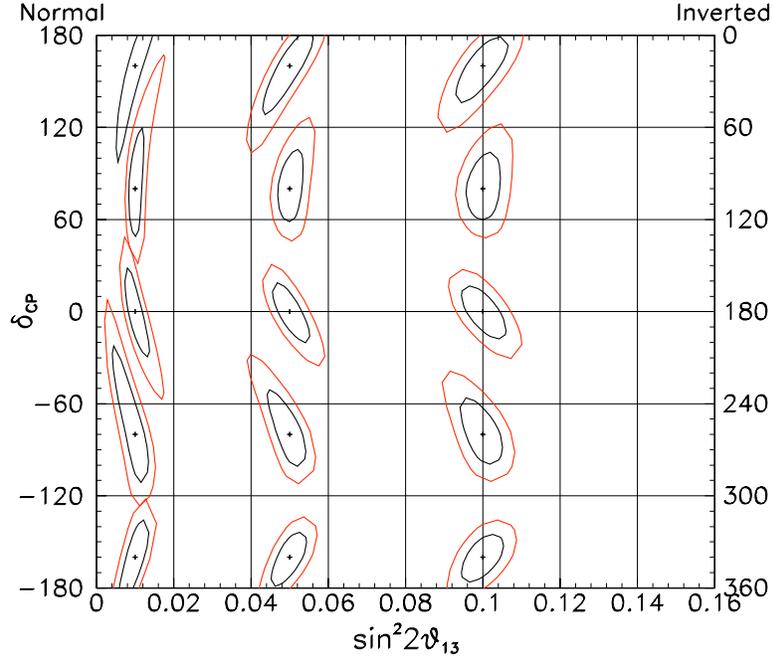}
} \end{center}
\caption{ One (inner contour) and two (outer contour) $\sigma$ sensitivities 
for DAE$\delta$ALUS for Phase 1 and 2 combined 
(10 years of running).  
DAE$\delta$ALUS is not sensitive to matter effects and,
therefore, has a degeneracy between the two mass hierarchies.   This can
be represented by showing the $\delta_{CP}$ scale for normal hierarchy on
the left and inverted on the right.  DAE$\delta$ALUS can
determine if there is CP violation (i.e. $\delta_{CP} \ne 0$ or
$180^\circ$) without this input.
Details are provided in Sec.~\ref{sse:CP}.
\label{deltaplot}}
\end{figure}

Fig.~\ref{deltaplot} shows the 1 and 2$\sigma$ contours for the
combined DAE$\delta$ALUS 2-phase running, where each point represents
a hypothetical true value of $\sin^2 2\theta_{13}$ and $\delta_{CP}$.
This figure presents the results for the normal hierarchy of neutrino
masses indicated along the left axis.  DAE$\delta$ALUS is a
short-baseline experiment with an inherent ambiguity between the two
hierarchies.  For the inverted hierarchy, the corresponding value of
$\delta_{CP}$ is shown on the right axis.  The expectation for the two
hierarchies can be presented in this way because DAE$\delta$ALUS does
not suffer from matter effects which introduce apparent
$CP$-violation, unlike long-baseline experiments. This means that the
value of $\delta_{CP}$ extracted by DAE$\delta$ALUS will have an
ambiguity until the hierarchy is known.  However, DAE$\delta$ALUS can
determine if there is CP violation (i.e. $\delta_{CP} \ne 0$ or
$180^\circ$) without this input.

As described in Sec.~\ref{sse:joint},
DAE$\delta$ALUS is designed to match the $\delta_{CP}$ and
$\theta_{13}$ sensitivity of LBNE, the planned experiment which uses a
conventional long-baseline neutrino beam from FNAL to DUSEL. This
experiment quotes capability based on $30 \times 10^{20}$ protons on
the FNAL target in neutrino and antineutrino mode, respectively
\cite{Barger:2007yw}.  We assume this will be delivered at a rate
of $6\times 10^{20}$ protons on target per year \cite{GinaPrivate},
making for a 10 year run, the same length as the
DAE$\delta$ALUS run.  While LBNE is designed to run for 5 years in
neutrino mode and 5 years in antineutrino mode, its statistical
strength is in neutrino mode.  Statistics in antineutrino running are
suppressed by a low production rate of $\pi^-$ compared to $\pi^+$ and
by a reduced cross section compared to that of neutrinos.

DAE$\delta$ALUS and LBNE are complementary experiments:
\begin{itemize}
\item The DAE$\delta$ALUS signal is entirely in antineutrino mode, while
the statistical strength of LBNE is in neutrino running.
\item DAE$\delta$ALUS is a short-baseline experiment with no matter effects,
while LBNE is a long-baseline experiment with matter effects.
\item DAE$\delta$ALUS events are at low energy and in a narrow energy-window 
from 20 to 52.8 MeV, while LBNE has a high energy, wide-band (300 MeV to 
about 10 GeV) signal.
\item DAE$\delta$ALUS has very low backgrounds, coming mainly from beam-off 
sources which can be well measured from beam-off running, while LBNE has 
a poorer signal-to-background ratio, but with very different systematics.
\end{itemize}

\begin{figure}[p]\begin{center}
{\includegraphics[width=4.5in]{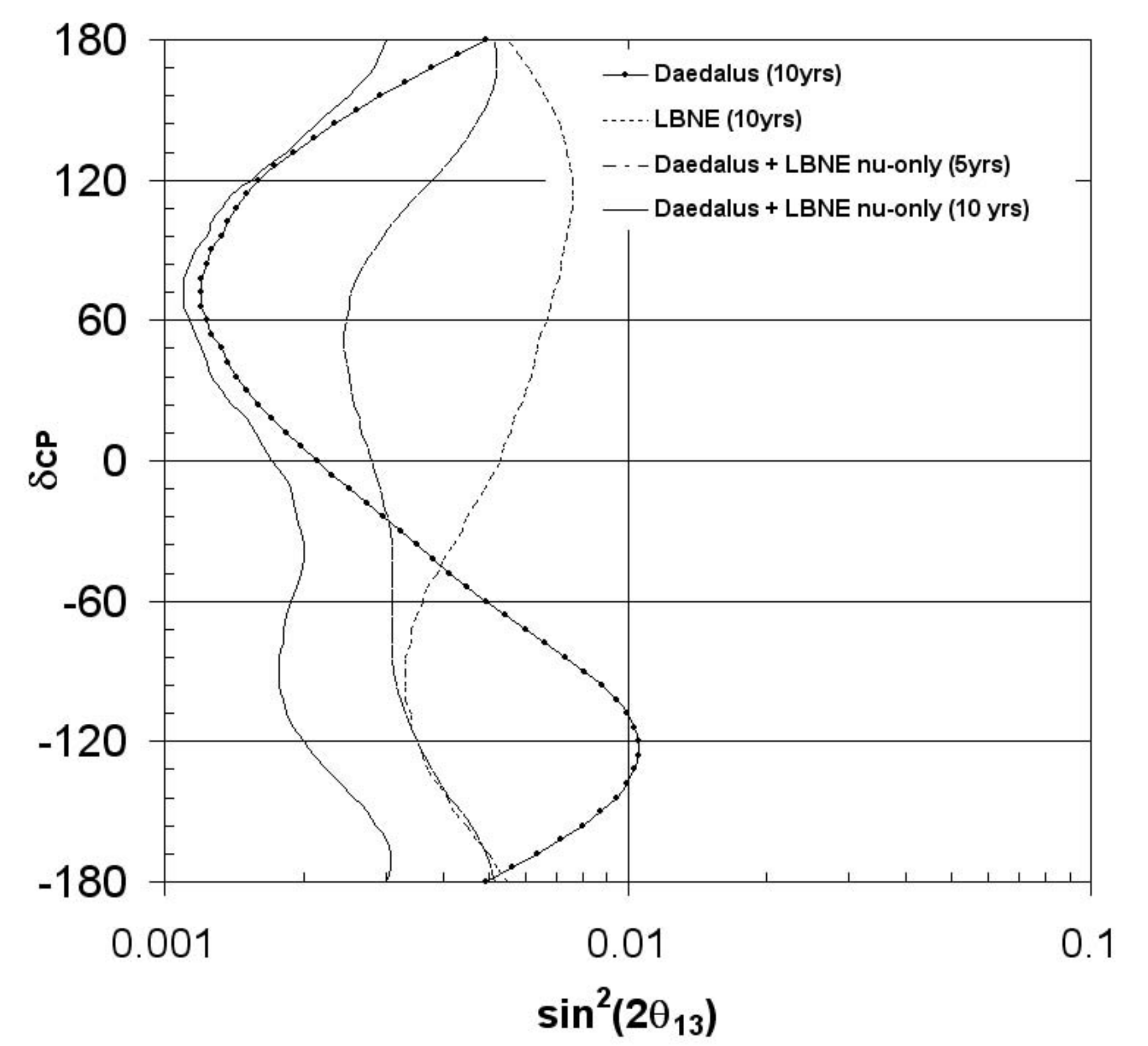}%
} \end{center}
\caption{The 3$\sigma$ confidence level sensitivity for determining a
non-zero value for $\theta_{13}$ as a function of $\sin^{2}2\theta_{13}$ and
$\delta_{CP}$. Solid-with-dots:
DAE$\delta$ALUS phase 1+2 result; Dashed: LBNE proposed running ($30\times 10^{20}$ POT in $\nu$ mode and $30\times 10^{20}$ POT in $\bar \nu$ mode);
Solid (Dot-dashed): the combined DAE$\delta$ALUS plus
LBNE $\nu$-only result for 10 years (5 years).  For the LBNE input, which 
is affected by matter effects, we assume normal hierarchy.
Details are provided in Sec.~\ref{sse:joint}.
\label{ScenarioResults_for_th13}}%
\end{figure}

\begin{figure}[p]\begin{center}
{\includegraphics[width=5.5in]{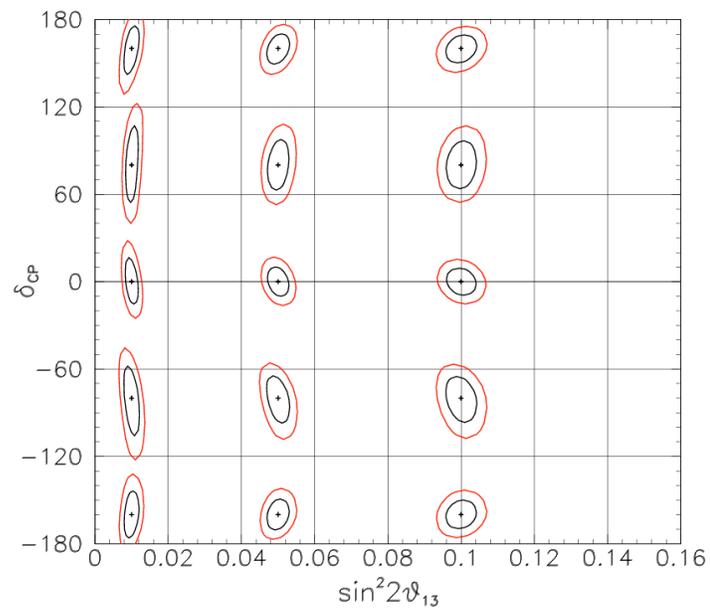}
} \end{center}
\caption{ One (inner contour) and two (outer contour) $\sigma$ sensitivities 
for the
DAE$\delta$ALUS$+$LBNE $\nu$---10 yr scenario. Normal mass hierarchy is assumed for LBNE. 
Details are provided in Sec.~\ref{sse:joint}.
\label{deltacombined}}
\end{figure}

As a result of the complementarity, when the two experiments are
combined, the sensitivity is substantially improved.  
Two scenarios which exploit the fact that LBNE's strength is in 
neutrino running are:
\begin{itemize}
\item  DAE$\delta$ALUS$+$LBNE $\nu$---$5$ yr: A five-year run of both experiments,  
combining DAE$\delta$ALUS Phase
1 with a $30\times10^{20}$ POT $\nu-$only LBNE data set.  
\item  DAE$\delta$ALUS$+$LBNE $\nu$---$10$ yr: A ten-year
run of both experiments,  with the Phase 1 + 2 DAE$\delta$ALUS sample combined
with a $60\times10^{20}$ POT $\nu$-only LBNE data sample. 
\end{itemize}
These can be compared to the standard 10-year running scenarios of
DAE$\delta$ALUS and LBNE. Fig.~\ref{ScenarioResults_for_th13} shows
the $3\sigma$ sensitivity for DAE$\delta$ALUS$+$LBNE $\nu$---$5$~yr 
and~10 yr scenarios.  The combined samples allow
exploration to very small values of $\sin^2 2 \theta_{13}$.
Fig.~\ref{deltacombined} provides the sensitivity, given the normal
hierarchy for the 10 yr scenario.  The expected errors are reduced
by a factor of two compared to running either LBNE or DAE$\delta$ALUS
alone.

\begin{figure}[p]\begin{center}
{\includegraphics[width=5.5in]{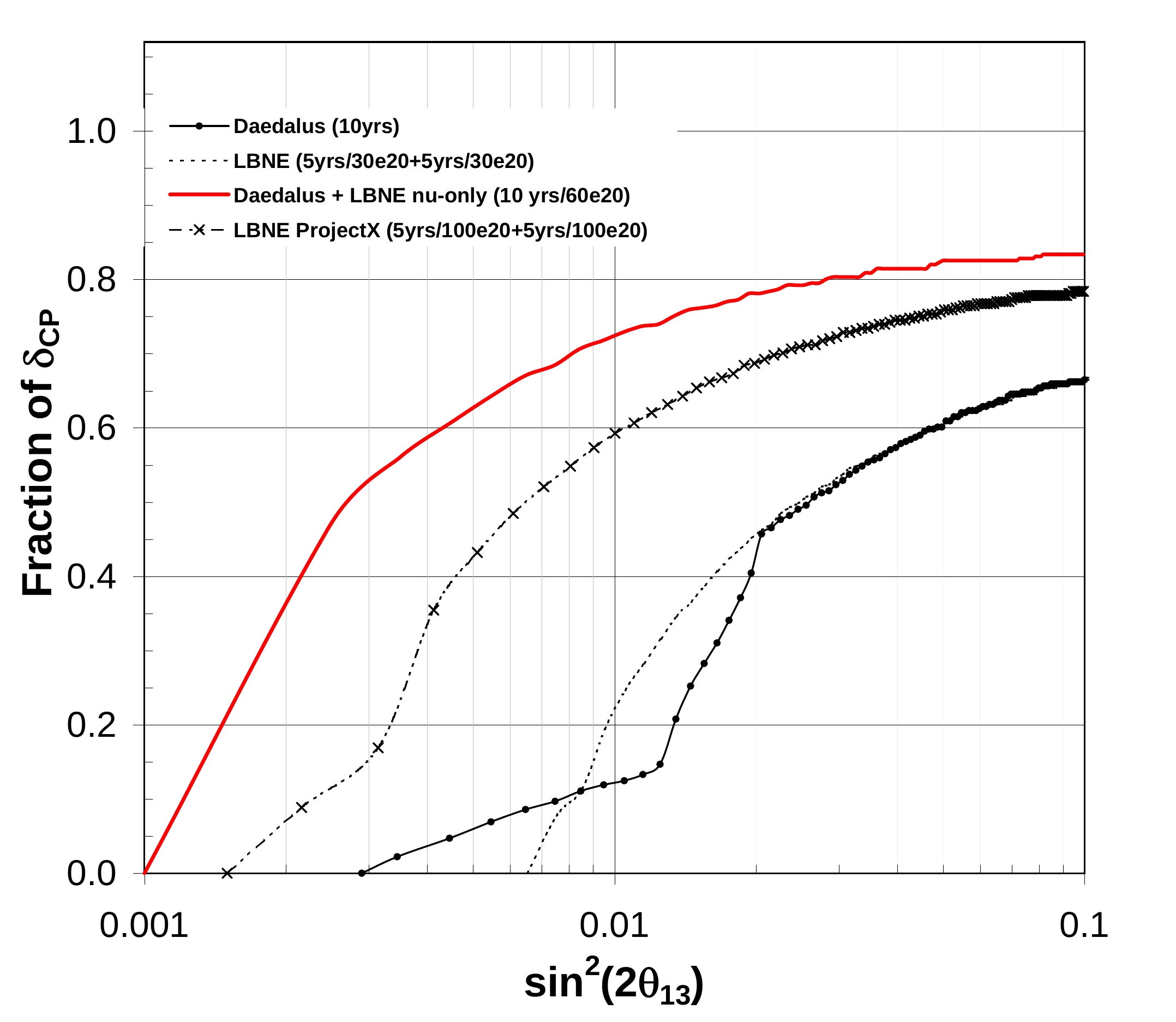}
} \end{center}
\caption{Fraction of $\delta_{CP}$ space over which a measurement can
  be differentiated from 0 or 180$^\circ$ at 3$\sigma$.  Red solid:
  Combined sensitivity for the DAE$\delta$ALUS$+$LBNE $\nu$---10 yr
  scenario.  Dashed with $\times$: Project X scenario for LBNE (5
  years of running with $100\times 10^{20}$ protons on target in
  neutrino mode followed by 5 years of running $100\times 10^{20}$
  protons on target in antineutrino mode).  Expectations for standard
  running for DAE$\delta$ALUS (solid line with dots) and LBNE dashed) are also
  shown.  Normal mass hierarchy is assumed for LBNE.  Details are
  provided in Sec.~\ref{sse:joint}.
\label{fraccoverage}}
\end{figure}

Fig.~\ref{fraccoverage} shows the strength of the
DAE$\delta$ALUS$+$LBNE $\nu$---10 yr scenario to LBNE running with
Project X.  This plot characterizes the capability of the experiments
in terms of the fraction of $\delta_{CP}$ space which is 3$\sigma$
from $\delta_{CP}=0$ or 180$^\circ$ after a 10 year run.  In the
Project X scenario, LBNE would receive $2\times 10^{21}$ POT per year,
with five years of running in neutrino mode and 5 years of running in
antineutrino mode \cite{JoachimPrivate}.  The Project X expectation is
shown by the dashed line with $\times$s.  The combined LBNE and
DAE$\delta$ALUS expectation, shown by the red line, is substantially
better than the Project X scenario for measurement of $\delta_{CP}$.\footnote{As this EOI reached its final 
draft, ref.~\cite{Agarwalla:2010nn} appeared on the arXiv, reporting the same
conclusion, though for slightly different design parameters.}
The combination also compares
favorably to second-generation super-beam facilities
\cite{Bandyopadhyay:2007kx}.

The near accelerator opens up opportunities for new
physics measurements at DUSEL, as discussed in Sec.~\ref{sse:near}.
The high intensity beam, with its well-measured flux normalization,
will allow precise measurements of neutrino cross sections in the
large water and liquid argon detectors up to 52 MeV.
This is the region of interest for supernova studies.  This beam will
also allow calibration of the large detectors.  Also, if the
accelerator is placed in proximity to the proposed 300 ft level campus,
a new program of small experiments could open up.   Potential 
sites are discussed in Sec.~\ref{se:schedule}.

We propose a Phase 0 run with only the near accelerator.  This will
allow us to learn to run the machines while at the same time allowing for
a near accelerator physics program.    This run would occur while the
water Cherenkov detector is under construction.  The installation of 
the other accelerators would be timed such that Phase 1 can begin
when the water Cherenkov detector is complete.

In summary, the 3-phase run-plan, explained in Sec.~\ref{se:schedule}, 
consists of:
\begin{enumerate}
\setcounter{enumi}{0}
\item {\bf Learn:}  Run the near accelerator to learn more
about operations, as well as to make useful preliminary cross section measurements.
\item {\bf Discover:}  Run in the 1-2-3 MW configuration to discover
the value of $\delta$, while maintaining flexibility of design.
\item {\bf Measure:}  Run for the remainder of the experiment with 
the most optimal accelerator design.
\end{enumerate}
This plan maximizes the physics capability of DAE$\delta$ALUS.  It
also dovetails well with timing of expected results on the mixing
angle $\theta_{13}$ from the reactor experiments and T2K
\cite{Mezzetto:2009cr}.  Along with a clear intellectual logic to the
plan, this three-phase design has the advantage of allowing for a
smooth funding profile over a period of about a decade.

%% file: physintro_v3.tex
The search for $CP$ violation in the light-neutrino sector is a priority
of the particle-physics community \cite{NUSAG, P5}.  Interest
has been sparked by models which invokes GUT-scale Majorana neutrino
partners which can decay, producing a matter-antimatter asymmetry in
the early universe through the mechanism of $CP$ violation
\cite{Murayama:1993em, Ma:1998dx, Davidson:2002qv}.  
This process is called ``leptogenesis.''  Observation of
$CP$ violation in the light neutrino sector would be a strong hint
that this theory is correct.  At the same time, one can argue that the
(dis)similarities of mixing in the lepton and quark sectors provide
clues about the theory at the highest energy scale, and information on
the relative level of $CP$ violation in the sectors can push this even
further (Refs. \cite{Harada:2005km,Xing:2006xa} provide examples).

DAE$\delta$ALUS, a short-baseline $\bar \nu_\mu \rightarrow \bar
\nu_e$ experiment, approaches this high-priority search in a novel
manner.  In this section, we describe the theory of $CP$
violation in the neutrino sector, with emphasis on those aspects
specific to the DAE$\delta$ALUS design.

\subsection{Introducing $CP$ Violation into the Light Neutrino Sector}

One can introduce three physically-meaningful CP-violating phases into
the model.  For convenience, these are presented within the context of
the product of two matrices: $U^{CP}= V K$.  In this case, $V$ is the
traditional 3-neutrino oscillation mixing matrix with the addition of
a $CP$ violating phase, $\delta_{cp}$:
\begin{equation}
V =
\begin{pmatrix}
c_{12}c_{13} & s_{12}c_{13} & s_{13} e^{-i\delta_{cp}} \cr
-s_{12}c_{23}-c_{12}s_{23}s_{13} e^{i\delta_{cp}} &
c_{12}c_{23}-s_{12}s_{23}s_{13} e^{i\delta_{cp}}
& s_{23}c_{13} \cr
s_{12}s_{23}-c_{12}c_{23}s_{13} e^{i\delta_{cp}} &
-c_{12}s_{23}-s_{12}c_{23}s_{13} e^{i\delta_{cp}} & c_{23}c_{13}
\end{pmatrix}.
\label{V}
\end{equation}
This is analogous to the CKM matrix of the quark sector.
The other term,
\begin{equation}
K~=~\mathrm{diag}\,(1, e^{i\phi_1},e^{i(\phi_2 + \delta_{cp})})
\end{equation}
has two further Majorana $CP$ violating phases, $\phi_1$ and $\phi_2$.

One would like to connect $\delta_{CP}$, $\phi_1$ and $\phi_2$ to $CP$
violation in heavy neutrinos at the GUT scale, since this might
motivate leptogenesis.  At present, there is no direct theoretical
argument which does this.  However, it is observed that in the
Lagrangian, these all come from a matrix of Yukawa coupling constants.
In principle, all of these phases can take on the full range of
values, including exactly zero.  However, it is difficult to motivate
a theory in which some are nonzero and some are exactly zero.  It is
expected that these parameters will either all have non-zero values or
all be precisely zero.  If the latter is the case, then the difference
between the lepton sector, with no $CP$ violation, and quark sector,
with clear $CP$ violation, must be motivated.
As a result, observation of $CP$ violation in the light neutrino
sector, through $\delta_{CP}$ or $\phi_1$ and $\phi_2$, is regarded as
the ``smoking gun'' for $CP$ violation in the heavy sector.  The
$\phi$ phases arise as a direct consequence of the Majorana nature of
neutrinos.  Therefore, in principle, the $\phi$ phase is
accessible in neutrinoless double beta decay. Even after neutrinoless
double beta decay is observed, observation of the $\phi$ phases is
expected to be extremely difficult because of uncertainties on the
Matrix Element for neutrinoless double beta decay.  Thus it seems
likely that measurement of the $\phi$ phases is in the distant
future.  On the other hand, $\delta_{cp}$, the ``Dirac'' $CP$-violating 
term in $V$ may be accessible through oscillation experiments
such as DAE$\delta$ALUS.

\subsection{$CP$ Violation in Neutrino Oscillations}

The parameter $\delta_{CP}$ is accessible through the muon-to-electron
neutrino flavor oscillation probability.  For oscillations in a vacuum, 
this is given by
\cite{Nunokawa:2007qh}:
\begin{align}
P_{\mu\rightarrow e} & =\sin^{2}\theta_{23}\sin^{2}2\theta_{13}\sin^{2}%
\Delta_{31} \nonumber\\
& \mp\sin\delta\sin2\theta_{13}\sin2\theta_{23}\sin2\theta_{12}\sin^{2}%
\Delta_{31}\sin\Delta_{21} \nonumber \\
& +\cos\delta\sin2\theta_{13}\sin2\theta_{23}\sin2\theta_{12}\sin\Delta
_{31}\cos\Delta_{31}\sin\Delta_{21} \nonumber\\
& +\cos^{2}\theta_{23}\sin^{2}2\theta_{12}\sin^{2}\Delta_{21} 
\label{equ:beam}
\end{align}
where $\Delta_{ij}=\Delta m_{ij}^{2}L/4E_{\nu}$.  In the second term,
the $-(+)$ refers to neutrino (antineutrino) running.  Traditionally,
searches for $CP$ violation through non-zero $\delta_{CP}$ have relied on
comparing neutrino and antineutrino oscillation probabilities,
exploiting this change of sign in order to isolate $\delta_{CP}$.
However, DAE$\delta$ALUS will be a search only in antineutrino mode:
$\bar \nu_\mu \rightarrow \bar \nu_e$, and so in our case, only the
$+$ sign applies.  Sensitivity to $CP$ violation comes about through the
interference between $\Delta_{12}$ and $\Delta_{13}$ transitions,
which has a distinctive $L$ dependence that we will exploit.
Eq.~\ref{equ:beam} depends on many parameters.   However, most are well-known.  
In Table~\ref{OscPar}, we provide a summary of the present level 
of knowledge of the parameters, as well as the improvement expected 
in the future, where ``NIN'' (No Improvement Needed) indicates
that any future improvement does not impact the DAE$\delta$ALUS analysis.
Along with $\delta_{CP}$, which is the focus of DAE$\delta$ALUS, two
other parameters, $\theta_{13}$ and the sign of $\Delta m_{31}^2$, are
unknown. 

With respect to $\theta_{13}$, global fits report a non-zero
value at the $\sim 1\sigma$ level \cite{Schwetz:2008er,Fogli:2009ce}.
This parameter drives the amplitude for the $CP$ violating terms in
Eq.~\ref{equ:beam} and therefore sets the level of technical
difficulty for observing $CP$ violation.  There is a clear road-map
toward discovery and measurement of $\theta_{13}$ within the present
neutrino program \cite{Mezzetto:2009cr} which dovetails well in time
with the DAE$\delta$ALUS search.   DAE$\delta$ALUS data will provide 
some constraint on $\sin^2\theta_{13}$; however, the external data remain
important to the analysis.     Because there is a dependence on
both $\delta_{CP}$ and $\sin^2 2\theta_{13}$, the DAE$\delta$ALUS
sensitivity must be expressed in a $\delta_{CP}$-$\sin^2 2\theta_{13}$
plane, as shown in Fig.~\ref{deltaplot}.

The unknown sign of $\Delta m_{31}^2$, referred to as ``the mass
hierarchy,'' will lead to an inherent degeneracy in the
DAE$\delta$ALUS analysis.  From Eq.~\ref{equ:beam}, one can see that
for the DAE$\delta$ALUS $\bar \nu_\mu \rightarrow \bar \nu_e$ search,
the probabilities for the combination ($\delta_{CP}$, $\mathop{\mathrm{sign}}(\Delta
m_{31}^2)=+1$) and ($180-\delta_{CP}$, $\mathop{\mathrm{sign}}(\Delta m_{31}^2)=-1$) are
equal.  In the text below, we will refer to 
$\mathop{\mathrm{sign}}(\Delta m_{31}^2)=+1$ as the ``normal hierarchy'' and 
$\mathrm{\mathrm{sign}}(\Delta m_{31}^2)=-1$ as the ``inverted hierarchy.''   
Given that the degeneracy is perfect, we can express the DAE$\delta$ALUS
sensitivity for normal and inverted hierarchy on the same plot.   
In Fig.~\ref{deltaplot}, the left vertical axis assumes the normal hierarchy
and the right vertical axis assumes the inverted hierarchy.
\begin{table}[tbp] \centering
%
\begin{tabular}
[c]{c|ccc|ccc}\hline
Parameter & Present: &  &  & Assumed & Future:  & \\
& Value& Uncert. & Ref. & Value& Uncert. & Ref.\\
&      & $(\pm)$ &  &          &  $(\pm)$ &   \\ \hline
$\Delta m_{21}^{2}\times 10^{-5} {\rm eV}^{2} $ & 7.65 & 0.23 &
\cite{Schwetz:2008er} & 7.65 & NIN & NIN\\

$\Delta m_{31}^{2} \times 10^{-3} {\rm eV}^{2}$ & 2.40 & 0.12 &
\cite{Schwetz:2008er} & 2.40 & 0.02 & \cite{Schwetz:2005jr}\\
$\sin^{2}(2\theta_{12})$ & 0.846 & 0.033 & \cite{Schwetz:2008er} & 0.846 & NIN &
NIN\\
$\sin^{2}(2\theta_{23})$ & 1.00 & 0.02 & \cite{Schwetz:2008er} & 1.00 & 
0.005 & \cite{Huber:2009cw}\\
$\sin^{2}(2\theta_{13})$ & 0.06 & 0.04 & \cite{Fogli:2009ce} & 0.05 & 
0.005 & \cite{McConnel:2004bd}\\\hline
\end{tabular}
\caption{Left: Present values and uncertainties for oscillation parameters, reported in the listed
 references. Right:  Future expectations used in this study, based on assumptions from 
 the associated references.  $NIN$ means ``No Improvement Needed'' for the DAE$\delta$ALUS analysis -- the present values are sufficiently precise.}\label{OscPar}
\end{table}%
\subsection{The Mass Hierarchy and Matter Effects}

External information will be required in order to break the mass
hierarchy degeneracy in DAE$\delta$ALUS.  
There are two sources of this information.  First, the
next generation neutrinoless double beta decay experiments, when
combined with neutrino mass measurements from cosmology or direct
searches, can, in principle, demonstrate the inverted mass hierarchy,
if neutrinos are Majorana \cite{Schonert:2010zz}.  On the other hand,
if no signal is seen, one does not know whether the hierarchy is
normal or if neutrinos are not Majorana.  A second approach is to use
``matter effects'' in muon-to-electron-neutrino, long-baseline,
oscillation searches.  These occur because neutrino and antineutrino
beams will have a different forward-scattering amplitude as the beam
propagates through the earth.  The result is an ``effective
$CP$ violation'' --- a difference in the rate of neutrino versus
antineutrino oscillations which arises from some effect other than
$CP$ violation in the lepton-$W$ coupling.  These arise regardless of
whether neutrinos are Majorana in nature or not.

Matter effects result in a modification of   
Eq.~\ref{equ:beam}:
\begin{eqnarray}
P_{\mathrm{mat}}  && =  \nonumber \\
&&\sin^{2}\theta_{23}\sin^{2}2\theta_{13}\frac{\sin
^{2}\left(  \Delta_{31}\mp aL\right)  }{\left(  \Delta_{31}\mp aL\right)
^{2}}\Delta_{31}^{2} \nonumber \\
&& \mp\sin\delta\sin2\theta_{13}\sin2\theta_{23}\sin2\theta_{12}\sin\Delta
_{31}\frac{\sin\left(  \Delta_{31}\mp aL\right)  }{\left(  \Delta_{31}\mp
aL\right)  }\Delta_{31}\frac{\sin\left(  aL\right)  }{\left(  aL\right)
}\Delta_{21} \nonumber \\
&& +\cos\delta\sin2\theta_{13}\sin2\theta_{23}\sin2\theta_{12}\cos\Delta
_{31}\frac{\sin\left(  \Delta_{31}\mp aL\right)  }{\left(  \Delta_{31}\mp
aL\right)  }\Delta_{31}\frac{\sin\left(  aL\right)  }{\left(  aL\right)
}\Delta_{21} \nonumber\\
&& +\cos^{2}\theta_{23}\sin^{2}2\theta_{12}\frac{\sin^{2}\left(  aL\right)
}{\left(  aL\right)  ^{2}}\Delta_{21}^{2}  \label{Pmatter}.
  \end{eqnarray}
  In this equation, $a=\frac{G_{F}N_{e}}{\sqrt{2}}$ 
and  $\mp$ refers to neutrinos
  (antineutrinos).  The same terms which are modified by the matter
  effects also will depend upon $\mathop{\mathrm{sign}}(\Delta m_{31}^2)$.  Because of
  this, the matter effects provide sensitivity to the mass hierarchy.
  Matter effects only appear when $L$ is large, because $a \approx (
  3500\text{ km}) ^{-1}$, for $\rho
  Y_{e}=3.0\text{ g/cm}^{3}$, is small.  Short-baseline experiments, such
  as DAE$\delta$ALUS, suffer negligible matter effects, and this is
  even true of moderate baseline experiments such as T2K
  \cite{McConnel:2004bd}, which is at 295 km.  However the new
  generation of proposed long baseline beams, starting with No$\nu$A
  \cite{Ayres:2004js}, at 730 km, and moving on to longer baselines in
  Japan \cite{Minakata:2007as} and the US \cite{Barger:2007yw}, at
  $>1000$ km, will be sensitive to matter effects.

Matter effects are interesting in their own right. However, if one
seeks to study $CP$ violation in the lepton-$W$ coupling in the same
experiment, they have to be measured and removed.  This is challenging
if $\sin^2 2\theta_{13}$ is small, as one can see from
Eq.~\ref{Pmatter}.  Also, if the hierarchy is inverted, the neutrino
oscillation probability is suppressed, substantially reducing the
statistics for neutrino oscillations.  The result is that the mass
hierarchy may be difficult to determine using this method, unless
$\sin^2 2\theta_{13}$ is large.

In summary, DAE$\delta$ALUS will have negligible matter effects.  As a
result, our experiment has sensitivity to $CP$-violation without
complications of an extra source of effective $CP$ violation.  This
offers a more straight-forward method of discovering and measuring
$\delta_{CP}$.  However, it means that a degeneracy with the mass
hierarchy will remain until $\mathop{\mathrm{sign}}(\Delta m_{31}^2)$ is measured in
some other experiment.

\subsection{Oscillation Probability Calculations}

The code used for calculating the oscillation probabilities presented
in this EOI was written by Stephen Parke of Fermi National Accelerator
Laboratory.  The calculation uses the equations provided in
Ref.~\cite{Nunokawa:2007qh}.  The full formula, including matter
effects, is used in all probability calculations presented in this
EOI.

The collaboration has also implemented the DAE$\delta$ALUS neutrino
fluxes in the GLoBES neutrino oscillation simulation program
\cite{Huber:2007ji}.  We have verified that the probability
calculations from the Parke code match the calculations from GLoBES.
In the near future, we plan to contact the GLoBES authors in order
to include these fluxes in the open-source GLoBES package.

%% file: CP_v3.tex
The primary goal of DAE$\delta$ALUS is to search for $CP$ violation in
a complementary manner to the present plans \cite{McConnel:2004bd,
 Barger:2007yw}.  The present suite of experiments is based on
long-baseline ``conventional'' neutrino beams.  These beams are
produced by impinging high-energy protons on a target, resulting in
pions and kaons which are sign-selected and focused by a magnet in the
direction of a neutrino detector located $\sim 1000$ km away.
$CP$ violation is explored by comparing the rates of $\nu_\mu
\rightarrow \nu_e$ to $\bar \nu_\mu \rightarrow \bar \nu_e$ These
experiments are hampered by a lack of antineutrino statistics and by
poor signal-to-background ratio.  These experiments have the added
complication of matter effects.

Given the high priority placed on a convincing measurement of
$\delta_{CP}$, should it be nonzero, we were motivated to develop a
design which addresses these issues.  We propose to use pion
decay-at-rest beams, which produce muon antineutrinos peaked at 50
MeV, at three locations at short baseline to a large detector.  This
design will provide a high-statistics data sample which explores $CP$
violation through the $L$ dependence of the interference terms of
Eq.~\ref{equ:beam}.  The measurement is novel in that it is done with
antineutrinos exclusively, while all existing proposals rely most
heavily on neutrino data. The antineutrino flux uncertainties are
different from the flux uncertainties of conventional beams and are
well-controlled. Because of the low beam energy, the interaction
systematics are also different from those of the present
program. Varying $L$, while employing a single detector, is novel for
an appearance experiment (though it has been successfully employed at
KamLAND \cite{Araki:2004mb} and Super-K \cite{Fukuda:1998ah} 
for disappearance) and reduces systematics. A two-phase
program which allows an optimized measurement strategy is powerful and
potentially cost-saving.

This chapter very briefly describes the design of the experiment.
This is followed by an extended discussion of event types, backgrounds
and systematics in order to justify the sensitivity which is presented
in Fig.~\ref{deltaplot}.  More information on design specifics and
issues is then provided in the following chapters.
\subsection{Overview of the DAE$\delta$ALUS Design}

DAE$\delta$ALUS searches for $\bar \nu_\mu \rightarrow \bar \nu_e$
oscillations using neutrinos from three stopped-pion, decay-at-rest
(DAR) beams, which interact in the 300 kton, Gd-doped water Cherenkov
detector at DUSEL.  This results in low systematics for the 
beam and detector. The shape of the DAR flux with energy is known to high
precision and is common among the various distances, thus shape
comparisons will have small uncertainties. The neutrino flux from the
three distances is accurately determined from the direct measurement
of the $\pi^{+}$ production rate using neutrino-electron scattering 
events from the near
accelerator. The interaction and detector systematic errors are low
since all events are detected in a single detector.  The
neutrino-electron cross section for normalization and inverse-beta-decay 
cross section for the signal are well-known. The fiducial
volume error on the IBD events is also small due to the extreme
volume-to-surface-area ratio of the ultra-large detector.

\begin{figure}[t]\begin{center}
{\includegraphics[width=4.5in]{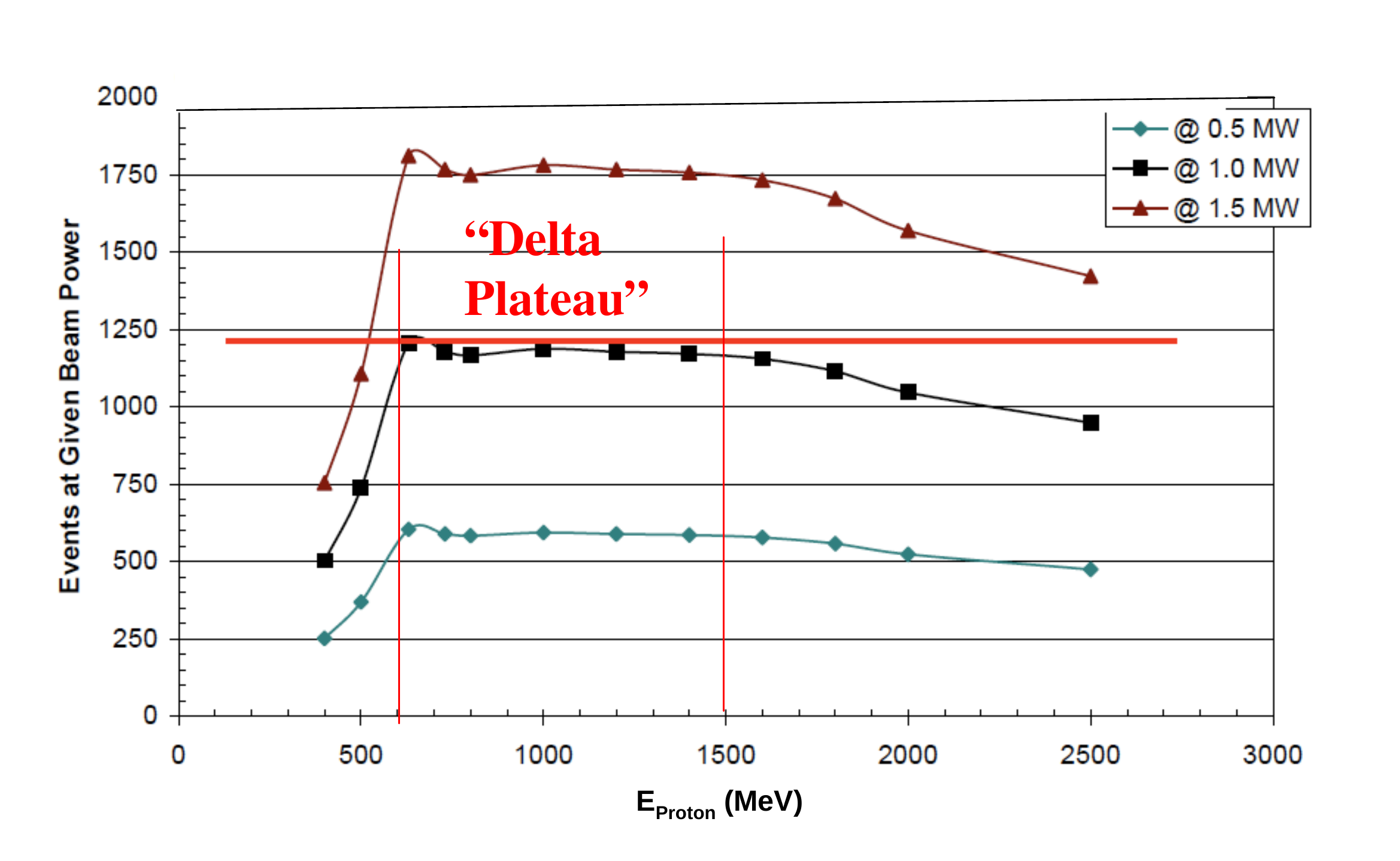}} 
\end{center}
\vspace{-0.25in}
\caption{$\bar \nu_\mu$  production (arbitrary units) as a function of proton energy for 
0.5, 1.0 and 1.5 MW on target
\label{plateau} }
\end{figure}

The beams are to be produced by proton accelerators with kinetic
energy in the 650 MeV to 1.5 GeV range.  In this range, the $\Delta$
resonance dominates $\pi^+$ production and little energy is lost to
the production of other particles, such as neutrons or $\pi^-$s.  As a
result, in this energy range, for a given total power on
target, the number of neutrino events produced by a DAR beam is flat
as a function of proton energy; see Fig.~\ref{plateau}.

Our proposal is to use low-cost cyclotrons, which are under development 
for commercial use, to provide the proton beams.    Development of high 
intensity cyclotrons which are $\sim 1$ GeV is being driven by interest
in accelerator driven systems (ADS) for thorium reactors and active 
interrogation for homeland security.  The work builds upon the 
recent development of cyclotrons for medical uses.   These cyclotrons 
are generally of two types:  $\sim 30$ MeV and high power for PET isotope 
production and $\sim 250$ MeV and low power for cancer therapy.  The evolution
from medical machines to high powered, $\sim 1$ GeV machines does require
further development, and there are none on the market today.   However,
we have found three groups interested in developing and commercializing 
cyclotrons which will satisfy our needs over the next few years.  Machine options are 
discussed in Chapter~\ref{sse:accels}.

\begin{figure}[t]\begin{center}
{\includegraphics[width=5.5in]{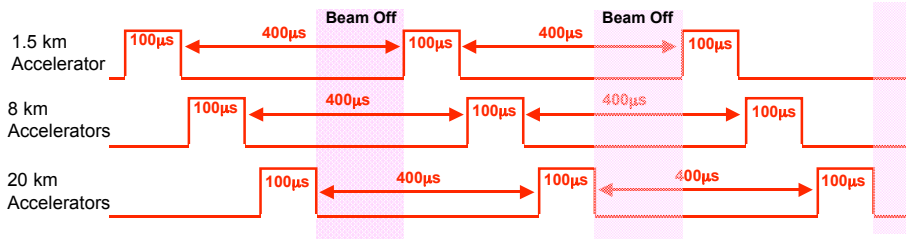}} 
\end{center}
\vspace{-2.5in}
\caption{The source accelerator for events is identified through timing.
We propose to run each accelerator with a 20\% DF.   Beam-off time allows
measurement of non-beam-backgrounds.  The timing structure presented here
is illustrative, and longer intervals are not an issue for the analysis.
\label{timing} }
\end{figure}

Neutrinos are produced through the decay chain  
\begin{eqnarray}
\pi^+ \rightarrow  \nu_\mu & \mu^+ & \nonumber \\
& \hookrightarrow & ~ \bar \nu_\mu e^+ \nu_e, \\
\end{eqnarray}
with a flux shown in Fig.~\ref{fluxplot}.  The shape of this energy
distribution is defined by the weak interaction, but the overall
normalization may vary between beam stops. We describe how we
normalize the accelerator sources using events in the detector in the
discussion below.  This beam is isotropic, so the orientation of the
accelerator with respect to the detector does not affect the
fluxes of these three flavors.   On the other hand, $\bar \nu_e$ is
produced at the $\sim 10^{-4}$ level, as discussed below, from
the chain of $\pi^-$ which decay in flight.  This background
can be significantly reduced by pointing the primary beam
at 120$^\circ$ or more with respect to the detector \cite{OscSNS}.
The $\bar \nu_e/\nu_e$ ratio also can be suppressed by
designing the beamstops with a low-$A$ target surrounded by a high-$A$
absorber \cite{Burman:1989dq, Burman:1996gt}.    
The oscillated neutrinos are detected through the inverse-beta-decay
(IBD) interaction, $\bar \nu_e + p \rightarrow e^+ + n$.  This
interaction has a high cross section at $\sim 50$ MeV, but requires a
detector with a high fraction of free protons.  The large water
Cherenkov detector at DUSEL, which is proposed to be 300 ktons, is
ideal for this.  Should DUSEL choose to install a large liquid-scintillator 
module, this could also be used, though we do not present
details here.  The large liquid-argon detectors cannot be used for
this analysis because there are no free proton targets.

In order to differentiate the signal IBD interactions from $\nu_e$
interactions, Gd-doping of the detector is required
\cite{Beacom:2003nk}. Fortunately, interest in studying supernova
relic neutrinos and in using the large detector for non-proliferation
studies has pushed forward techniques for Gd-doping water
\cite{Watanabe:2008ru, Dazeley:2008xk, Kibayashi:2009ih}.  Gd-doping
is within the scope of the S4 development. We discuss the issue of
Gd-doping further in Chapter~\ref{sse:detector}.

The accelerators will be positioned at 1.5, 8, and 20 km from the large
water Cherenkov detector.  The 1.5-km, ``near accelerator'' position is
4850 feet from the detector --- that is, the accelerator remains above
ground.  The complexity of running the accelerators is sufficiently
high to make above-ground running desirable.  This accelerator will be
on the DUSEL site.  The 1.5-km accelerator allows measurement of the
beam-on backgrounds and the normalization.  The 8-km site is at an
oscillation wavelength of about $\pi/4$ at 50 MeV and the 20-km site is
at oscillation max for this energy.  These two accelerators are
referred to as the ``signal accelerators,'' below.  The location for
these accelerators are off the DUSEL site and will need to be negotiated.
We have in mind other mining sites which are no longer in use, but which
have access roads and power.

We plan to run each site for 20\% of the time; see Fig.~\ref{timing}.
This allows us to use the time-stamp to identify which events come
from each accelerator site.
IBD events do not have a strong angular dependence \cite{Vogel:1999zy},
and so event-pointing cannot be used to connect events to a given accelerator.
This also allows time for beam-off running, in order to measure non-beam-related
backgrounds.   Our initial proposal is to run each accelerator for 100 $\mu$s,
with 400 $\mu$s beam-off.   However, other timing patterns can be considered,
up to intervals of minutes and even hours.   During the interval that 
a given accelerator is on, it is run continuously (``$cw$'').
We propose a phased running plan over 10 years.  In Phase 1, we begin
with 1 MW, 2 MW and 3 MW at the respective 1.5-km, 8-km, and 20-km
sites.  The purpose of this run is to discover $CP$-violation at the
$3\sigma$ level, which is a 5-year run.  At that point, more
accelerators can be added in the proportion which is most advantageous
to measuring the signal.  Fig~\ref{deltaplot} is for a 2-phase run
where the second 5-year period has 1, 2 and 7 MW at the three
respective sites.

\subsection{Events in the Detector}
\label{eventsindetector}

\subsubsection{Neutrino Interactions of Interest to the Analysis}

\bigskip%
\begin{table}[tbp] \centering
\begin{tabular}
[c]{l|ccc}\hline
Event Type & 1.5 km & 8 km & 20 km\\\hline
IBD Oscillation Events (E$_{vis}>20$ MeV) &  &  & \\
$\delta_{CP}=0^{0}$, Normal Hierarchy & 763 & 1270 & 1215\\
\quad\quad" \quad, Inverted Hierarchy & 452 & 820 & 1179\\
$\delta_{CP}=90^{0}$, Normal Hierarchy & 628 & 1220 & 1625\\
\quad\quad" \quad, Inverted Hierarchy & 628 & 1220 & 1642\\
$\delta_{CP}=180^{0}$, Normal Hierarchy & 452 & 818 & 1169\\
\quad\quad" \quad, Inverted Hierarchy & 764 & 1272 & 1225\\
$\delta_{CP}=270^{0}$, Normal Hierarchy & 588 & 870 & 756\\
\quad\quad" \quad, Inverted Hierarchy & 588 & 870 & 766\\\hline
IBD from Intrinsic $\overline{\nu}_{e}$ (E$_{vis}>20$ MeV) & 600 & 42 & 17\\
IBD Non-Beam (E$_{vis}>20$ MeV) &  &  & \\
\multicolumn{1}{r|}{atmospheric $\nu_{\mu}p$ \textquotedblleft invisible
muons\textquotedblright} & 270 & 270 & 270\\
\multicolumn{1}{r|}{atmospheric IBD} & 55 & 55 & 55\\
\multicolumn{1}{r|}{diffuse SN neutrinos} & 23 & 23 & 23\\\hline
$\nu-$e Elastic (E$_{vis}>10$ MeV) & 21570 & 1516 & 605\\\hline
$\nu_{e}-$oxygen (E$_{vis}>20$ MeV) & 101218 & 7116 & 2840\\\hline
\end{tabular}
\caption{Event samples for the combined two-phase run for $\sin^2 2\theta_{13}=0.05$ and
parameters from Table \ref{OscPar} (future).
}\label{events_123_127}%
\end{table}%
\begin{figure}[t]
{\hspace{-0.35in}\includegraphics[width=6.25in]{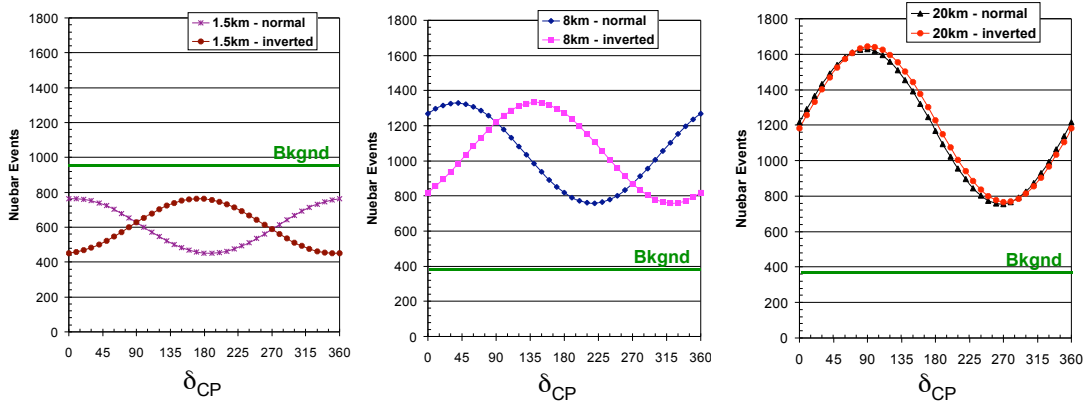}} 
\vspace{-2.25in}
\caption{The oscillation event distribution as a function of $\delta_{CP}$ from each accelerator for the 10 year run, assuming $\sin^2 2\theta_{13}=0.05$.   The green line indicates the level of background for each data set.
 \label{wiggle} }
\end{figure}

\begin{figure}[t]\begin{center}
{\includegraphics[width=4.5in]{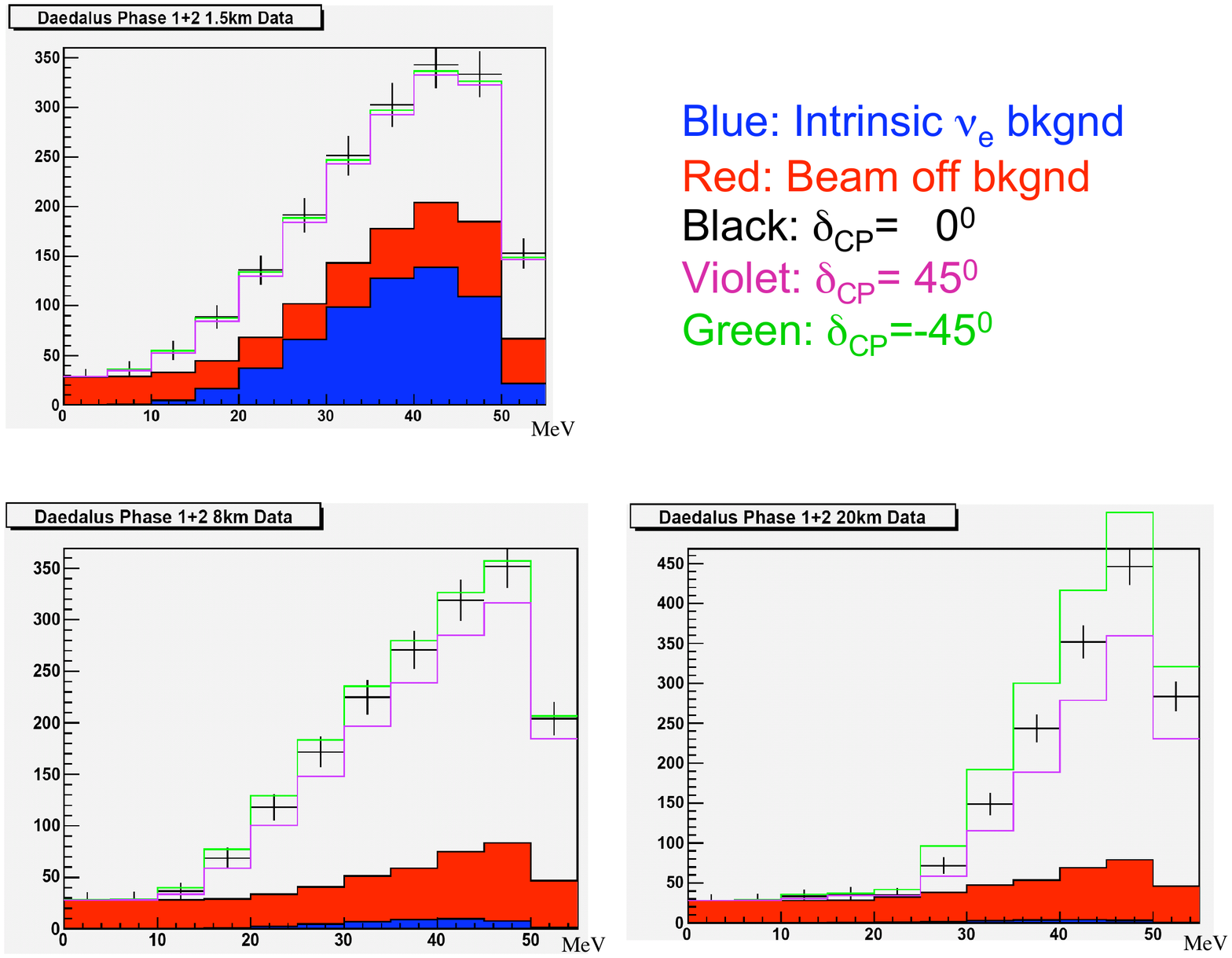}} 
\end{center}
\caption{The event energy distributions for signal and background 
at $\sin^2 2\theta_{13}=0.04$.   Black, green and violet histograms
show signals for $\delta_{cp}=0$, 45$^\circ$ and -45$^\circ$.  The
blue histogram shows the intrinsic $\bar \nu_e$ beam-on background. The
red histogram shows the beam-off backgrounds.   Top row:  the near
accelerator events.   Bottom row: events in the signal accelerators
\label{eventsbkgds} }
\end{figure}

There are three important event types in this analysis: neutrino-electron
scatters, $\nu_e$-oxygen scatters, and IBD Events.    The event rates
for the 10 year run are shown in Table~\ref{events_123_127}.
The energy ranges for the analysis are listed below.
The lower energy bound for the event samples is chosen to maintain a
region of high signal-to-background.  The upper bound is just above
the 52.8 MeV endpoint of the neutrino spectra.

\paragraph{Neutrino-electron scatters} During the 10-year run,
21.5k events will be collected from the near accelerator.  Because the
cross section for these events is known to better than $1\%$
\cite{Conrad:2004gw}, these events can be used to obtain a precise
measure of the near-accelerator flux normalization.  The
neutrino-electron analysis uses events in the visible energy range of
10 to 55 MeV.  Neutrino-electron scatters can be separated from
$\nu_e$-oxygen scatters by angular cuts
\cite{Auerbach:2001wg,Haxton:1988mw}.  We estimate the reconstruction
efficiency of these events as $\epsilon_{\mathrm{recon}}=75\%$ based on
Ref.~\cite{Auerbach:2001wg}
\paragraph{$\nu_e$-oxygen scatters} During the 10-year run a large
sample of these events are collected from all three accelerators.  The
cross section is suppressed compared to the IBD cross section, because
the target nucleons are bound \cite{Lazauskas:2007bs, Haxton:1987kc}.
The cross section is not well-known, and we use a parameterization
from Ref.~\cite{Lazauskas:2007bs}, which gives the smaller predicted
data set.  (We note that an outcome of DAE$\delta$ALUS will be a precise
measurement of this cross section from the near accelerator data set,
as discussed in Sec.~\ref{sse:near}).  The sample from the 20-km
accelerator is on the order of 2800 events, implying a 2\% statistical
error, for visible energy between 20 and 55 MeV.  Comparing the rates
between the three accelerators, adjusting for the $1/r^2$ dependence
of the flux, allows determination of the normalization across the
sites.  We estimate the reconstruction
efficiency of these events as $\epsilon_{\mathrm{recon}}=75\%$.

\paragraph{IBD Events} These are the signal events for $\bar
\nu_\mu \rightarrow \bar \nu_e$ oscillations.  The number of events,
therefore, depends upon the oscillation parameters.  Fig.~\ref{wiggle}
shows how the event rate from the three accelerators varies for the 10
year run, as a function of $\delta_{CP}$, assuming $\sin^2
2\theta_{13}=0.05$.  One can see that the design leads to roughly
equal signal samples from the 8-km and 20-km accelerator.  One can also
see that the 1.5-km accelerator will have signal events, and this
must be included in the fits.  Events in the energy range of 20 to 55
MeV will be used in the analysis, and the distribution of events is
strongly peaked at about 50 MeV; see Fig.~\ref{eventsbkgds}.  As a
result, DAE$\delta$ALUS may be thought of as a narrow-band beam
experiment.  In this energy range, three types of interactions must be
considered. First is the IBD signal, with an estimated reconstruction
efficiency of $\epsilon _{recon}=67\%$, based on studies for Super-K
\cite{Watanabe:2008ru}.

\subsubsection{Background Events}

{\begin{figure}[t]\begin{center}
{\includegraphics[width=4.5in]{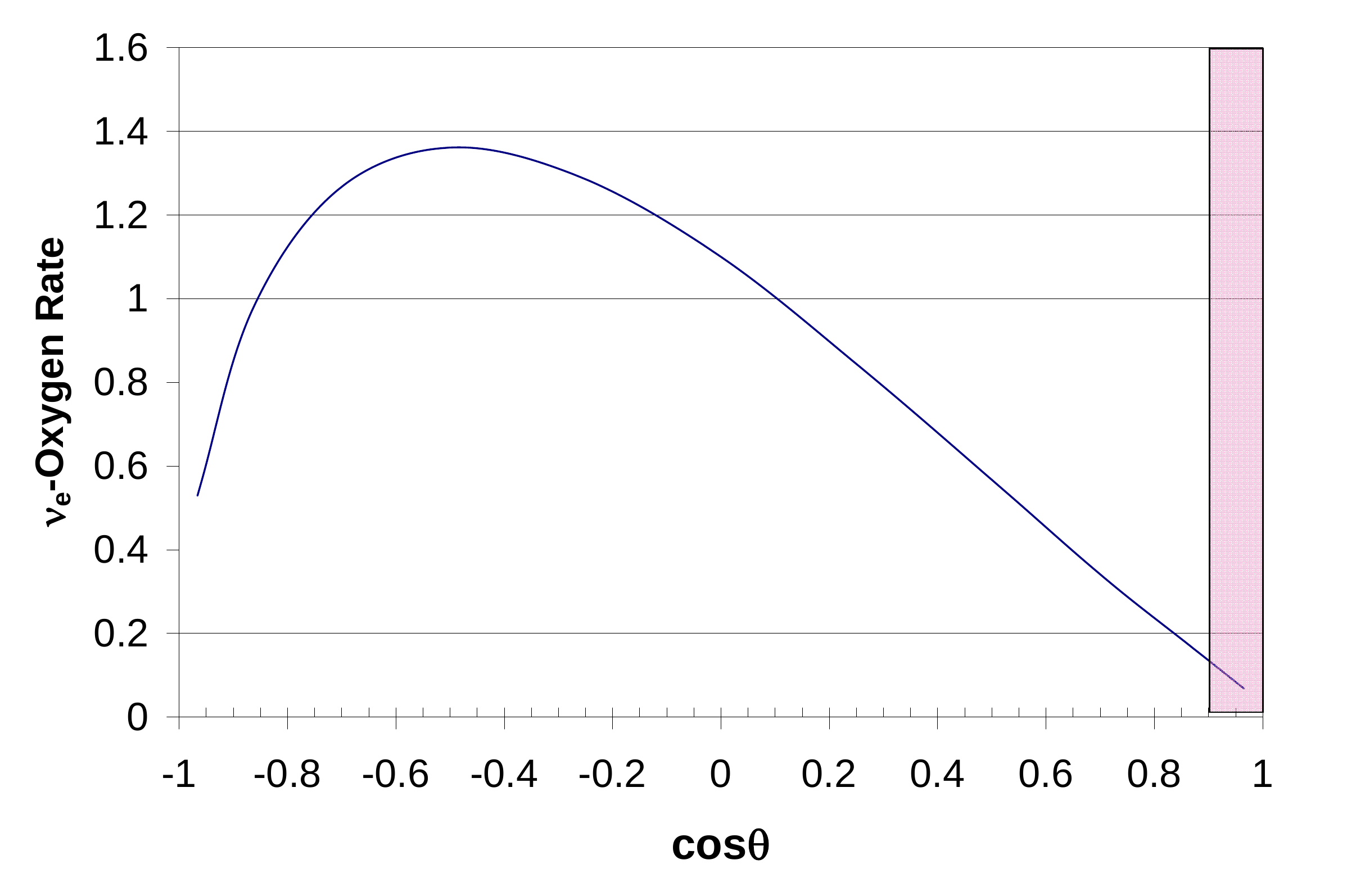}} 
\end{center}
\caption{Rate of $\nu_e$-oxygen events as a function of scattered
electron angle.  The shaded region, with $\cos\theta > 0.90$, is the 
region which the neutrino-electron scatters will populate.
\label{nueO} }
\end{figure}
The total background from each accelerator is indicated by the green
lines on Fig.~\ref{wiggle}.  The energy distribution of the
backgrounds compared to the expected events for $\sin^2
2\theta_{13}=0.04$ and three values of $\delta_{CP}$ are shown in
Fig.~\ref{eventsbkgds}.  One can see that the near accelerator (left
in Fig.~\ref{wiggle} and top in Fig.~\ref{eventsbkgds}) provides a
high-statistics sample for the beam-on background measurement.
Because the isotropic beam produces backgrounds that fall as 
$1/r^2$, the signal accelerators have very little beam-on background.
In the signal accelerators, the beam-off background dominates.

The beam-on backgrounds are dominated almost entirely by intrinsic
$\bar \nu_e$ events in the beam.  In principle, correlated beam-on
backgrounds may be produced by $\nu_e$ charged current (CC) scatters
in which a neutron is emitted.  This is of special concern for
interactions with $^{17}$O and $^{18}$O, which have natural abundances
of 0.04\% and 0.20\%, respectively.  We find that the rate of neutron
production is negligible from excited $^{16}$F \cite{Tilley16},
$^{17}$F \cite{Ajzenberg17}, and $^{18}$F \cite{Ajzenberg18}.
However, excited $^{18}$F will decay to nitrogen plus an $\alpha$.
Data from the ($\alpha$, n) interaction
on oxygen
indicate about $1 \times 10^{-8}$ neutrinos per 3 MeV $\alpha$
\cite{alphan}.  Hence, we expect this background to be negligible.
The accidental backgrounds arise from the $\nu_{e}$ in the beam that
are followed by a neutron-like event.  This background is estimated to
be very small using the measurements from the Super-K Gd-doping study
\cite{Watanabe:2008ru}.  As a result, in Figs.~\ref{wiggle} and
\ref{eventsbkgds} we only show the intrinsic $\nu_e$ background, as
the blue histogram.  However, in the final data analysis all 
background sources are included.

Beam-off backgrounds arise from atmospheric $\nu_{\mu}p$ scatters with
muons below Cherenkov threshold that stop and decay (``invisible
muons''), atmospheric $\bar \nu_e$ IBD events, and supernova relic
neutrinos. These are all examples of correlated backgrounds.  The
rates of these correlated backgrounds are scaled from analyses for the
GADZOOKS experiment \cite{Beacom:2003nk}.  As, this study was done for the
latitude of Super-K, not for DUSEL, in the future the level of
this background will need to be adjusted; however, this study is 
sufficient for 
illustrative purpose in this EOI.  The interaction rates of the
beam-off backgrounds will be well-measured during the 40\% beam-off
running fraction.  Thus the dominant beam-off error is the statistical
error from the bin-by-bin background subtraction.
\subsection{Systematic Errors Before Constraints \label{sys}}

\bigskip%
\begin{table}[tbp] \centering
\begin{tabular}{|l|c|}\hline
\multicolumn{2}{|c|}{Systematics on the Normalization} \\ \hline
{\it On all events } & {\it Fractional Uncertainty} \\ \hline
$\pi^+$ production & 0.100 \\ \hline
{\it Neutrino-electron scattering events}  & {\it Fractional Uncertainty} \\ \hline
2.1\% energy scale uncertainty & 0.010 \\
Cross section error from NuTeV $\sin^2 \theta_W$ & 0.005 \\
Background subtraction systematics & 0.000 \\ \hline 
{\it $\nu_e$-oxygen scattering} & {\it Fractional Uncertainty }\\ \hline
Cross section & 0.100 \\ \hline \hline
\multicolumn{2}{|c|}{Systematics on the Signal} \\ \hline
{\it Oscillation IBD events} & {\it Fractional Uncertainty} \\ \hline
Efficiency of neutron detection & 0.005 \\
Fiducial volume & 0.000 \\ \hline \hline 
\multicolumn{2}{|c|}{Systematics on the Backgrounds} \\ \hline
{\it Intrinsic $\bar \nu_e$ IBD Events} & {\it Fractional Uncertainty} \\ \hline
$\pi^-$ production & 0.100 \\
$\pi^-$ decay-in-flight & 0.100 \\
$\mu^-$ decay-in-flight & 0.050 \\ \hline \hline
{\it Beam-off subtraction} & {\it Fractional Uncertainty} \\ \hline
Statistical error for Phase 1 period & 0.054\\
Statistical error for Phase 2 period & 0.038\\ \hline \hline
\end{tabular}
\caption{Systematic errors which are inputs to the fits.  The systematics
 are then further constrained by the event samples in the fit.
}\label{syslist}%
\end{table}%

In the analysis, initial systematic uncertainties are assigned to
quantities, which are then further constrained by fits to the
DAE$\delta$ALUS data set.  Tab~\ref{syslist} provides the inputs for
the systematics in the fit.  The resulting final systematic error, 
after constraints, is at
the level of 2\%.

Because the flux chain begins with $\pi^+$ production, this systematic
error is common to all types of events in the normalization and
signal samples.   This uncertainty is not well known, but becomes highly
constrained when we use the neutrino-electron sample to set the 
overall normalization.   Thus the analysis is insensitive to the level
we assume as an input systematic uncertainty, which we take to be
10\% here.

For the neutrino-electron sample, the largest systematic uncertainty comes 
from knowledge of the energy scale at $E_{\mathrm{vis}}=10$ MeV, the position of 
the cut in this analysis.  This estimate comes from Super-K
\cite{Ashie:2005ik} and leads to a 1\% error on the DAR flux
normalization.  However, we think that this is an overestimate.  As we
discuss in Sec.~\ref{calib}, the energy dependence of the
neutrino-electron sample is very well predicted.  Demanding that the
ratio of measured to predicted energy distribution be flat should
allow for energy calibration to better than 1\%.  The next largest
error is due to the error on the cross section normalization from our
knowledge of $\sin^2 \theta_W$.  The LEP and SLAC results provide a
precision measurement of $\sin^2 \theta_W$ which is in $3\sigma$
disagreement with the NuTeV deep inelastic neutrino scattering results
(for discussion see Ref.~\cite{Adams:2008cm}).  Taking an agnostic
viewpoint, we assign a systematic error which covers this
disagreement, leading to the cross section error in
Table~\ref{syslist}.

The analysis relies on obtaining a pure sample of neutrino-electron
scatters.  The $\nu_e$ events on oxygen and IBD events with a missing
neutron represent potential sources of background.  However, these can
be separated from the neutrino-electron sample since the angular
distribution of neutrino-electron events is very forward-peaked, while
$\nu_e$-oxygen scatters have a broad distribution \cite{Haxton:1988mw}, as
shown in
Fig.~\ref{nueO}.  Only 0.8\% of the $\nu_{e}-$oxygen events
have $\cos\theta>0.90$ ({\it i.e.}, $\theta < 25^\circ$), as shown by
the shaded region, 
giving a less than 4\% background to the IBD
sample.  These backgrounds can be measured in the $\cos\theta<0.90$
region and extrapolated into the neutrino-electron signal region,
allowing them to be subtracted with
negligible error, as we explicitly list in Table~\ref{syslist}.
\bigskip%
\begin{table}[t] \centering
\begin{tabular}
[c]{c|ccccc}\hline
sin$^{2}2\theta_{13}$/$\delta_{CP}$ & --180 & --90 & 0 & 90 & 135\\\hline
0.01 & 52.5 & 47.2 & 29.0 & 38.8 & 48.5\\
0.05 & 21.6 & 21.9 & 19.5 & 25.6 & 30.4\\
0.09 & 18.3 & 19.9 & 17.6 & 23.8 & 26.3\\\hline
0.01 & 51.3 & 45.5 & 26.9 & 36.7 & 46.9\\
0.05 & 19.9 & 21.2 & 18.2 & 24.2 & 29.6\\
0.09 & 16.8 & 19.2 & 16.5 & 22.6 & 25.3\\\hline
\end{tabular}
\caption{The 1$\sigma$ measurement uncertainty on $\delta_{CP}$ for various values of
$\sin^{2}\theta_{13}$ for the combined two-phase data.
Top:  Systematic and statistical errors.
Bottom: Statistical error only.}\label{deltaCP_123_127}%
\end{table}%

The largest input uncertainty on the IBD signal events arises from
neutron-tagging and is taken to be 0.005.  We also considered a fiducial
volume error.  Because the volume-to-surface area is so high in
these large detectors, we find this error to be negligible.

The largest systematics associated with the intrinsic
$\bar \nu_e$ background are related to the $\pi^-$ production
and decay chain.   However the final analysis is
insensitive to these effects for two reasons.  First, the near accelerator
provides a high precision {\it in-situ} measure of the background.
Second, because of the $1/r^2$ dependence of this flux in the signal
accelerators, the rate is small (see blue histograms in
Fig.~\ref{eventsbkgds}, bottom).

The beam-off rates are measured rather than predicted.  Therefore
the systematic error associated with this source comes from the statistical 
error on the background subtraction.
\subsection{Oscillation Sensitivities}
\begin{figure}[t]
\begin{center}
{\includegraphics[width=4.in]{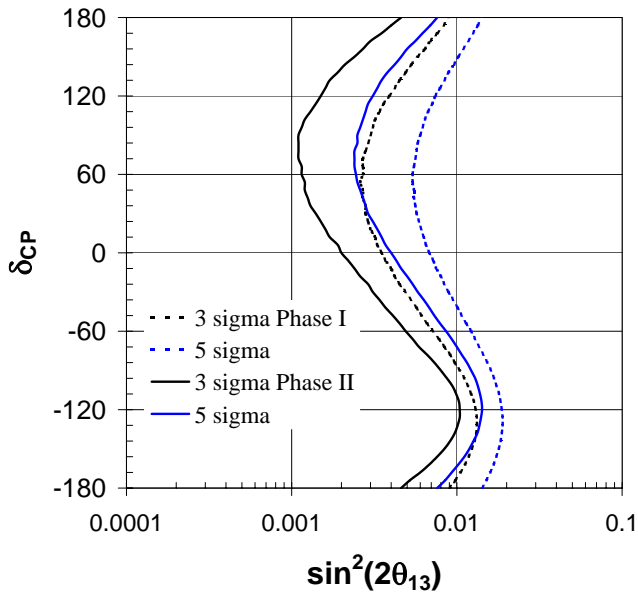} \vspace{-1.25in}
}
\end{center}
\vspace{-0.5in}
\caption{Phase-1 and Combined-phase sensitivity to $\theta_{13} \neq0$ at
3$\sigma$ and 5$\sigma$. }%
\label{thetaplot}
\end{figure}}

\begin{figure}[t]\begin{center}
{\includegraphics[width=5.in]{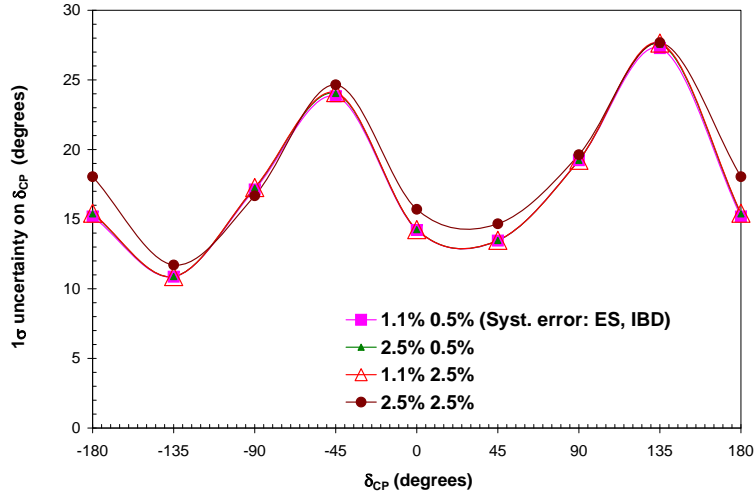}}
\end{center}
\caption{   The DAE$\delta$ALUS result is insensitive to the 
input systematic errors.
In this study,  we vary the magnitude of the input systematic uncertainties
(these are the systematic uncertainties prior to the fit).  For clarity,
we consider only two of the errors ES and IBD (see text for details).  
\label{exploresys} }
\end{figure}

\begin{figure}[t]\begin{center}
{\includegraphics[width=5.in]{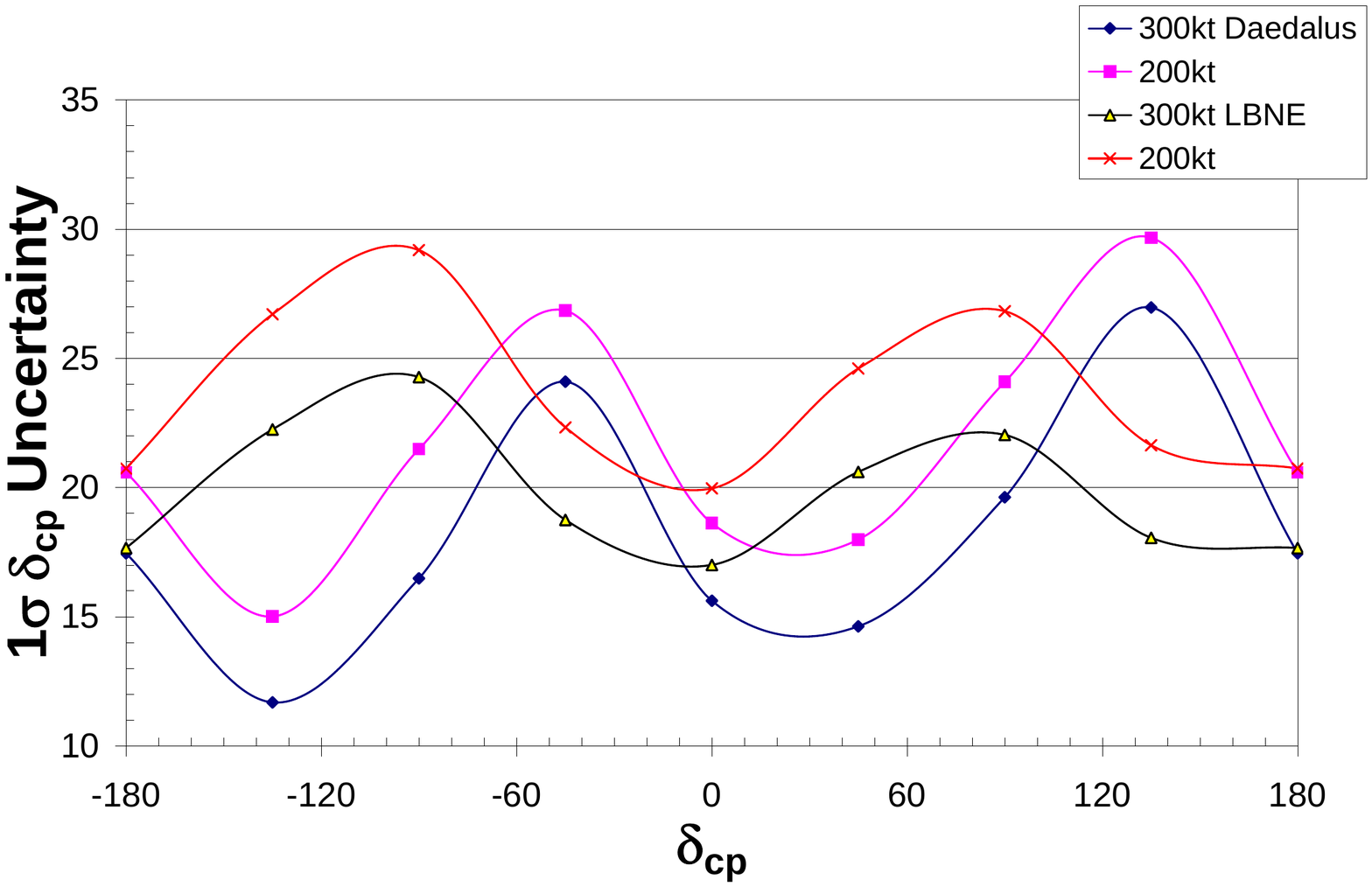}} 
\end{center}
\caption{The DAE$\delta$ALUS result is sensitive to the
statistical uncertainty in the experiment.  In this study, we show
how the analysis result changes with 
detector mass.  Blue (Magenta) --- DAE$\delta$ALUS 300 (200) kton, Black (Red) 
--- LBNE 300 (200) kton.
\label{explorestat} }
\end{figure}

Sensitivity estimates were made using a method similar to that described in 
Ref.~\cite{McConnel:2004bd}.   Data are
generated according to the experimental expectations with assumed
underlying parameters.  For a given set of new parameters, a 
standard $\chi^2$
value is found by comparing the prediction with these parameters to
the originally generated data. The difference between this $\chi^2$
value and the one calculated with the original parameters is the
$\chi^2$ value.  Systematic uncertainties are  
constrained by pull-term contributions
of the form $\left( k_{i}-1\right) ^{2} /\sigma_{i}^{2}$, where
$\sigma_{i}^{2}$ are the uncertainties given in 
Table~\ref{OscPar}, for the oscillation parameters 
and Table~\ref{syslist}, for the experimental expectations. 
The $\chi^2$ minimization 
is performed using the MINUIT program \cite{James:1975dr}.

The DAE$\delta$ALUS sensitivity is shown in Fig.~\ref{deltaplot}.
Information on the sensitivity is also tabulated in
Table~\ref{deltaCP_123_127} (top), while Table~\ref{deltaCP_123_127}
(bottom) tabulates the sensitivity considering only the statistical uncertainty.Thus, one
can see that the DAE$\delta$ALUS measurement is statistics-limited.
Another way to present the DAE$\delta$ALUS sensitivity is to consider
the sensitivity for observing a non-zero value for $\theta_{13}$ at
the 3 and 5$\sigma$ confidence level (CL) for Phase-1 (5-year) running and Phase-1+2 (10-year) 
running.  This result depends upon $\delta_{CP}$ because the
level of $CP$ modulates the number of events in the sample (see
Fig.~\ref{wiggle}).  The DAE$\delta$ALUS expectation is shown in
Fig.~\ref{thetaplot}.  This sensitivity meets that of LBNE, but is
inverted with respect to its $\delta_{CP}$ dependence
\cite{Barger:2007yw}.  This result is because DAE$\delta$ALUS is an
antineutrino experiment while the strength of LBNE is in its neutrino
data set.

\begin{figure}[t]\begin{center}
{\includegraphics[width=5.in]{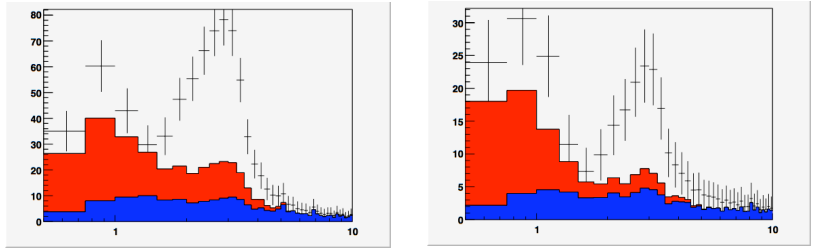}} 
\end{center}
\vspace{-1.75in}
\caption{Expected events, as a function of energy in GeV, for the LBNE
experiment with a 300 kton water Cherenkov detector at 1300 km for
$\nu~(5~{\mathrm{yr}}) +\bar \nu~(5~{\mathrm{yr}})$ running (see text for details). Black
points --- rate with
statistical error for $\sin^{2}2\theta_{13}=0.04$, $\delta_{CP}=0^{\circ}$,
and a normal hierarchy. Red --- total background.  Blue ---
intrinsic electron-flavor neutrino background.
Left: $\nu_\mu \rightarrow \nu_e$ running;  Right: $\bar \nu_\mu \rightarrow \bar \nu_e$ running.
}%
\label{LBNE_events}%
\end{figure}

The final result, after fitting, is insensitive to the input
systematic errors.  This is because the constraints from the fit to
$\nu$-electron and $\nu_e$-oxygen events from the three accelerators
are strong.  We illustrate this in Fig.~\ref{exploresys}, which
presents the 1$\sigma$ uncertainty on $\delta_{CP}$ from the final fit
as a function of the value of $\delta_{cp}$.  The magenta curve (with
squares) shows the result using the standard input errors (reported in
Tab.~\ref{syslist}). To examine the sensitivity of the result to the
magnitude of each input uncertainty, we arbitrarily increase each
systematic error, in turn.  For example, the effect of increasing the
elastic scattering (ES) rate error from 1.1\% to 2.5\% is indicated in
Fig.~\ref{exploresys} by the green line (with solid triangles).  In
fact, the new fit result, is so similar to the original fit (magenta
line) that it is difficult to see on the plot.  This is because the
data strongly constrains the fit, so that the input systematic error
is not very important.  As a second example, we return the ES error
back to its original value of 1.1\% and increase the error on IBD
event rate from 0.5\% to 2.5\%.  The resulting new fit, indicated by
the red line (with open triangles), is also in excellent agreement
with the original (magenta line) result.  In order to see a small but
clear effect in Fig. \ref{exploresys}, one must substantially increase
both the ES and the IBD errors to 2.5\%.  This slightly weakens the
constraint from the data and the result is the brown curve (with
closed dots).

In contrast to the case of the systematic error, the DAE$\delta$ALUS
result is very sensitive to the statistical errors in the experiment.
In order to illustrate this sensitivity, Fig.~\ref{explorestat}
compares the DAE$\delta$ALUS 1$\sigma$ errors in $\delta_{CP}$ for the
300 kton detector to a 200-kton detector.  Even at 200 ktons,
DAE$\delta$ALUS maintains $>3\sigma$ capability across a wide range of
values of $\delta_{CP}$.  For comparison, LBNE is less affected by a
loss of statistics, because it is  limited by systematics.  However, the
$>3\sigma$ coverage at 200 ktons is less.

%% file: joint_v5.tex
{\begin{figure}[p]%
\vspace{-1.in}
\centering
\includegraphics[
width=3.5in
]%
{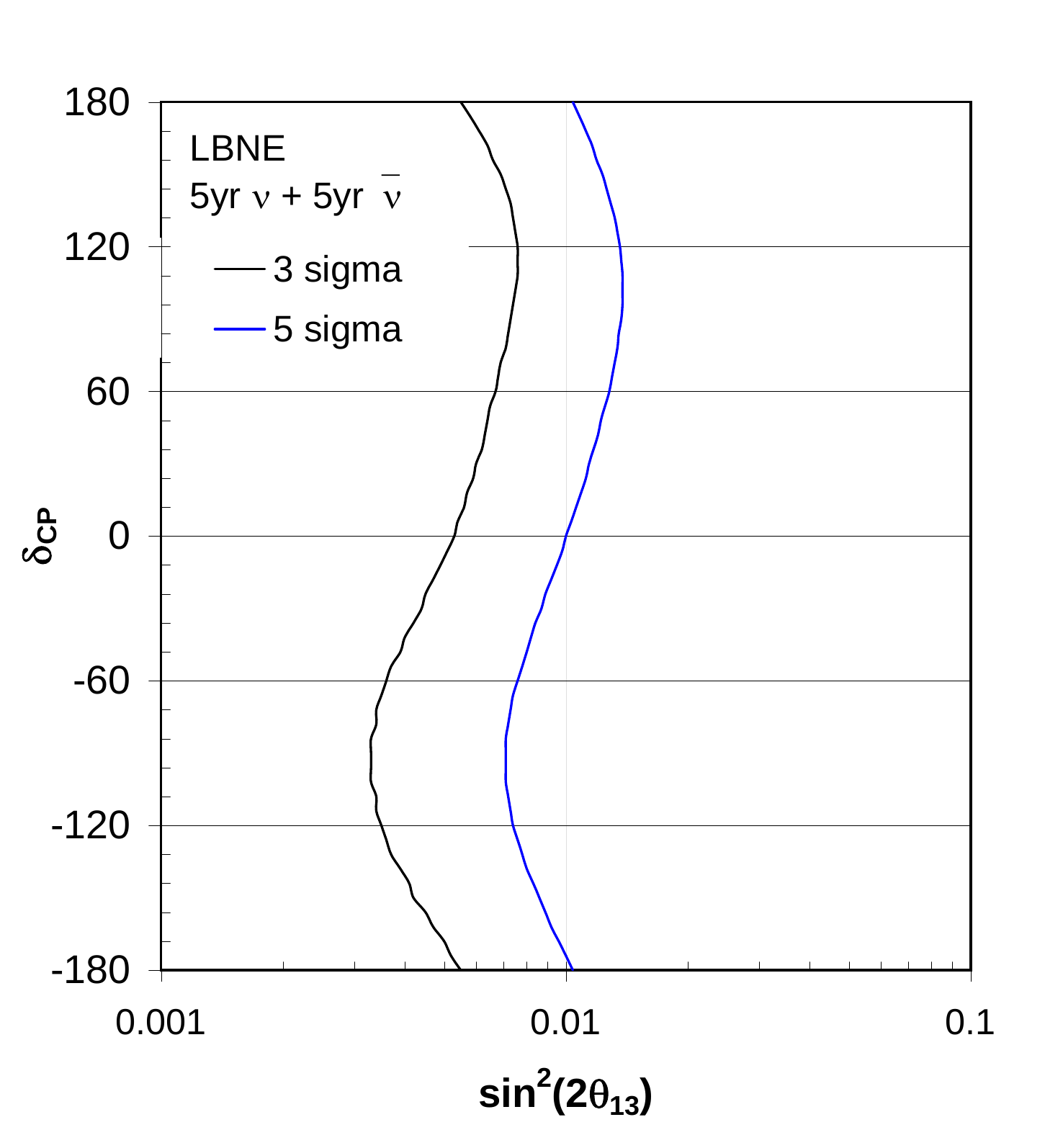}
\caption{The 3$\sigma$ and 5$\sigma$ confidence-level exclusion limits for
determining a non-zero value for $\theta_{13}$ as a function of $\sin
^{2}2\theta_{13}$ and $\delta_{CP}$, normal hierarchy, assuming  
$\nu~(5~\mathrm{yr}) +\bar \nu~(5~\mathrm{yr})$,
with the efficiency and uncertainties given in the text.%
\label{LBNE_only_th13}}%
~\\
\centering
\includegraphics[
width=3.5in
]%
{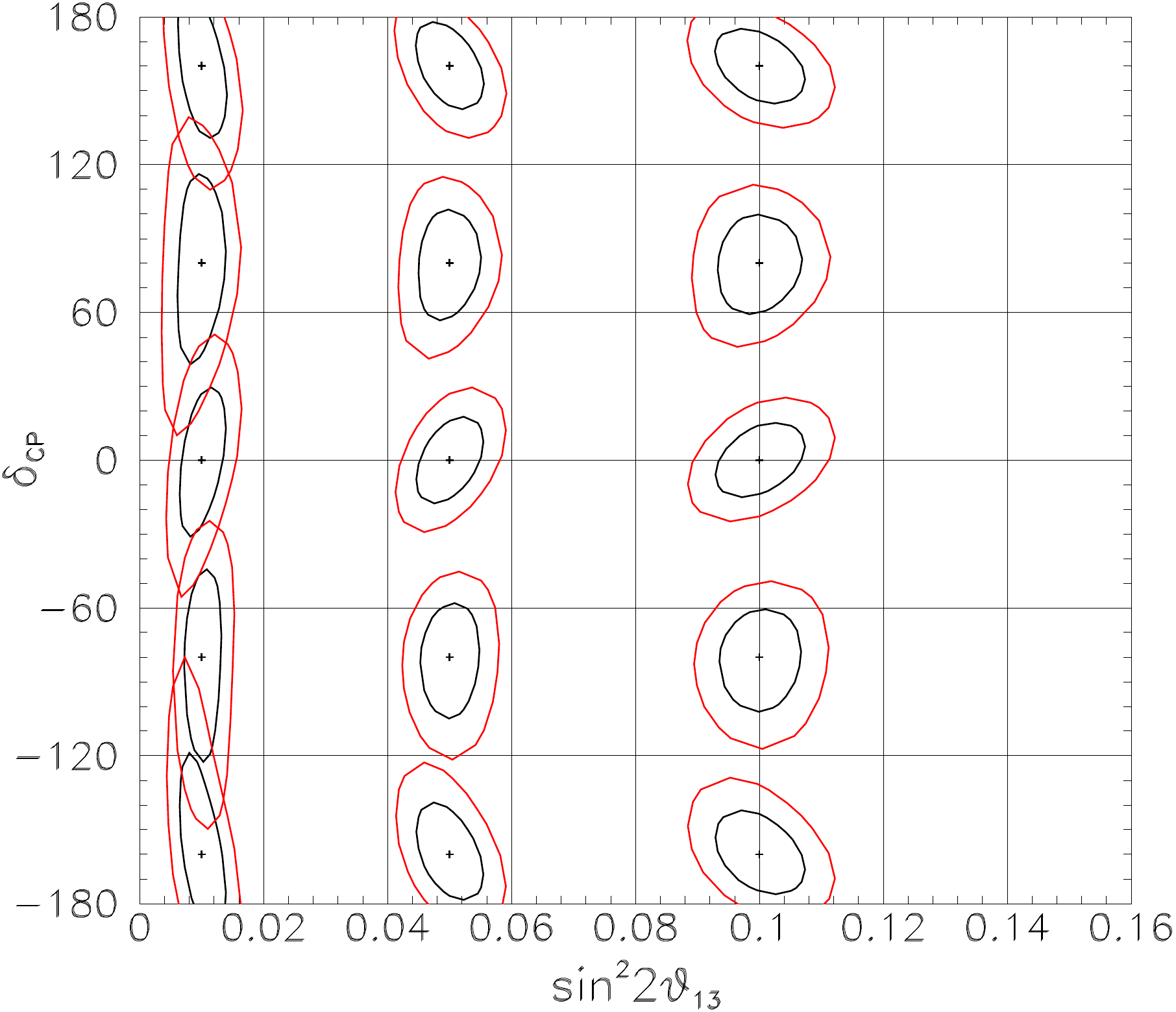}
\caption{Estimates of the uncertainty for a correlated measurement
of $\sin^{2}2\theta_{13}$ and $\delta_{CP}$ at 1$\sigma$ and 2$\sigma$,
normal hierarchy,
assuming  $\nu~(5~\mathrm{yr}) +\bar \nu~(5~\mathrm{yr})$.  The efficiency and
uncertainties are given in the text.  Sensitivity is for the optimized 120-GeV beam, 
discussed in the text. %
\label{LBNE_only_dcp_vs_th13}}%
\end{figure}}

The DAE$\delta$ALUS high-statistics antineutrino data can be combined
with LBNE neutrino-only running data to give a sensitivity for
observing and measuring $\theta_{13}$ and $\delta_{CP}$ which far
exceeds either experiment alone, as well as the Project X expectation
\cite{JoachimPrivate}, and approaches that of the most sensitive
superbeam facilities \cite{Bandyopadhyay:2007kx}. 
In this section, we describe the presently planned LBNE sensitivity,
which is based on equal neutrino and antineutrino running.  
For simplicity, we use the normal hierarchy as our example.  We then
explain what is gained by running LBNE in neutrino-only mode
simultaneously with DAE$\delta$ALUS.  The results presented here are
in agreement with Ref.~\cite{Agarwalla:2010nn}, once differences in
the proposed design are taken into account.

\subsection{LBNE Current Plans: $\nu~(5~\mathrm{yr}) +\bar \nu~(5~\mathrm{yr})$ 
Running}

The current plans for the long baseline neutrino experiment from
Fermilab to DUSEL (LBNE) are as follows.  LBNE proposes to use a wide
band neutrino beam, in the range of about 300 MeV to about 10 GeV
\cite{MaryPrivate}.  This allows the LBNE beam, which is produced at a
distance of 1300 km from the detectors at DUSEL, to potentially observe
both the oscillation maximum at about 3 GeV and the second maximum
near 0.9 GeV.  To be specific, the LBNE flux files used in this
discussion are:
\begin{itemize}
\item dusel120e250i002dr280dz1300km\_flux.txt (neutrino flux) 
\item dusel120e250ni002dr280dz1300km\_flux.txt (antineutrino flux)
\end{itemize}
This is a 120-GeV, proton-on-target, on-axis, NuMI-like beam with a 280-m
decay, designed to reduce the high-energy tail above the first
oscillation maximum.  This optimization is important because high-energy 
neutrinos produce single and multiple neutral-current $\pi^{0}$
mesons, which decay to produce electromagnetic showers that are the
primary misidentification background in a beam of this type.

We refer to the standard LBNE run as 
``$\nu~(5~\mathrm{yr}) +\bar \nu~(5~\mathrm{yr})$.'' 
This is defined to be
$3\times10^{21}$ POT in neutrino mode and $3\times10^{21}$ POT in
antineutrino mode.   If Project X is not in place for the initial
run of LBNE, which is planned to begin in 2021, then Fermilab
is expected to provide $6\times 10^{20}$ POT per year.   As a 
result,  this running will require 5 years in 
neutrino mode and 5 years in antineutrino mode.   

Figure~\ref{LBNE_events} shows the event rates for LBNE $\nu~(5~yr)
+\bar \nu~(5~yr)$ for the 300 kton water Cherenkov detector.  The
points show the expected signal for $\sin^2 2\theta_{13} = 0.04$, and
$\delta_{CP} = 0^\circ$ for neutrinos (left) and antineutrinos (right).
As seen from these figures, the mis-identification background,
shown in red, is substantial, especially for the
lower energy second maximum. In addition, there is also a sizeable intrinsic
electron (anti-)neutrino contamination, shown in blue,  
under the first maximum peak.

An issue for the LBNE $\nu~(5~\mathrm{yr}) +\bar \nu~(5~\mathrm{yr})$ plan 
is that the antineutrino event rate is low and has a very poor
signal-to-background ratio.  The rates are low due to the low production
rate of the $\pi^{-}$ mesons from the primary protons and because the
antineutrino cross section is a factor of two lower than the neutrino
cross section at 3 GeV. The antineutrino analysis is further
complicated and diluted by the high neutrino content of the beam
during antineutrino running. For example, in the region between 2 and
5 GeV, 20\% of the oscillation events are from
$\nu_{\mu}\rightarrow\nu_{e}$ instead of
$\overline{\nu}_{\mu}\rightarrow\overline{\nu}_{e}$.

\begin{figure}[pt]%
\centering
\includegraphics[
width=5.5in
]%
{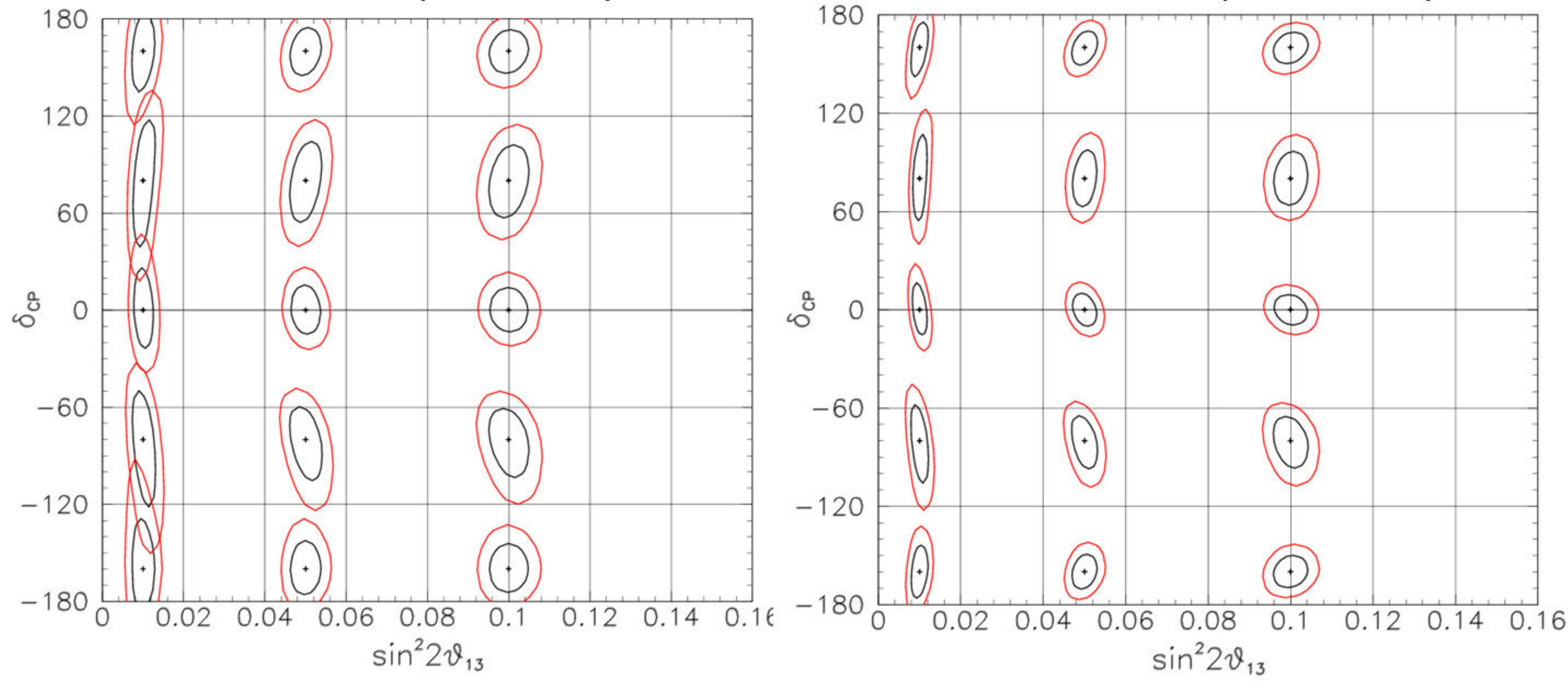}%
\caption{The estimated measurement uncertainty for a correlated measurement of
$\sin^{2}2\theta_{13}$ and $\delta_{CP}$ at $1\sigma$ and $2\sigma$ for the
DAE$\delta$ALUS-plus-LBNE, $\nu$-only scenario. Compares a 5-year (left) and
10-year (right) run. Normal mass hierarchy is assumed for LBNE.  (Right plot is the same as Fig.~\ref{deltacombined}).%
\label{Combined_Scenario_dcp_vs_th13}}%
\end{figure}

In order to estimate the sensitivity for measuring $\theta_{13}$ and
$\delta_{CP}$ for the $\nu~(5~\mathrm{yr}) +\bar \nu~(5~\mathrm{yr})$ scenario, we
assume a 15\% reconstruction efficiency and systematic
uncertainties of 1\% for the neutrino/antineutrino flux, 10\% for the
background, and 5\% for the density along the flight path.  With these
assumptions, the $\theta_{13}$ observation sensitivity is given in
Fig.~\ref{LBNE_only_th13}.  The correlated measurement precision for
$\theta_{13}$ and $\delta_{CP}$ is shown in
Fig.~\ref{LBNE_only_dcp_vs_th13}.  These sensitivities can be compared
to those of DAE$\delta$ALUS shown in Figs.~\ref{thetaplot} and
\ref{deltaplot}.  This leads to the conclusion that the two experiments,
DAE$\delta$ALUS and LBNE, are comparable in sensitivity and 
complementary in their regions of best precision.%

\subsection{DAE$\delta$ALUS$+$LBNE ($\nu$ Only)}

The complementarity of the two experiments suggests that a combination of the
two should give improved sensitivity and precision. In fact, combining the
DAE$\delta$ALUS antineutrino data with an LBNE, neutrino-only data set gives a
significantly improved oscillation sensitivity and measurement capability. The
high-statistics, low-background DAE$\delta$ALUS antineutrino data sample takes
the place of the limited LBNE antineutrino sample. In addition, the
DAE$\delta$ALUS samples bring in a sample with no matter effects and multiple
distances that exploit the on and off-maximum terms in Eq.~\ref{Pmatter}.
The LBNE data are sensitive to mass hierarchy through matter effects, and
for the study below, we assume a normal hierarchy.

LBNE and DAE$\delta$ALUS can take data simultaneously because the 
two data sets are in well-defined energy ranges that do not overlap.
Two scenarios are considered:
\begin{itemize}
\item  DAE$\delta$ALUS$+$LBNE $\nu$---$5$ yr: A five-year run of both experiments,  
combining DAE$\delta$ALUS Phase
1 with a $30\times10^{20}$ POT $\nu-$only LBNE data set.  
\item  DAE$\delta$ALUS$+$LBNE $\nu$---$10$ yr: A ten-year
run of both experiments,  with the Phase 1 + 2 DAE$\delta$ALUS sample combined
with a $60\times10^{20}$ POT $\nu$-only LBNE data sample. 
\end{itemize}

The $1\sigma$ measurement uncertainties on $\delta_{CP}$ for various
scenarios are given in Table~\ref{ScenarioTable} for an assumed value
of $\sin^{2}2\theta_{13}=0.05$. As seen from
the table, the DAE$\delta$ALUS$+$LBNE$ \nu$---$5$-yr scenario does
better that either the DAE$\delta$ALUS Phase 1$+$2 or LBNE $\nu~(5~\mathrm{yr})
+\bar \nu~(5~\mathrm{yr})$ measurements.  The DAE$\delta$ALUS$+$LBNE$ \nu$---$10$-yr
combination reduces the measurement error by a
factor of two with respect to the single technique measurements. 

The 3$\sigma$ $\theta_{13}$ discovery potential for the various
scenarios is shown in Fig.~\ref{ScenarioResults_for_th13}.  The
complementarity of DAE$\delta$ALUS and LBNE leads to an excellent
$\sin^{2}2\theta_{13}$ sensitivity down to 0.002, that is almost
independent of $\delta_{CP}$. Finally, the outstanding correlated
measurement capability of the combined experiments is as shown in 
Fig.~\ref{Combined_Scenario_dcp_vs_th13}.  This capability 
is comparable to the most ambitious
superbeam facilities (see \cite{Bandyopadhyay:2007kx}, Table~5 and 
Fig.~60). 
\begin{table}[t] \centering
\begin{tabular}
[c]{rccccc}\hline
$\delta_{CP}$ & $-160^{\circ}$ & $-80^{\circ}$ & $0^{\circ}$ & $80^{\circ}$ &
$160^{\circ}$\\\hline
\multicolumn{1}{l}{LBNE $\nu~(5~\mathrm{yr}) +\bar \nu~(5~\mathrm{yr})$} & 24.5 & 31.6 &
21.3 & 30.8 & 21.6\\
\multicolumn{1}{l}{DAE$\delta$ALUS Phase 1+2} & 17.7 & 25.3 & 19.6 & 23.6 &
27.2\\
\multicolumn{1}{l}{DAE$\delta$ALUS$+$LBNE$\nu$ --$5$ yr} & 16.8 &
23.7 & 15.3 & 25.5 & 15.0\\
\multicolumn{1}{l}{DAE$\delta$ALUS$+$LBNE$\nu$ --$10$ yr} &
10.6 & 16.2 & 10.1 & 17.3 & 10.4\\\hline
\end{tabular}

\caption{The $1\sigma$ measurement uncertainty on $\delta_{CP}$ for various scenarios
and $\sin^2 2\theta_{13} = 0.05$, normal hierarchy. The uncertainties are given in degrees and a normal hierarchy is
assumed.}\label{ScenarioTable}%
\end{table}%

\subsection{Comparing to Project X}

The Project X proposal at Fermilab would supply $100 \times 10^{20}$ protons
on target to the LBNE beamline in a 5 year time period \cite{JoachimPrivate}.
We assume a ``Project-X scenario'' for LBNE which has a 
5 year run in neutrino mode, followed by 
a 5 year run in antineutrino mode.   We use a normal mass hierarchy
for this study.

\begin{figure}[p]\begin{center}
{\includegraphics[width=5.5in]{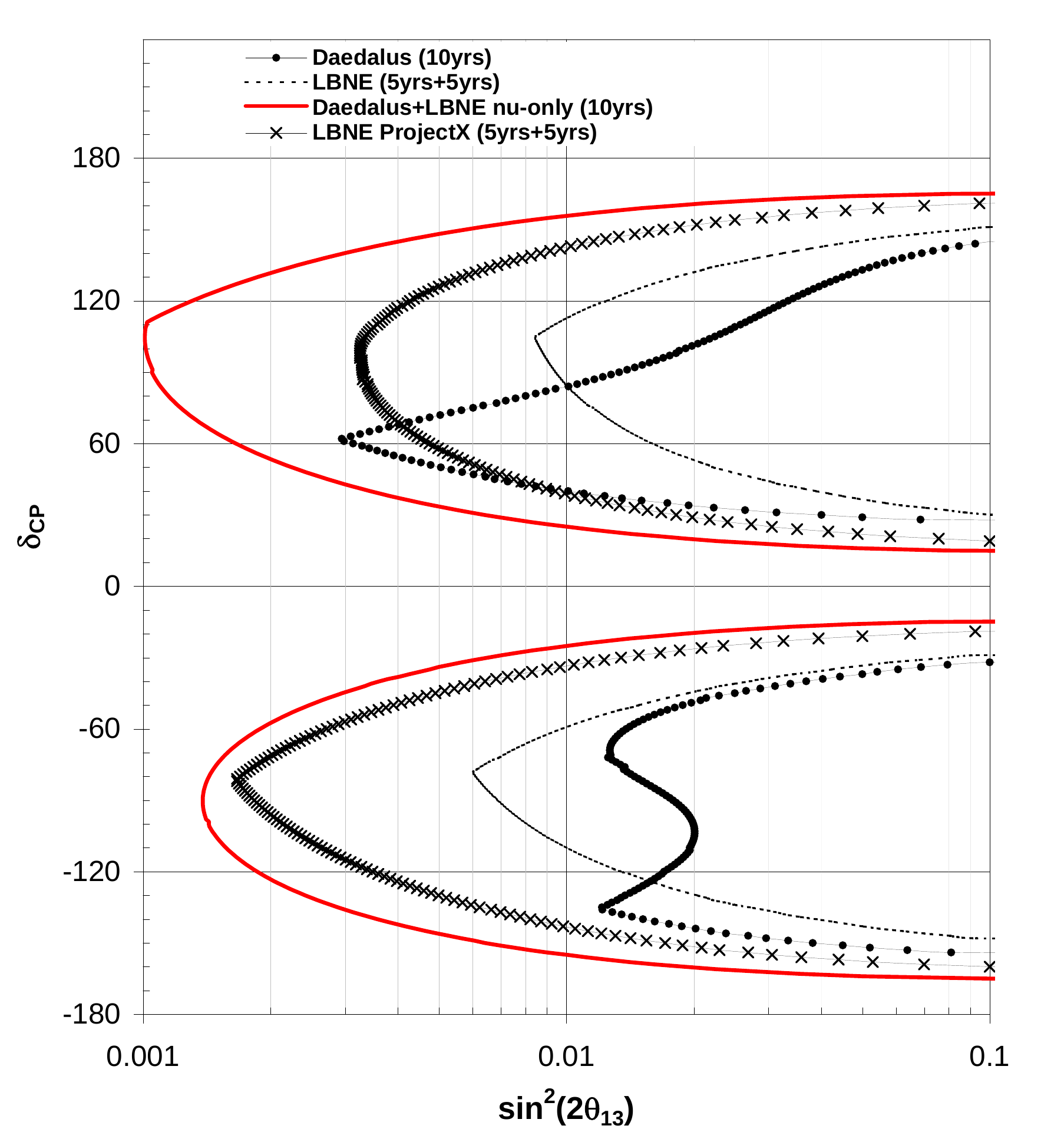}
} \end{center}
\caption{
The region in $\delta_{CP}$ and $\sin^2 2 \theta_{13}$ 
space over which a measurement 
can be differentiated from 0 or 180$^\circ$ at 3$\sigma$.
Red solid:  
DAE$\delta$ALUS$+$LBNE $\nu$---10 yr scenario, Dashed with $\times$: 
Project X scenario.  Expectations for standard running for DAE$\delta$ALUS (solid line with dots) and LBNE (dashed) are also shown.   Normal mass hierarchy is assumed for 
LBNE.
\label{pxtotalcov}}
\end{figure}

The DAE$\delta$ALUS$+$LBNE$ \nu$---$10$-yr expectation is substantially
better than the Project X scenario.  This can be seen in Fig~\ref{pxtotalcov},
which gives the total coverage over which a measurement of $\delta_{CP}$ can be
distinguished from 0 or 180$^\circ$.  The solid red line is the 
DAE$\delta$ALUS$+$LBNE$ \nu$---$10$-yr scenario.   The solid line with 
$\times$s is the Project X scenario.  Also shown are the standard DAE$\delta$ALUS and standard LBNE sensitivities.  Fig.~\ref{fraccoverage} provides
similar information, in the form of the fraction of $\delta_{CP}$ space 
which is covered by the four scenarios.

%% file: near_v6.tex
The DAE$\delta$ALUS near-accelerator will provide a high intensity
source of decay-at-rest neutrinos in the 0-50~MeV energy
range. Because of the intensity of the neutrino source, many
additional physics topics besides the search for CP violation in the
lepton sector become available.  There are two basic experimental
situations possible: physics with the gigantic Gd-doped water
Cherenkov detector that will be used for the CP violation measurement
and physics that could be done with a dedicated detector on the 300
foot level.

We briefly consider several physics topics accessible with the
DAE$\delta$ALUS near accelerator flux.  For each topic, we discuss
briefly the physics issue and in some cases consider briefly the
detector requirements.  This is not meant to be a rigorous study but
rather a sampling of the breadth of topics made available by the
intense DAE$\delta$ALUS flux.

\subsection{Coherent neutrino-nucleus scattering}

Although the process has never been observed before, coherent
neutrino-nucleus elastic scattering has the highest theoretical cross
section in this energy range by about an order of magnitude. The
non-observation is a result of the low momentum transfer
characteristic of this type of interaction. Typical nuclear recoil
energies for this interaction fall in the $\sim$0-100~keV range for
medium-A nuclei.

Neutrino-nucleus coherent scattering is an unobserved Standard Model
process and as such, is interesting to pursue. Furthermore, any
measured deviation from the well-predicted Standard Model expectation
could be a sign of new physics. Neutrino-nucleus coherent scattering
is also relevant for supernova-burst neutrino detection in terms of
oscillation physics and supernova dynamics and
evolution. Approximately ten neutrino-nucleus coherent events on argon
in a ten second window per ton for a galactic core-collapse supernova
at ten kiloparsecs are expected \cite{Horowitz:2003cz}. These
neutral-current, flavor-blind, events could provide much-needed
information about the $\nu_{\mu}/\overline{\nu}_\mu$ and
$\nu_{\tau}/\overline{\nu}_\tau$ supernova flux that is out of reach
for conventional neutral-current-blind (at low energy) neutrino
detectors.

A dedicated coherent neutrino-nucleus scattering detector will be
required in order to make precision measurements of this process. A
noble liquid detector is envisioned at a distance of 10-50~m
from the DAE$\delta$ALUS near-accelerator neutrino source. Assuming
4$\times$10$^{22}$/flavor/year, a nuclear recoil energy window from
20-120 keV (30-160~keV), a 1000 (857)~kg LAr (LNe) target, and a
baseline of 30~m, there will be about 9050 (3700) neutrino-nucleus
coherent events/year in the detector. These event rates have been
found scaling from the CLEAR proposal~\cite{Scholberg:2009ha}.

As for direct dark matter searches, cosmic-ray and intrinsic
steady-state backgrounds are a significant worry for a coherent
neutrino-nucleus measurement. A dedicated coherent detector will enjoy
the advantage of a beam window for background estimation and rejection
and employ standard WIMP detector background mitigation
techniques. The DAE$\delta$ALUS accelerator cycle allows a study of
the steady state rate and reduction of the background by 80\% simply
by the duty factor. The detector design and underground location can
further improve background rejection.

\subsubsection{Measurement of $\sin^2\theta_W$}

Measuring the absolute cross section of neutrino-nucleus coherent
scattering can provide sensitivity to the weak mixing angle. A cross
section measurement with $\sim$10\% uncertainty gives a
$\sin^{2}\theta_{W}$ uncertainty at a typical Q value of 0.04~GeV/c of
$\sim$5\%~\cite{Scholberg:2005qs}. Although a first generation
experiment may not be competitive with precision atomic parity
violation and e-e scattering experiments, it is worth noting that
there are no other neutrino-based measurements near
Q$\sim$0.04~GeV/c. Furthermore, the only neutrino-based
$\sin^{2}\theta_{W}$ measurement (NuTeV) found a value 3$\sigma$ away
from the Standard Model prediction~\cite{Zeller:2001hh}. As for 
the weak mixing angle, planned and existing precision measurements are
not sensitive to new physics specific to neutrino-nucleus(on)
interactions. An absolute cross section measurement consistent with
the Standard Model prediction with $\sim$10\% uncertainty would
provide non-standard neutrino interaction limits more sensitive than
current limits by more than an order of
magnitude~\cite{Scholberg:2005qs}.

\subsubsection{Non-standard interactions}

As mentioned previously, coherent neutrino-nucleus scattering can be
used to search for new physics in the neutrino
sector\cite{Barger:1989rk}.  One can add to the standard model
Lagrangian an extra term of the form\cite{Davidson:2003ha}:
\begin{equation}\label{lagrangiannsi}
-{\cal L}^{eff}_{\rm NSI} =
\epsilon_{\alpha \beta}^{fP}{2\sqrt2 G_F} (\bar{\nu}_\alpha \gamma_\rho L 
\nu_\beta)
( \bar {f} \gamma^\rho P f ), 
\end{equation}
where $f$ is a first generation SM fermion: $e,u$ or $d$, and $P=L$ or
$R$.  Such non-standard interaction (NSI) terms can naturally appear
when adding neutrino mass into the standard
model\cite{Schechter:1980gr} or from super-symmetry
\cite{Barger:1989rk}.  Thus a diverse range of new physics can be
included through the effective parameter $\epsilon_{\alpha
 \beta}^{fP}$ and constrain using neutrino scattering data.

.

\subsubsection{Additional processes}

There are additional motivations for new measurements.
For example, the nuclear neutron form factor can be linked to neutron
star radii, and one might measure this through neutrino-nucleus
coherent scattering~\cite{Amanik:2009zz}.  Also coherent scattering is a process
that occurs during stellar collapse and precise measurements would be
useful input to models.

Electron capture on nuclei occurs during stellar collapse, and
measurements of the reverse process, neutrino capture on nuclei,
should elucidate this mechanism~\cite{Langanke:2003ii,Hix:2003fg,Juodagalvis:2009pt}.  
However, electron capture occurs at lower energies than are 
available from a stopped pion source, and extrapolation to lower 
energies could be problematic.

\subsection{Measurements useful for astrophysics}

Because of their copious production rates in astrophysical bodies,
neutrinos play a large role in many astrophysical processes.  However,
poor neutrino interaction cross-section measurements on many nuclei
inject significant uncertainty into predictions for rates and
kinematics of many astrophysical models, especially stellar processes
including supernova explosion and nucleosynthesis\cite{nu-sns-proposal}.

\subsubsection{Supernova detectors}

Supernova neutrino detectors currently suffer from large uncertainties
in neutrino-nucleus cross section measurements, and would benefit
significantly from new measurements in the relevant energy region on
appropriate nuclear targets.  Such measurements are needed to
understand the distribution of strength in the nucleus.  Often, the
lower lying strength, e.g. IAS and Gamow-Teller, can be studied using
other methods, but the higher order multipoles, e.g. first forbidden,
are more uncertain~\cite{McLaughlin:2004va}.  This resonance can have a substantial
contribution to the cross section.

Any nucleus that is used as a target in supernova neutrino detectors
would be useful to measure.  In addition, measurements on nuclei in
roughly the same mass region would improve supernova neutrino event
rate predictions.

\subsubsection{Nucleosynthesis}

There are two types of nucleosynthesis for which cross section
measurements are needed.  The first is the neutrino process.  In this
process supernova neutrinos spall neutrons and protons off
pre-existing nuclei in the outer layers of the star that is undergoing
a supernova explosion.  Therefore, all the nuclear yields depend
sensitively on the neutrino spallation cross sections.  Some of the
relevant nuclei are light, so measurements on light nuclei would
improve the model predictions.

The other process for which new measurements are needed is
nucleosynthesis from hot outflows, e.g.  supernovae, gamma ray bursts
or compact object mergers.  There has been some recent work on light p
nucleosynthesis showing that neutrino interactions on protons in
relatively late stages have a large impact~\cite{Frohlich:2005ys,Kizivat:2010ea}.  Therefore, the
neutrino flux will cause some rearrangement in the abundance pattern
through neutrino-nucleus interactions. For another type of
nucleosynthesis, the r-process, something similar was shown in the 
mid--nineties: the abundance pattern is rearranged due to neutrino nucleus
interactions~\cite{McLaughlin:1996eq, Qian:1996db}.  For both these processes the nuclear
astrophysics community is interested in nuclei much larger than iron,
so measurements on heavy nuclei are desired.

\subsection{Neutrino magnetic moment}

A more specific example of neutrino NSI is the case of electromagnetic
couplings~\cite{Vogel:1989iv}.  Electromagnetic interactions in
neutrino-electron elastic scattering, $\nu e\rightarrow\nu e$, can be
written in terms of the neutrino energy, $E_{\nu}$, and the recoil
energy of the electron, $T$:\[ \left(\frac{d\sigma}{dT}\right)^{EM}
=
\frac{\pi\alpha^2\mu^2_{\nu}}{m^2_e}\left[\frac{1-T/E_{\nu}}{T}\right]
\]
where $\mu_{\nu}$ is the neutrino magnetic moment, which is usually
expressed in units of Bohr magnetons, $\mu_B = e/2m_e$.  The current
best limit on the muon neutrino magnetic moment comes from the LSND
experiment,
$\mu_{nu}(\nu_{\mu})<6.8\times10^{-10}\mu_B$~\cite{Auerbach:2001wg}.
With the DAE$\delta$ALUS flux, a 1 ton detector at a baseline of 20~m
could expect to observe 3 $\nu e$ elastic scattering events from the
weak interaction and three from EM interactions near 1~MeV recoil
energy in a one year run if
$\mu_{nu}(\nu_{\mu})=1.0\times10^{-10}\mu_B$.

\subsection{Measurement of $\Delta s$}

The contribution of strange quark and antiquark spins ($\Delta s$) to
the nucleon spin continues to be an open question. In QCD, $\Delta s$
is connected to matrix elements of axial operators between nucleon
states with 4-momentum $P$ and spin $S$: \[ \langle
P,S|\bar{q}\gamma^{\mu}\gamma_{5}q|P,S\rangle=2MS^{\mu}\Delta q,\]
where the right-hand side is understood to be at the asymptotic limit,
$Q^{2}\rightarrow\infty$, while the matrix element on the left-hand
side is calculated at zero 4-momentum transfer, $Q^{2}=0$. The same
matrix elements also occur in the expressions for the cross sections
of elastic lepton-nucleon scattering and in particular play a
significant role in neutral-current, neutrino-nucleon elastic
({}``NCEL'') scattering.  The axial term in this cross section can be
written as
\begin{equation}
G_{A}^{NC,p(n)}(Q^{2})=\mp\frac{1}{2}G_{A}(Q^{2})+\frac{1}{2}G_{A}^{s}(Q^{2}),\label{eq:AxialFF}
\end{equation}
where the minus sign in the first term is for scattering off protons
and the plus sign is for neutrons.

The first form factor above is known, in the $Q^{2}=0$ limit, as the
axial coupling constant in neutron $\beta$-decay: \[ g_{A}^{(3)}\equiv
G_{A}(Q^{2}=0)=g_{A}\approx1.26.\] Measurement of the NCEL $\nu N$
scattering cross section can be used~\cite{Garvey:1993sg} 
to extract the strange axial form
factor $G_{A}^{s}(Q^{2})$ which, extrapolated to zero, gives the
strange axial matrix element $\Delta s$:\[ G_{A}^{s}(0)\equiv
g_{A}^{s}=\Delta s.\]

In practice, it is necessary to consider appropriate ratios of cross
sections, in order to minimize uncertainties from the neutrino beam
flux and detector efficiencies.  New efforts to measure this
quantity must determine NCEL cross sections at low-enough $Q^{2}$ to
minimize uncertainties from the extrapolation to zero, and must have
adequate shielding and active vetoing of cosmic rays,
especially in an accelerator with high duty factor. In the following,
we consider two possible options for extracting this important
quantity.

\subsubsection{$\Delta s$ in a mineral oil scintillator detector}

The method proposed here follows~\cite{Tayloe:2002dz}.
The quantity of interest is the NCEL cross-section ratio off protons
and neutrons \begin{equation}
R_{p/n}\equiv\frac{\sigma(\nu p\rightarrow\nu p)}{\sigma(\nu n\rightarrow\nu n)}.\label{eq:Rpn}\end{equation}
Scattering occurs primarily off nucleons in the C nuclei in the liquid
scintillator. As can be seen from Eq.~\ref{eq:AxialFF}, $G_{A}^{NC}$
has a different dependence on $G_{A}^{s}$ for protons and neutrons,
therefore this ratio is sensitive to the value of $\Delta s$. The
ratio is clearly insensitive to uncertainties in the neutrino flux.
Knockout neutrons can be identified via their capture by a proton
in the liquid scintillator with emission of a 2.2-MeV photon, which
then converts and produces scintillation light that can be detected
in a photomultiplier-tube array. A capture likelihood ratio can be
formed using information on the distance in time and space of the
photon signal from the primary hadron and the PMT multiplicity. This
is then compared to a Monte Carlo simulation for the same variable
and for different values of the quantity of interest $\Delta s$,
allowing a determination of the most likely value of $\Delta s$.

Since the target protons and neutrons are bound, nuclear effects must
be considered. However, nuclear corrections have been shown
theoretically to be small, expected to contribute about 0.03 to the
extracted value of $\Delta s$~\cite{Barbaro:1996vd}. This is due to the
isoscalarity of the target nucleus, implying that nuclear corrections
largely cancel in the ratio.

\subsubsection{$\Delta s$ in a Gd-doped water Cherenkov detector}

One can measure the proton-to-neutron ratio of
Eq.~\ref{eq:Rpn} by detecting the nucleon-knockoff reactions
\[\nu+^{16}\textnormal{O}\rightarrow\nu+p+^{15}\textnormal{N}+\gamma\]
and 
\[\nu+^{16}\textnormal{O}\rightarrow\nu+n+^{15}\textnormal{O}+\gamma\]
in a large water Cherenkov detector, as previously proposed for 
ORLaND~\cite{Avignone:2001ti}. Events are observed by triggering
on the emitted nuclear gamma rays. By doping the water with gadolinium
salts, it will be possible to identify neutron-knockoff events via
neutron capture in Gd and the accompanying (delayed) gamma rays, while
proton-knockoff events will have no delayed $\gamma$s. Again, the
ratio method cancels most uncertainties due to beam flux, nuclear
binding effects ($^{16}$O also being isoscalar), and final-state
interactions.

%% file: designintro_v1.tex
This chapter provides information on the preliminary design of 
DAE$\delta$ALUS.  We specifically note the challenges.  We are in the process of a cost and design study of
the accelerators and report three cyclotron-based options here.  The detector follows
the presently proposed Gd-based design.  
Lastly, we consider how the accelerators will be
deployed, assuming the detector is available in 2021.

%% file: accel_intro_v2.tex
DAE$\delta$ALUS requires Megawatt-class $\sim 1$ GeV accelerators.
While superconducting linacs provide the most conservative technology
option, space and cost constraints suggest that high-power cyclotrons
could be developed that would meet our goals.  Three possible options
have been identified: the Compact Superconducting Cyclotron (CSC),
which accelerates protons; the Multi-Megawatt Cyclotron (MMC), which
accelerates H$_2^+$, and the Stacked Cyclotron (SC), which accelerates
multiple proton beams in one accelerator.  All three are designs which
are being developed for commercial purposes -- the first for active
interrogation for homeland security and the second two for accelerator
driven systems used to drive subcritical reactors.  We have now embarked on
a year-long study of these options, as they apply to DAE$\delta$ALUS,
to understand the cost, schedule and and R\&D needs of each
machine. At the same time, we do not wish to exclude high power linacs
as an option, considering the synergy with other communities and the
possibility of cost-sharing for multi-purpose installations.

Several desirable aspects inherent to cyclotrons attract us to this
option.  First is compactness, minimizing costs for shielding and
space, of particular value for the near site where footprint will be
an important consideration.  Second is that the fixed-energy and
continuous beam character of cyclotrons are desirable features for
DAE$\delta$ALUS neutrino production, reducing peak-power loads on targets and
providing good compatibility with the short-duty-factor beam structure
from Fermilab.

In the sections below, we discuss synergistic uses for the
accelerators.  However, here we not that these accelerators are not
well-suited two common applications for cyclotrons at present -- PET
isotope production and proton therapy.  For the former, one needs
intense beams but only at $\sim 30$ MeV -- the optimal energy for
isotope production.  For the latter, one wants about 250 MeV and very
low power compared to the DAE$\delta$ALUS machines.

\subsection{Overview of Cyclotron Subsystems}

In this section, we consider each component of a MegaWatt class
cyclotron, to point out the important issues that drive the design. In
following sections we describe the specific design options being
considered and how each addresses these issues. At the end we discuss
a linac that could be an interesting option.

\subsubsection{Ion Sources and Injection}

Producing milliamperes of protons is not an issue. Modern ECR and
multi-cusp plasma sources with CW currents up to 100 mA are available
with the requisite brightness, duty factor and reliability.

The injection channel into the cyclotron is usually axial at the
radial center with magnetic or electrostatic inflection through 90$^\circ$
into the cyclotron plane. At high currents, coupling in all
phase-space planes is difficult to avoid because of space-charge
forces, potentially leading to large emittance growth and hence
greater difficulty in minimizing beam losses. Acceptable emittance
growth depends in large measure upon the extraction strategy, as
discussed below.

If the magnetic field in the central (injection) region is high, the
low-energy beam will turn on a small radius, and designing the
inflector so beam is not lost on the first turn is a challenge. Some
beam loss is unavoidable, but the design goal will be to limit this
loss, to enable high-power beam at extraction and to minimize heating
of inflector components.

\subsubsection{The Magnetic Field}

Compactness is desirable, hence the attraction of higher magnetic
fields available from superconducting magnets. However, the higher
fields also bring complexity. Vertical focusing and isochronicity
require spatial variations (referred to as``flutter'') in the magnetic
field of factors of two or more. In lower field machines these
variations are produced by shaping the iron poles with hills and
valleys.  However, in compact machines which must go to higher fields
($\sim 10$T, well above the saturation of iron) the iron pole shapes
play only a minor role, geometry of the superconducting coils must
provide the required field variations.  Even the H$_2^+$ ring
cyclotron will use large, aperture, superconducting magnet coils at
fields as high as 4 to 5 T.  Using materials with higher saturation
points, such as rare-earth metals, may help in the optimization
process.  This alternative is currently being studied at MIT
\cite{AntayaPrivate}.

\subsubsection{RF}

The accelerating RF source for high-field cyclotrons will fall within
the broadcast band, so that basic generating equipment, which is
relatively inexpensive and available, can be used. Because of geometric
considerations, the MMC and the SC designs each have ``interesting''
configurations for the RF cavities.

\subsubsection{Extraction}

The question of clean extraction is a design driver for the
cyclotrons.  Conventional designs place a thin metallic (or carbon)
septum between the beam circulating and the beam emerging from an
extraction channel. However, if the machine does not good
turn-separation\footnote{{\it i.e.}, sufficient separation of the beam
orbit on successive turns.}, losses on the septum will be unacceptably
high, leading to high activation and even melting of the septum. Good
turn separation comes from high momentum-gain per turn (large change
in orbit radius) and very low emittance (effective beam size). The
former requires a high voltage (and expensive) RF system and the
latter depends crucially on tight control over space charge at
injection.  For very high extraction efficiency, the beam size plus
width of the septum must be smaller than the turn separation.  Note
that turn separation is related to $\Delta p/p$, thus as $p$ grows,
since typically $\Delta p$ is relatively constant, it becomes increasingly
difficult to preserve good turn separation for higher-energy machines.

As a result of these considerations, the designs described here each
step away from conventional practice to take novel approaches to
extraction. The CSC proposes ``resonant self-extraction,'' a technique
in which the beam is conducted through the edge of the magnetic field via
a notched-channel in the magnetic poles and perturbation coils.  The
MMC accelerates H$_2^+$ because of the simplicity of extraction via
stripping foils that dissociate the molecular hydrogen into two
protons, whose orbits can be designed to easily leave the inside of
the magnet plane.  The stacked cyclotron circulates many lower current
beams, thus while the whole is high-intensity, each extracted beam is,
in itself, low enough intensity for conventional extraction.

\subsubsection{Vacuum}

The vacuum system is not a major design issue. Preliminary evaluation
shows that with a vacuum of $2\times10^{-8}$ Torr, the beam losses
should be sufficiently low along the total acceleration path, even 
for the molecular hydrogen beam. The
design goal for beam losses is lower than the estimated beam losses at
the TRIUMF cyclotron (520 MeV) which accelerates the very weakly-bound H- ion.

\subsubsection{The Beam Stop}

The design of the beam stop, which is the neutrino source, will be
partly driven by the extraction scheme.  Our design goal is to have
each extracted beam be no more than 1 MW (though there may be more
than one extracted beam).  This limitation keeps the beam stop design
within the range of existing stops, including LAMPF/LANSCE and the 3 GeV
hadron line at JPARC.

A side benefit of DAE$\delta$ALUS targeting is production of $^3$He in
the beam stop region.  The $^3$He can be separated by cooling the air
and sold.  This has already been done -- Los Alamos sells the $^3$He
produced by LANSCE \cite{MacekPrivate}.

%% file: compact_v4.tex
Compact superconducting cyclotrons are a potentially low-cost
accelerator option for DAE$\delta$ALUS.  A class of small-footprint,
single stage, mA-current, high magnetic field, superconducting
cyclotrons (CSCs) are under development at MIT for the Defense Threat
Reduction Agency (DTRA).

\begin{figure}[t]\begin{center}
{\includegraphics[width=3.5in]{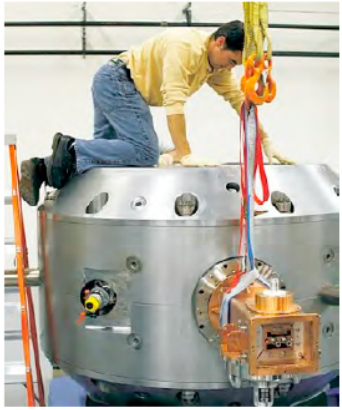}} 
\end{center}
\vspace{-0.25in}
 \caption{The Monarch 250 Cyclotron is an example of a compact (250 MeV) superconducting 
cyclotron which can run at 10 T and which uses the technique of resonant self-extraction.   This was designed at MIT for proton therapy applications. 
\label{fig:CSC}}
\end{figure}

While the CSC represents an advance of the state of cyclotron technology, many
aspects of its design and operation with $\sim$5 to 10 Telsa  
fields have already been proven:
\begin{itemize}
\item The fundamental feasibility of accelerated beams of $\sim$10 mA has
been established \cite{tenma1, tenma2}.  

\item Superconducting compact cyclotrons exist \cite{sccexist}, including the the K500 and K1200 at Michigan State University, where the ``K'' parameter
resents the maximum proton energy that could be obtained based on magnetic field
size and strength (though it should be noted that these machines are used
for heavy ions and so, while in principle the K1200 accelerates protons to
1.2 GeV, this has not been demonstrated in practice). The
Monarch 250 (running at 250 MeV) was designed at MIT and is now being
commissioned.  The Monarch 250 is shown in Fig.~\ref{fig:CSC}.

\item The self-extraction technique has been developed successfully
  for a 10-MeV conventional cyclotron, the Cyclone
  SEC\cite{SEC1,SEC2}, marketed by IBA.  The Monarch 250, marketed by
  Still River also relies on a self-extracted beam.

\item The high current-density Nb$_3$Sn superconducting cable, capable 
of carrying more than
3 kA/mm$^2$, is commercially available \cite{Nb3Sn}.   This conductor can be 
used for fabricating compact magnets  exceeding 10 Tesla field strength.

\end{itemize}

The limiting intensity in isochronous cyclotrons depends on 1) the
ability to capture a high current beam at low energy, 2) the
suppression of beam loss due to resonant instabilities during
acceleration, and 3) the ability to extract the beam from the
cyclotron without high losses. In a single stage cyclotron, low
velocity ions are captured into stable orbits on every RF cycle to
yield a continuous wave (CW) accelerator. 

The CSC design uses a proton beam.  The attractive option of using
H$^-$ ions, common now for low-energy isotope-producing cyclotrons (a
stripper foil converts the H$^-$ to a proton, which bends in the opposite
direction, facilitating 100\% extraction efficiency), is not available for
the CSC because of Lorentz stripping.  The extra electron is very
loosely bound (0.7 eV), and the relativistic transformation of the
magnetic field produces a transverse electric field that, if higher
than the binding energy over the size of the ion, will cause loss of
the extra electron.  The highest-energy H$^-$ cyclotron is the 18 m 
radius TRIUMF machine, with a beam energy of 
500 MeV and a maximum field strength of 0.5 Tesla.  The other possible
ionic form for protons, molecular hydrogen (H$_2^+$) also allows for
high-efficiency extraction via stripping, but requires, as discussed in
the next section, larger machine sizes because of the higher rigidity
of the beam (with charge-to-mass ratio of 1/2 instead of 1 for
protons.)

In any accelerator, space charge forces are largest when the ion
velocity is low; but emittance growth continues, though at a reduced
rate as the energy increases, throughout the many orbits required to
reach the full beam energy. The growth of transverse emittance (and
therefore beam size) may be mitigated and somewhat controlled 
by the horizontal and vertical focusing generated by the ``flutter
field'' design. The longitudinal emittance growth, on the other hand,
leads to an energy spread in the beam which after many turns produces
filamentation in longitudinal phase space\footnote{filamentary
  structure (or ``stringiness'' with substantial empty space between)
  is a problem because it represents a loss of ``phase-space
  density.''}. In addition to increasing beam size, an undesirable
consequence of these emittance growths is the production of beam
halos, particles occupying areas in phase-space well outside the
central core of the beam.

The halos are especially troublesome at the extraction channel of the
cyclotron, where they will most likely be scraped, depositing their
energy on the channel edges, causing unacceptable activation or, possibly,
mechanical destruction of components. Loss of even 1\% of a
1-MW beam will deposit 10 kilowatts of power, enough to cause
substantial damage to components.

The most effective way of mitigating these effects is to inject beam
at the highest practical energy.  This ensures that injected beam has
lowest possible emittance.  One also aims to accelerate the beam as
quickly as possible with a very powerful RF system.

External, axial injection using a well-designed Electron Cyclotron
Resonance (ECR) source offers the best option for yielding high-quality
beam.  ECR sources produce the required beam current
with very good emittance, having this source on a platform external to
the cyclotron provides good control over injection beam quality and
energy.  Limits on injection energy will come from the design of the
inflection channel; it must bend the beam through 90$^\circ$ into the plane
of the cyclotron; and must be small enough so the beam orbit does not
strike it on the first turn -- a challenge for a high-field cyclotron
for which the first radius will be very small.

A high-voltage RF system will not only speedily accelerate the beam
through the low energy region where space-charge forces have the
greatest effect, but will also provide the best possible turn
separation at high energy, necessary for extraction efficiency.  For
this reason, the conventional approach to building a high power
cyclotron is to accelerate the beam as quickly as possible with a very
powerful RF system.  This strategy has been employed successfully by
the world's currently most powerful cyclotron (1.2 MW), the PSI 600
MeV, 15-meter diameter machine in Switzerland \cite{PSI}.

The self-extraction concept has been demonstrated at low energies (~10
MeV), and is being developed for the 250 MeV superconducting Monarch
machine.  This approach must be extended to the 1 GeV region needed for the
DAE$\delta$ALUS machine, and must be refined to provide the extremely high
extraction efficiency needed (well in excess of 99\%).

All of the relevant design issues will be addressed in DTRA-sponsored
research at MIT that is aimed at beam parameters very similar to the
DAE$\delta$ALUS parameters.  This study will determine whether the CSC
approach can be driven to high beam energy and high intensity.  In a
separate project members of the DAE$\delta$ALUS collaboration have
proposed to build a 1 MeV electron model of a high intensity (2.5 mA)
cyclotron.  This very low cost device would use an eight sector design
to explore all the space-charge physics and extraction issues of a
single-stage CSC.

In the case of the CSCs, we would envision implementing cyclotrons
which are 1 MW each.  Thus, for the Phase 1 scenario, we would install
1, 2 and 3 cyclotrons at the 1.5 km, 8 km and 20 km sites,
respectively.

%% file: h2plus_v5.tex
\begin{figure}[t]\begin{center}
{\includegraphics[width=3.5in]{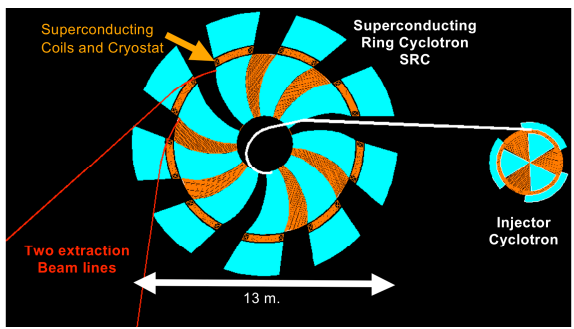}} 
\end{center}
\vspace{-0.25in}
\caption{Schematic of the Multi-MegaWatt Cyclotron
\label{fig:MMC} }
\end{figure}

A two-cyclotron system -- an injector followed by a superconducting
ring -- which accelerates H$_2^+$ ions provides elegant solutions to
the injection and extraction issues outlined in the introduction to
this chapter.  This idea is under development at INFN, Catania, for
use as ADS technology \cite{Calabretta}.  The group at INFN is now involved in
adapting their design for DAE$\delta$ALUS use. The concept of the Multi
Mega Watt Cyclotron (MMC), which can supply a 1.5 MW beam, for
DAE$\delta$ALUS is shown in Fig.~\ref{fig:MMC}. Preliminary parameters
of the injector cyclotron and of the superconducting ring cyclotron
are also presented in Tab.~\ref{tab:MMC}.


The cyclotron complex consists of an injector cyclotron, which will
accelerate the H$_2^+$ beam (a hydrogen molecule with just one
electron) from the injection energy of about 100 keV (50 keV/n) up to
a maximum energy of about 68 MeV (34 MeV/n). An electrostatic deflector
performs the beam extraction, from the injector cyclotron. The beam is
then injected into the large cyclotron, the Separated Ring Cyclotron
(SRC), by a special bending magnet and a further injection
electrostatic deflector. The SRC will accelerate the H$_2^+$ beam up to
the maximum energy of about 800 MeV/n. The use of a thin carbon foil
(stripper foil) placed at the radius of maximum energy breaks
the hydrogen molecule into two independent protons.

\bigskip%
\begin{table}[tbp] \centering
\begin{tabular}
[c]{|c|c||c|c|}\hline
\multicolumn{4}{|c|}{Superconducting Ring Cyclotron} \\ \hline
$E_{inj}$ & 	34 MeV/n  &	$E_{max}$ &	800 MeV/n \\ \hline
$R_{inj}$ &	1.4 m	  &     $R_{ext}$ &     4.5 m \\ \hline
$\langle B \rangle$ at $R_{inj}$ &  1.2 T & $\langle B \rangle$ at 
                                    $R_{ext}$ & 2.17 T \\ \hline
sectors	 &          9	   &   Accel. Cavities	& 6 \\ \hline
RF 	 &    53.7 MHz	   & Harmonic	 & 6th \\ \hline
V-peak   &	220 kV	   & $\Delta E$/turn	& 1.950 MeV  \\ \hline
$\Delta R$ at $R_{inj}$	& 15 m & $\Delta R$ at $R_{ext}$ &2.7 mm  \\ \hline
\hline
\multicolumn{4}{|c|}{Injector Cyclotron} \\ \hline
$E_{inj}$	& 50 keV/n	   & $E_{max}$	& 34 MeV/n \\ \hline
$R_{inj}$	& 5.5 cm	   & $R_{ext}$	& 1.4 m \\ \hline
$\langle B \rangle$ at Rinj & 1.2 T & $\langle B \rangle$ at $R_{ext}$ &	2.17 T \\ \hline
sectors	 & 3	           & Accel. Cavities &	3 \\ \hline
RF	& 26.85 MHz	   & Harmonic	& 3rd \\ \hline
V-inj	& 70 kV    	   & V-ext	& 180 keV \\ \hline
$\Delta E$/turn	 &1080 keV & $\Delta R$ at $R_{ext}$ &	11 mm \\ \hline
\end{tabular}
\caption{Parameters of the SRC and of the injector cyclotron .
Superconducting Ring Cyclotron  \label{tab:MMC}}%
\end{table}%

Because the H$_2^+$ and the protons have significantly different magnetic rigidities, the protons escape quite easily from the magnetic
field of the cyclotron.  Extraction by stripper foil does not require
well-separated turns at the extraction radius and allows use of a
moderate energy gain per turn during the acceleration process, 
with a
significant simplification of the RF
accelerating system. Stripping
also allows the extraction of beams with large energy spread (0.5 to
1\%), so the beam bunch elongation produced by space charge effects can
be more readily tolerated. Lower requirements for power transferred
inside each cavity increases both the reliability and the conversion
efficiency from electrical- to beam-power, as well reducing the overall
cost of the RF system.

The special duty cycle (100 $\mu$s Beam On, 400 $\mu$s Beam Off) and
the high beam power required by the DAE$\delta$ALUS experiment will
produce a serious beam loading effect on the accelerating cavities.
That is, when the leading edge of the 100 ms beam pulse enters the
cavity the sudden draw of power drops the voltage and affects the
balance between the electrode and the supply.  This beam-loading will
produce a voltage and phase instability that must be compensated.  
This level of compensation must be high if the beam is extracted
by electrostatic deflector (PSI solution).  This is a much less
serious issue if the extraction is, instead, performed by a stripper
foil.

The layout shown in Fig.~\ref{fig:MMC} is an evolution of the previous
INFN design, presented at EPAC 2000 \cite{Calabretta}. The main
difference is the reduction of number of sectors, from 12 to 9, and
also of the extraction radius, from 6 m to 4.5 m, with a relevant cost
reduction. These changes are feasible for two reasons.  First, the maximum
proton energy for DAE$\delta$ALUS is 800 MeV, compared to 1000 MeV in
the ADS design.  Second, since 2000, RF cavities with double gaps have
been designed which achieve voltages as high as 200kV. According to
Fig.~\ref{fig:MMC}, these accelerating cavities have a spiralled shape
that matches the shape of the sectors. At the same time the use of
only 9 sectors allow to use of RF cavities with an electrode width of about
16$^\circ$, with a convenient 6th harmonic driving frequency. In this
condition the accelerating voltage across a double gap cavity
is$V_{max}*2*\sin(96/2)=1.48 V_{max}$. Therefore, a double gap cavity
driven with a voltage of 200 kV is equivalent to a single gap cavity
driven at 300 kV. The greatest advantage of the double gap cavities
versus the single gap cavity or pill box cavity is that their
ends exceed the injection and extraction radius by only about 10 cm.
The pill box cavity, for example, needs an extra length at the entrance and at
the extraction side of about 1 m. Moreover, the double gap cavities
can be designed to produce an acceleration voltage that increases
with radius, while the pill box cavities have their maximum voltage
just at the middle between the injection and extraction radius.

The acceleration of H$_2^+$ requires a magnetic field two times higher
than for protons; fortunately this is not a serious problem if
superconducting coils are used.  Generally, it is quite difficult to
achieve the required isochronous field for a ring using
superconducting coils wrapped around the iron pole.  The more elegant
solution is the use of the so-called ``S-coils.''  These are a pair of
superconducting coils wrapped around the pole/yoke and perpendicular
to the median plane, outside of the extraction radius. This solution
allows for more free space among the sectors, and for the cryostat of
the superconducting coils to be completely in the outer region, without
any interference with the cavities and the accelerated
beam. Preliminary evaluations by 3D Mafia code were
done to verify the feasibility of this Cyclotron.

The injector cyclotron will be designed scaling up the commercial
cyclotron TR-30 of the EBCO company (1.2 mA) or the C-30 of the IBA company
(1mA).  Both of these cyclotrons are able to deliver proton beams at
30M eV and were designed to be compact, and highly efficient. In our
case we face the problem of accelerating a beam current which is about
5 time higher, so the space charge effects will be serious. To overcome this
problem we plan to inject the H$_2^+$beam at energy of 100 keV and to
use 3 RF accelerating cavities to minimize the turn number in the
injection cyclotron. The use of 3 sectors and of an extraction radius
about double that of the commercial proton cyclotrons, allows more
room for the RF cavities as well as use of an acceleration voltage
higher than 100 kV even at the injection radius. We expect that this
cyclotron should be able to deliver a beam current of H$_2^+$ as high
as 5 mA, assuming a 100\% duty cycle, or 1 mA with a 20\% duty
cycle. With respect to the space charge effects, we have to consider
that the perveance\footnote{The value of perveance is a measure of
the significance of space charge effects on the motion of the beam.} of a proton beam and an H$_2^+$beam with the same
energy is the same if we work at low energy ($E<$100 MeV). Therefore, the
emittance of a H$_2^+$ beam at 100 keV is the same or better than the
emittance of a proton beam at the same energy. Moreover, proton
sources are generally able to produce high beam currents of
H$_2^+$. In the present case, we need an H$_2^+$ source able to deliver
a beam current of about 33 mA. The injector cyclotron,
equipped with a simple buncher, should be able to catch and accelerate
about 15\% of the injected beam. This means that about 28 mA of
H$_2^+$ beam with an energy of 100 keV (2.8 kWatt) will be lost in the
first turn.  The effect of this power loss on the superconducting
materials needs to be considered, but because of the low energies, this does
not result in high activation.

At DAE$\delta$ALUS, for Phase 1, we could install 1, 1 and 2
accelerators at the 1.5 km, 8 km and 20 km locations, respectively.
This is slightly different from our base design in that it supplies
1.5 MW,1.5 MW and 3 MW, respectively.

%% file: stackedcyclo_v2.tex
\bigskip%
\begin{table}[tbp] \centering
\begin{tabular}{|c|c|c|c|}\hline
 & injection & extraction & \\ \hline
Energy & 150 & 800 & MeV \\ hline
Radius & 3.0 & 5.0 & m \\ \hline
Magnetic Field & 1.2 & 1.4 & T \\ \hline
RF Frequency (6$^{th}$ harmonic) & 48 & & MHz \\ \hline
Number of magnet sectors & 8 & & \\ \hline
Number of RF Cavities & 4 & 6 & \\ \hline
Energy gain per turn & 4 & 6 & MeV \\ \hline
Radial separation of turns & & 7 & mm \\ \hline
\end{tabular}
\caption{Parameters of the Stacked Cyclotron  \label{tab:stacked}}%
\end{table}%

The stacked cyclotron proposal, with parameters summarized in 
Tab.~\ref{tab:stacked}, is inspired by the success of the
isochronous cyclotron (IC) at the Paul Scherer institute (PSI).  The
PSI machine was built 35 years ago, and after two generations of upgrades and
improvements it routinely delivers 2.2 mA of proton beam at 650 MeV
energy.  As stated earlier, the challenge in improving upon this
performance is in increasing the beam current.  The space charge tune
shift at injection drives rapid emittance growth, and the separation
of orbits at extraction after emittance growth produces large beam
loss at the extraction septum.

\begin{figure}[t]\begin{center}
{\includegraphics[width=3.5in]{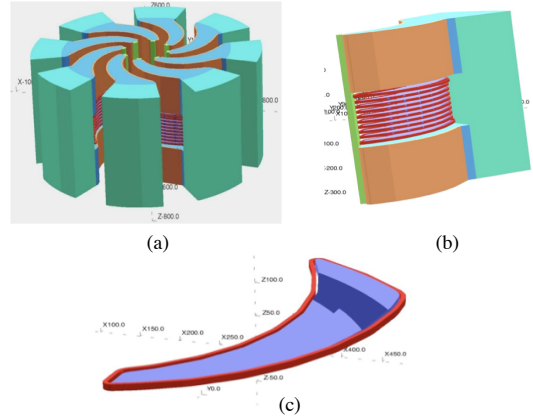}} 
\end{center}
\vspace{-0.25in}
\caption{The design of the stacked cyclotron.  a)  The overall design; 
b)  detail of one sector magnet;  c) detail of one pole piece.  
\label{stacked} }
\end{figure}

The Texas A\&M University (TAMU) group has invented a novel method to
solve these problems, inspired by the needs of the ADS application.
They take the established performance of PSI and replicate it in each
layer of a flux-coupled stack of superconducting ICs, shown in
Figure~\ref{stacked} (a).  The magnetic configuration follows that
developed for the single ring cyclotron at RIKEN\cite{RIKEN}, in which the field
in each aperture of a sector magnet is produced by a pair of cold iron
pole pieces on which superconducting windings are bonded.  The pole
pieces are suspended within a warm-iron flux return, with vacuum gaps
above and below, so that the huge Lorentz image forces between the
poles and between pole and flux return cancel one another.  This
approach makes it possible to enjoy the benefits of a superconducting
magnet without having to cool the immense thermal mass of the flux
return.  Figure \ref{stacked} shows a detail of one sector magnet (b)
and of one pole piece (c).  Assuming the same performance as PSI, a
stack of five ICs operating with 800 MeV, 2.2 mA would provide the
proton beam needed for DAE$\delta$ALUS.  Targeting the five beams in a
symmetric pattern on each beam dump should significantly simplify the
design and operation of the beam dump by limiting the local energy
density.  Each element of the flux-coupled IC is, in effect, a close
cousin of the PSI IC, except with higher injection and extraction
energies, and the use of superconducting magnet coils and
superconducting RF cavities.

\begin{figure}[t]\begin{center}
\vspace{-0.5in}
{\includegraphics[width=3.5in]{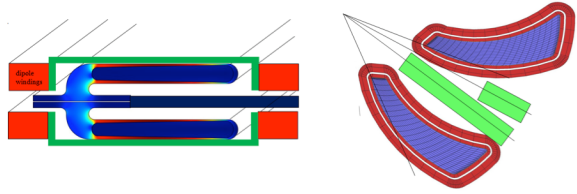}} 
\end{center}
\vspace{-1.in}
\caption{Left: Dielectric loaded superconducting cavity (1 MV/turn each);
Right Placement showing additional cavity to boost orbit separation at 
extraction.
\label{stackreson} }
\end{figure}

The design of the superconducting RF cavity was a particular
challenge.  It operates at 58 MHz (6th harmonic of the cyclotron
frequency), and if it were an empty cavity it would be much larger
than what is needed to fit within the space between ICs in the
flux-coupled stack.  The TAMU group invented a dielectric-loaded resonator that
fits in the required space, as shown in Fig.~\ref{stackreson}.  It
should be capable of 1 MV accelerating voltage and excellent mode
stability. This is the one element of the flux-coupled IC stack that
requires long-lead R\&D to design, build, and test a prototype to
validate the design.

In related research, the Cyclotron Institute at Texas A\&M University
operates a superconducting cyclotron complex for nuclear research, and
is currently upgrading the complex to re-accelerate radioactive nuclei
using its original Cu-coil cyclotron.  The Nuclear Engineering
Department at Texas A\&M University is interested in collaborating
with the Accelerator Research Lab to develop ADTC (Accelerator Driven
Thorium Cycle) fission. The TAMU group proposes a development scenario
in which the accelerator would be built and commissioned at a site
near Texas A\&M University.  Because the parameters of the accelerator
are consonant with our long-term objective of ADTC fission, the TAMU
group envisages forming a collaboration with our colleagues at the
University of Texas and seek funding from the State of Texas with
which to construct the physical facilities needed to house and operate
the accelerator during development in Texas.  Once it were complete
and operating, having demonstrated technological feasibility, this
prototype could be dismantled and moved to the DUSEL site for use by
DAE$\delta$ALUS.

%% file: otheraccel_v1.tex
A superconducting linac would be a very conservative option for
achieving the high currents needed for good neutrino production.
Superconducting RF cavities are now used in many high-current,
high-energy accelerators, and while the first of these were electron
machines (JLAB, TESLA), the final stage of the SNS Linac constitutes a
good example of a proton linac in the GeV range.  Furthermore, design
studies at Fermilab for Project X, and at MSU and Argonne for the FRIB
project have further optimized engineering and costing for this class
of linacs.

Injection into a linac is much simpler than for a cyclotron, and
front-end designs with over 100 mA of continuous proton current have
been demonstrated (the LEDA project at LANL).  Apertures inside
superconducting cavities are very large, typically 10 cm or more, so
beam loss during acceleration is a much easier problem to solve.
Extraction is also certainly not the problem it is with cyclotrons.
One drawback is the large footprint required.  The second, which
accelerator physicists are working to address, is the high cost of the
systems.

An interesting opportunity might exist for working with the
medical-isotopes community for developing a mutually-beneficial
design.  A serious problem exists in the US today regarding a
steady, reliable supply of $^{99}$Mo, an isotope used in approximately
50,000 nuclear medicine procedures per day in the US alone.  This
isotope is abundantly produced as a fission fragment from neutron
irradiation of highly-enriched uranium (HEU).  There are no reactors
in the US capable of supplying the year-round quantity of the isotope
required, and considering the problems in the US associated with HEU
reactors, this is not likely to change anytime soon.  As a
consequence, the supply of the isotope comes from three aging
reactors, one in Canada (currently shut down) and two in Europe (both
facing significant maintenance problems).

An interesting accelerator design has been proposed by the Physics
Division of Argonne National Lab, which would use a multi-megawatt proton
beam (at an energy close to 1 GeV) to produce neutrons in a
heavy-metal blanket that would then bombard a small core of HEU from
which the 99Mo would be extracted.  The production from this
accelerator would be capable of satisfying the entire US demand for
the isotope, using less than half of its beam capacity.

%% file: nu-source_v1.tex
The neutrino source for the DAE$\delta$ALUS
experiment will be a beam stop for the proton beam.  The beam stop
will have a series of important functions, all of which require 
design considerations.  First, the beam stop produces the mesons which
in turn either capture without decay, come to rest and decay, or
escape the beam stop altogether and decay;  all decays produce
neutrinos.  The design of the beam stop can affect the neutrino flux
by altering the relative likelihood of these processes.  Second, the
beam stop dissipates the $\sim$1~MWatt energy of the beam, dissipating
it into heat in a controlled fashion.  Third, the outer portions of
the beam stop contains the significant flux of neutrons and gamma rays
produced by interactions of the proton beam with the inner core, of
importance not only for personnel protection but also for limiting
dose on nearby equipment and the water table.  High-power beam stops
for proton accelerators have been designed for a variety of
experiments \cite{jparc-nu,jparc-meson,numi,lanl,isis}, providing
significant guidance in all three of these considerations.  Each
design goal of the beam stop is discussed in turn.

\subsubsection{Neutrino Flux}
The beam stop produces positive and negative pions and no kaons due to
the below-threshold kinetic energy of the incident proton beam.  With
appropriate geometric design, 98\% of pions come to rest in the beam
stop, with only $\sim2$\% escaping laterally or longitudinally
\cite{Burman:2003ek}.  Neutrinos result predominantly from the decays
$\pi^+\rightarrow\mu+\nu_\mu$,
$\mu^+\rightarrow\overline{\nu}_\mu\nu_e e^+$ for pions and muons
coming to rest within the beam stop.  The charge-conjugate decays are
suppressed because all but $\sim 5\times10^{-4}$ of $\pi^-$ are
captured on nuclei without decaying, and even the $\mu^-$ from the few
remaining $\pi^-\rightarrow\mu^-\overline{\nu}_\mu$ decays have an
increased likelihood for capture without decay for heavy nuclei
\cite{Burman:1997wc}.  To control the neutrino flux, it is important
to reduce or understand the meson leakage from the beam stop which
results in decays-in-flight to neutrinos, albeit at higher neutrino
energies than those from the stopped pion decay chain, and it is
important to reduce or understand the residual decays of non-captured
$\mu^-\rightarrow e^-\overline{\nu}_e\nu_\mu$ which produce an anti
electron-neutrino background to the oscillation search.

We propose to study a beam stop with an inner core composed of
graphite slabs stacked longitudinally along the beam axis.  A
significant amount of data exists
\cite{Crawford:1980,cochran1972,denes1983,digiacomo1985} for pion
production on nuclei, particularly for carbon.  These
data, along with information from Monte Carlo simulations
\cite{Burman:1989dq, Burman:1996gt}, indicate that, at a beam energy
of 800~MeV, the number of neutrinos produced is $\nu_\mu/p \approx
0.172$, while the ratio $\pi^+/\pi^-\approx4.5$.  Furthermore, the
moderate atomic number $Z=6$ produces a significant capture rate for
$\mu^-$, reducing the $\overline{\nu}_e$ contamination in the beam.
With better optimization, one should be able to further reduce the
wrong-sign neutrino contamination relative to the LANL beam stop.

Design considerations for the beam stop will include possible
consideration of larger $Z$ absorber plates embedded in the beam stop
to reduce wrong-sign neutrino contamination, and the specification of
the lateral dimensions of the beam stop which controls meson leakage
and the decay-in-flight component of the beam.  Past experiments
suggest that the uncertainty in these components is 2-5\% and in the
overall neutrino flux is of order 11\% \cite{allen1989}, so it is of
some use to reduce the background fluxes.

\subsubsection{Thermal Analysis}

\begin{figure}[p]
\vskip -3.cm
  \centering
  \includegraphics[width=5.in]{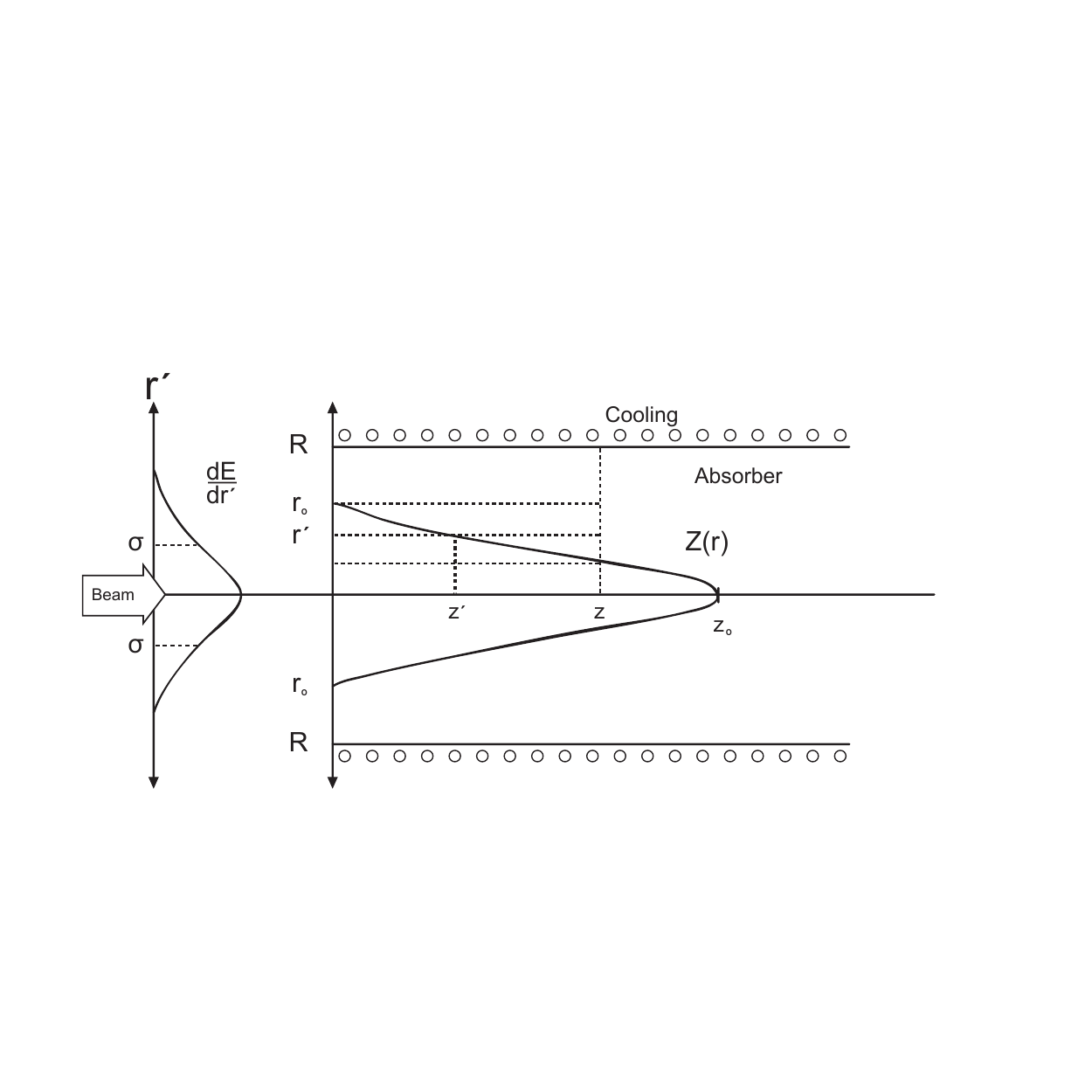}\\
\vskip -1.cm
  \includegraphics[width=2.75in,angle=90]{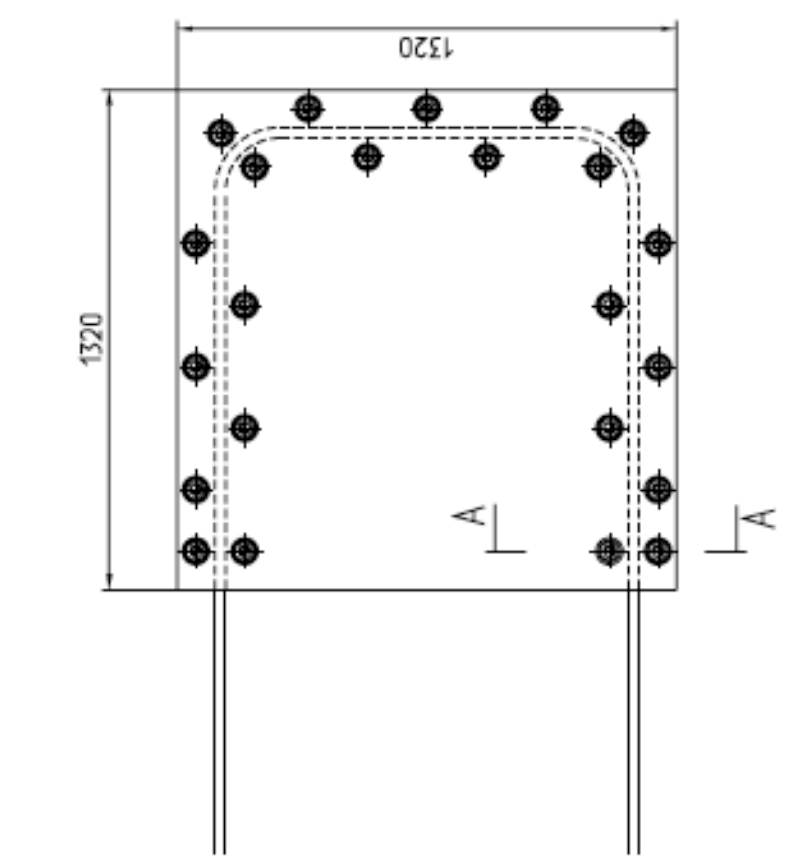}
\vskip 0.cm
\caption{(top) Conceptual design of a beam stop with tapered entrance
  hole for the proton beam.  The shape of the entrance hole produces a
  constant temperature profile along the length of the beam stop.
  (bottom) Conceptual design from the NuMI beam for establishing
  thermal contact between the slabs of the beam stop core to water
  cooling lines kept external to the beam profile which involves
  sandwiching the cooling lines into channels between adjacent bolted
  slabs of the beam stop.}
  \label{fig:tapered-dump}
\end{figure} 

The conceptual design of a beam dump for pion production from a
high-current proton beam has been studied. It consists of a
water-cooled stainless-steel or copper tube centered on the beam axis
which contains a set of graphite disks or slabs. The disks near
the beam entrance have concentric holes forming a re-entrant hole for
the beam to spread the beam energy loss into the depth of the dump
(see Figure~\ref{fig:tapered-dump}, left).  
The energy deposited by
the beam in the graphite disks is conducted radially to the
surrounding water-cooled lines. Such lines can be shrink-fit to the
exterior of the graphite disks, or, as in the case of the NuMI beam
stop, the water cooling lines are clamped into the slabs of graphite
for thermal contact (see Figure~\ref{fig:tapered-dump}, right).  The
re-entrant hole is shaped in such a way as to keep the temperature
constant over the re-entrant hole surface assuming constant
temperature at the cooling tube.  A more sophisticated design in which
the cooling tube is tapered radially with respect to the beam axis
such as to keep the heat flux density constant over the tube between
the beam entrance and the end of the re-entrant hole in addition to
keeping the re-entrant hole temperature uniform was also studied.

\begin{figure}[t]
\vskip 0.cm
  \centering
  \includegraphics[width=4.in]{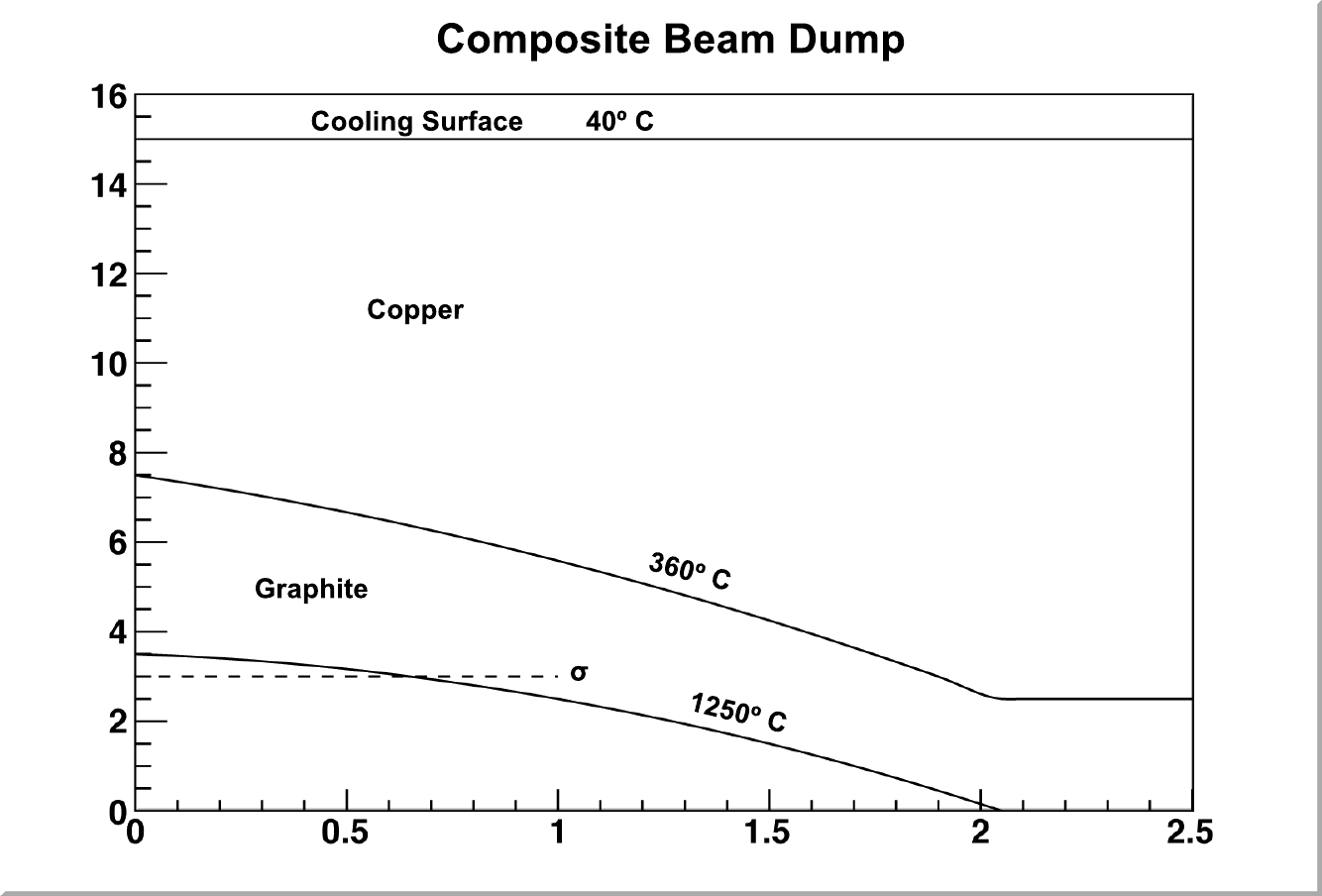}
\vskip 0.cm
\caption{Graph of the temperature profile in the beam stop.  The
  vertical axis is the radius away from the beam axis and the
  horizontal axis is longitudinal depth within the beam stop.  Two
  curves are shown, one at the inner edge of the re-entrant hole which
  is maintained at a temperature 1250$^\circ$C and the second at a
  line of constant temperature 360$^\circ$C. }
  \label{fig:stop-profile}
\end{figure}

Assuming a circular double-Gaussian proton beam profile of
r.m.s. radius $\sigma$ and exponential proton energy deposition in
graphite with a interaction length $L$, examples for a cylindrical and
a tapered beam dump were evaluated for the following parameters
typical for a DAE$\delta$ALUS beam stop: beam power = 1 MWatt, beam radius
R.M.S. $\sigma=5$~cm, graphite density = 1.75g/cm$^3$, $L=50$~cm,
diameter of cooling channel = 30~cm, diameter of re-entrant
hole=11.4~cm, cooling water = 100$^\circ$C.  The resulting shapes of a
cylindrical and a tapered beam dump are shown in
Figure~\ref{fig:stop-profile}. The total length of the beam stop is
constrained by the depths of the re-entrant holes plus the range of
protons in graphite (125cm for 700MeV protons).
Figure~\ref{fig:stop-profile} shows the result of thermal calculations
for the beam stop.  As with the NuMI design, the present concept has
the advantage of keeping water cooling lines out of the shower profile
of the beam \cite{wehman1999,abramov2000}.  A significant advantage
over other beam stop designs, however, is that the re-entrant hole
reduces the temperature at the center of the core by spreading out the
interactions of the proton beam.  To minimize sublimation from the hot
graphite slabs, they could be sputter-coated with a few microns of
Molybdenum or Tungsten eliminating the need for a beam window.

\subsubsection{Radiation Protection}
A significant design consideration will be the protection of
personnel, equipment, and ground water from the radiation produced in
the beam stop.  It is envisioned that the primary proton beam will be
brought to the beam stop through a penetration in a shield wall which
separates the proton extraction line from the target station.  It is
further envisioned that the water-cooled graphite core of the beam
stop is embedded in a larger structure of cast steel and concrete
shielding blocks for attenuation of gamma rays and neutrons produced
in the beam stop.  While the containment of the core inside such a
larger structure accentuates the need for thermal analysis, it is
clear that shielding thicknesses of $\sim1$~m of steel and $\sim$1~m
of concrete is required just to keep residual activity levels to
30~mR/hour after a year of 1~MWatt operation of the beam line
\cite{wehman1999}.  Such will be the subject of simulation and
calculation work during the first year of DAE$\delta$ALUS R\&D.  An
additional concern will be the mitigation of humidity and water in the
target station area.  A significant experience from operation of the
NuMI beam was that high humidity levels due to the below-grade
location of the target station causes moisture capture and irradiation
in the concrete shielding leading to high levels of tritium
\cite{hylen2000}.  Such humidity requires mitigation by
dehumidification systems.

%% file: detectorintro_v1.tex
 In designing DAE$\delta$ALUS, we have adhered closely to the proposed
large water Cherenkov detector design.  As with many of the low energy
($<$100 MeV) analyses, this experiment requires Gd-doping and benefits
from good photocathode coverage.  We consider progress on these issues in this
section.

%% file: Gd_v2.tex
In order to detect IBD events, the DAE$\delta$ALUS detector must be doped
with the neutron capture agent gadolinium (Gd). Gd offers two
essential advantages over the competing process of hydrogen capture:
it reduces the capture time for the neutron generated in the IBD
interaction from 200 $\mu sec$ to about 30 $\mu sec$ , and the energy
release from the capture interaction is higher - ~8 $MeV$ compared
with 2.2 $MeV$ for hydrogen. These twin benefits provide a clean tag
for the time correlated positron and neutron which make up the IBD
signal.

Gd-doping is a familiar technology for antineutrino detection, having
been implemented in numerous liquid scintillator experiments,
including Double Chooz \cite{chooz}, SONGS1 \cite{nuresults}, and
others. Stable operation over 1 year time scales has been demonstrated
in these 1-10 ton scale liquid scintillator detectors.

Though well understood in small (by DAE$\delta$ALUS standards) scintillator
detectors Gd-doping has been only partially demonstrated for large
water Cherenkov detectors.  LLNL has built and successfully operated
1-5 ton scale water detectors doped with $0.1\%-0.2\%$ concentrations
of Gd \cite{LLNLH2Odet}. These detectors have shown both the expected
time correlations and the increased energy output from the neutron
capture event using tagged neutron (Americium-Beryllium, AmBe)
calibration sources. A 250 kg detector also empirically demonstrated
the long term compatibility (1 year time scale) of GdCl$_3$ compounds
with acrylic\footnote{In this detector, the entire detector interior
  was constructed of acrylic in order to reduce the likelihood of
  contamination/clouding from leached materials.}.  Implementation of
these detectors is underway for nonproliferation applications that
require large solid angle and efficient neutron detection.

In addition to these small scale performance tests, a test was
performed at the SuperKamiokande experiment, in which a small sealed
vessel of Gd-doped water was lowered to the center of the detector
\cite{SuperkGdtest} . The vessel also contained a tagged AmBe source,
so that the neutron emission time was known to within a few ns. The
significance of this test is that the large water volume guaranteed
full containment of the Gd-capture gamma-ray shower. By comparing with
the well-tuned Superkamiokande energy calibration, it was found that
approximately 50\% of the shower energy is detectable in the Cherenkov
light emission channel. ( 4.3 $MeV$ of energy out of an available 8
MeV from the capture gamma-ray shower.) The missing energy is a
limitation imposed by physics that would be present even with perfect
light collection. It occurs because not all of the Compton electrons
generated by the gamma-ray interactions are above the 0.78 $MeV$
threshold for Cherenkov light production. In a set of soak tests
related to this experiment, the compound Gd$_2$(S0$_4$)$_3$ was found
to be most compatible with the structural materials exposed to the
water-Gd solution.

\begin{figure}[p]
\centering
\includegraphics[height=6.25in]{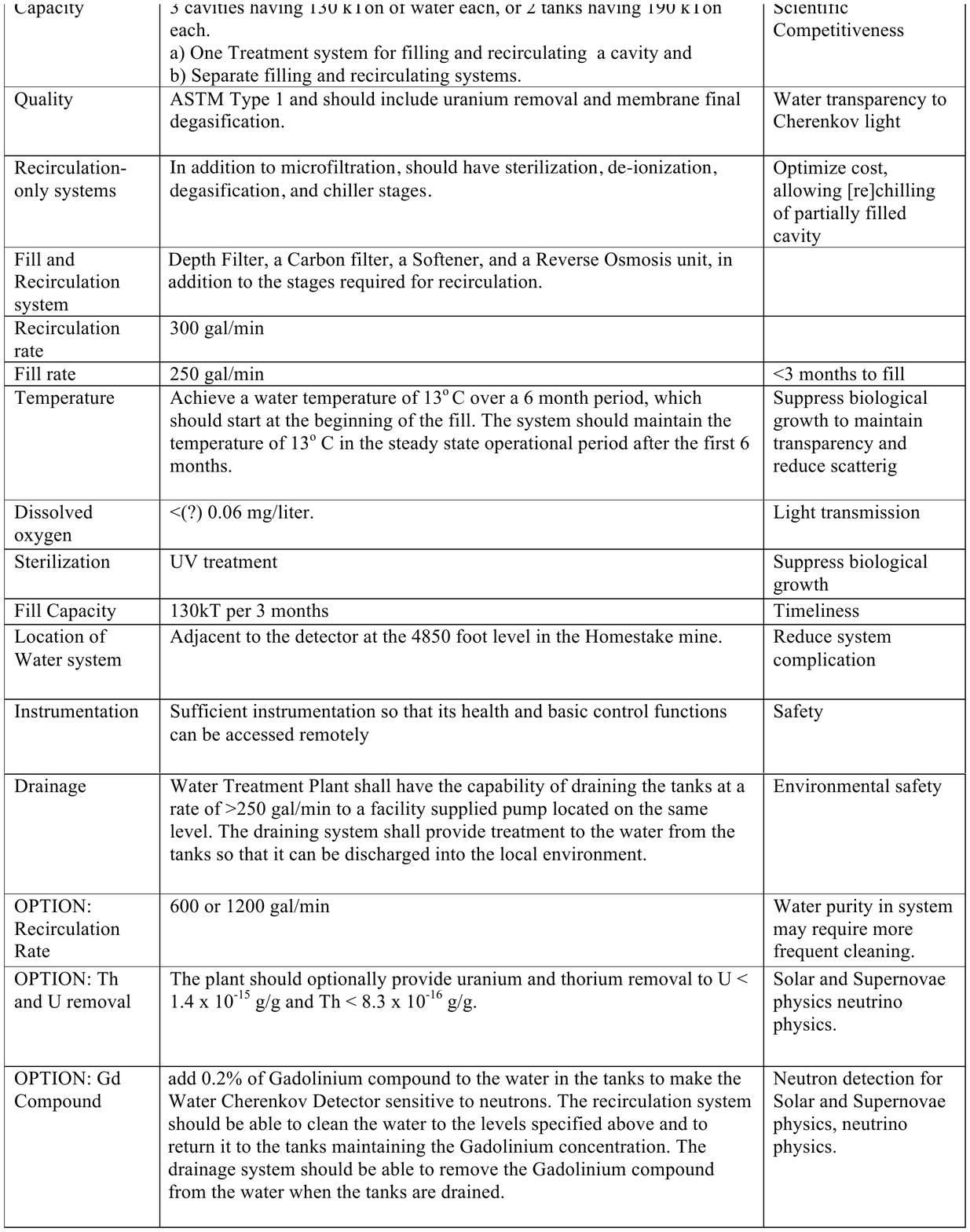} \vspace{-0.25in}
\caption{ A list of basic requirements for the LBNE water systems, including the Gd doping option. This table is extracted from \cite{KadelWaterReq} }
\label{WaterTable}
\end{figure}

While these tests are all encouraging for a large scale Gd-doped
detector, important R\&D steps remain in order to demonstrate
successful operation at the 150,000 ton scale required for
DAE$\delta$ALUS (and LBNE).  The R\&D program is derived from a set of
physics requirements defined for the LBNE water detectors, which
include a Gd-doping option. The table in Figure
\ref{WaterTable}. taken from an LBNE project planning document
\cite{KadelWaterReq}, summarizes these requirements. The requirements
are essentially identical to those needed for DAE$\delta$ALUS, though
some relaxation of constraints could arise in DAE$\delta$ALUS, due to
the different event rates. DAE$\delta$ALUS will consider which if any
elements of the LBNE Gd doping strategy would need to be modified on
signal to background or cost grounds.

From the physics requirements, a set of research and development
priorities have been identified. The steps listed below are part of
the larger work plan for the overall LBNE water detector development
program. DAE$\delta$ALUS will benefit from the overlap in detector
needs of the two proposed experiments.

\begin{itemize}
\item demonstration that the water can be purified sufficiently even
  in the presence of the Gd compound, which would normally be removed
  in the purification process, and
\item demonstration of long term compatibility of all exposed
  structural materials with the Gd-water solution, and completion of a
  downselection process for the choice of Gd-bearing chemical
  compound;
\item demonstration that the attenuation length of the water is not
  compromised directly by the Gd-solution or by leaching from
  structural materials;
\item demonstration of scaling of the Gd purification process,
  including demonstration that Gd can be kept uniformly dissolved
  throughout the detector on time scales compatible with the water
  circulation and purification rate.
\end{itemize}

The LBNE work plan elements relating to Gd-doped detector R\&D are
described briefly in the following sections. DAE$\delta$ALUS will join
forces with the rest of the LBNE team to support studies of the Gd
doping option for the DUSEL WC detectors.

In addition to the US LBNE effort, a 200 ton scale Gd-doped detector
and accompanying filtration system is being built in Japan
\cite{vaginswebref}. This detector, known as EGADS, will help answer
several of the outstanding questions regarding Gd-doping of detectors.

\subsubsection{Water Purification in the Presence of Gd}

\begin{figure}[t]
\centering
\includegraphics[height=3in]{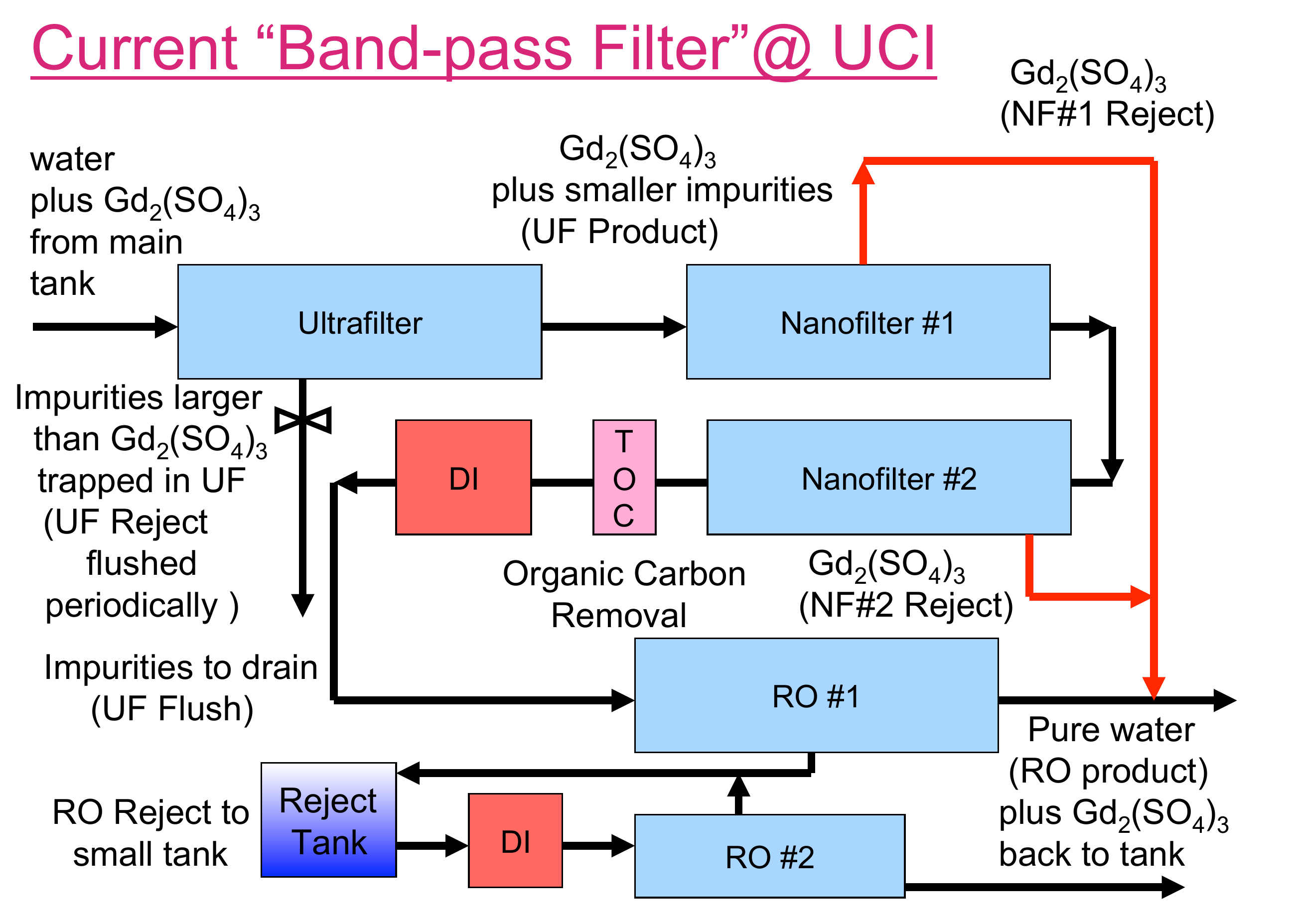}
\caption{ A flow chart of a proposed Gd purification and mixing procedure for the DUSEL WC detectors. The basic concept has been developed at UC Irvine. }
\label{gdflow}
\end{figure}

The LBNE collaboration has a significant effort invested in developing
a large scale purification system that is compatible with the use of a
Gd dopant. The 'standard' purification system for large water
detectors, such as that used for the SuperK detector, must be modified
to allow introduction of purified Gd after cleaning of the water.  A
flow chart for one proposed purification method, developed at UC
Irvine, is shown in Figure \ref{gdflow}.  The basic idea is to remove
a Gd concentrate from the partially cleaned water, pass the Gd-free
water through additional purification steps, separately remove Uranium
and Thorium contaminants from the Gd concentrate, and recombine the
purified Gd concentrate with the purified water.  The first elements
of this process have been demonstrated in a lab-scale system at UC
Irvine. Research is ongoing at BNL, UC Irvine and elsewhere on the
U/TH removal from the Gd concentrate.

\subsubsection{Materials Compatibility and Attenuation Length Studies}

As part of its LBNE activities, Brookhaven National Laboratory (BNL)
has developed a list of structural materials likely to be directly
exposed to the Gd water in the DUSEL water detectors.  LLNL is
conducting long term tests of the integrity of this set of materials,
and is making $~1\%$ accurate measurement of changes in the water
attenuation length that may be caused by the introduction of a Gd
compound.  This precision is sufficient to confirm whether the $~100 m$
attenuation length targets for LBNE and DAE$\delta$ALUS can be met.
GdCl$_3$ and other compounds will be tested. Initial test results
demonstrating possible leaching of $Fe$ from stainless steel in the
presence of GdCl$_3$ may be found in \cite{LLNLGdatten}

\subsubsection{Scaling of Water Purification Systems in the Presence of Gd}

In addition to defining a process for maintaining water purity in the
presence of Gd, the LBNE collaboration is working on the engineering
aspects of scaling the system to work with 150 kT mass
detectors. Important elements include the throughput of filtration
systems and pumps, the specific layout of the DUSEL caverns and water
supply lines, and uniformity of mixing of the Gd in the water.  Leaks
in the detector are also far more important in a Gd-doped detector,
because the underground sites will not allow direct discharge of
Gd-doped water to drainage systems.  Since prior detectors such as
Superkamiokande have had water losses, leak tightness must be
improved.

The EGADS experiment in Japan will address several of the above R\&D
questions simultaneously, using a fully functional 200 ton scale
detector and selective filtration system.  In particular, EGADs goals
are to study the rate at which Gd can be introduced into the water,
the uniformity of Gd mixing in the water, the Gd removal process, the
selective water filtering process, and materials effects in an
integrated detector. (It will not directly address the U/Th removal
process).

\subsubsection{Conclusions for Gd Doping and Detector Scaling}
The underlying physics of Gd-doped water detectors is now well
established.  The R\&D program described above, is being pursued by
LBNE and by the EGADS experiment in Japan. The program directly
addresses the outstanding engineering questions that must be answered
in order to achieve 150 kton operation of a Gd-doped detector.

%% file: photocover_v1.tex
In terms of photocathode coverage, DAE$\delta$ALUS' requirements 
are motivated in principal by the need for efficient triggering at 
low visible energies, accurate energy reconstruction, and accurate reconstruction 
of the out-going lepton scattering angle. Efficient triggering 
and vertex reconstruction is necessary to observe the 8 MeV gamma
cascade from the neutron capture on Gd characteristic of the IBD 
signal. Further, since the size of the $\delta_{CP}$ conversion effect in 
Eq. \ref{equ:beam} is coupled to the amount of interference between the 
two neutrino mass splittings, accurate energy reconstruction 
is important to extract $\delta_{CP}$.   
Good angular resolution is essential for
distinguishing the two dominant normalization event 
samples from one another. The Super-Kamiokande experiment's
first (SK-I) and second (SK-II) run periods provide an instructive 
example of the capabilities of a large water Cherenkov detector 
at $<100$ MeV energies with  differing photocathode 
coverages: 40\% and 19\%, respectively. Numbers in the discussion 
below for SK-I are taken from~\cite{Hosaka:2005um} and those for 
SK-II are from~\cite{Cravens:2008zn}.

Super-Kamiokande has demonstrated efficient triggering at low energies 
in both of the above run periods. At 40\% photocathode coverage, SK-I 
collected 7 PMT hits per MeV of total energy (kinetic energy plus rest mass energy)
and triggered with 100\% efficiency down to 4.5 MeV. SK-II similarly
triggered at 100\% efficiency above 6 MeV with half as many hits per MeV. 
Note that these thresholds are below the minimum visible energy requirement for both IBD 
positrons (20 MeV) and recoil electrons (10 MeV) from $\nu-e$ scattering.  This is 
below the total energy of the gamma cascade from neutron capture on Gd, often the cascade energy is partitioned
among several gammas, with one or more receiving less than 2 MeV. 
The least energetic gammas are not likely to result in Compton-scattered
electrons above Cherenkov threshold, and not all of the gamma energy is
observed. For this reason, 
the mean observed energy of Gd capture events is shifted down to 4.3 MeV \cite{Watanabe:2008ru},
where the trigger efficiency of SK-II was only 30\%.  

As discussed in Sec \ref{eventsindetector} the recoil 
electron direction in $\nu-e$ scattering events is strongly peaked
in the forward direction. Electrons emitted in $\nu_{e}-\,O$ scattering,
on the other hand, have a weaker angular dependence. 
At DAE$\delta$ALUS energies the $\nu-e$ scattering peak can be isolated by selecting 
scattering angles within $26^{\circ}$ of the beam direction 
(see Sec. \ref{calib}), so angular 
resolution in the water Cherenkov detector that is better than this value is desirable.
During the SK-I run period the angular resolution of recoil electrons  
was found to improve from $27^{\circ}$ at  
total energies of 10 MeV to $23^{\circ}$ at 15 MeV. With decreased 
photocathode coverage in SK-II the resolution at 10 MeV was 
$30^{\circ}$ and fell to $25^{\circ}$ at 13 MeV. 
Both photocathode coverages lead to resolutions near the required 
threshold in the energy range of interest and generally 
improve to a plateau near the IBD positron minimum energy

Despite the relative stability of the above variables against lower 
photocathode coverages, the vertex and energy resolution of the detector 
are more sensitive to less photon collection. 
At 5 MeV of total electron energy, the SK-I (SK-II) resolution is 
18\%(28\%) and improves to 12\%(16\%) at 20 MeV. The vertex resolution 
at 5 MeV was 150~cm (210~cm) and 70~cm (100~cm) at 10 MeV.

Current designs for the DUSEL water Cherenkov detector propose photocathode coverages between 
30 and 40\%.  The latter is expected to provide sufficient performance for DAE$\delta$ALUS
while SK-II level coverage is likely insufficient for reconstructing the Gd capture gammas 
from IBD events. Choosing the middle ground, 30\% coverage, though providing improved   
vertex and angular resolution, may also suffer from inefficient triggering of these 
signal events.

%% file: schedule_v1.tex
\begin{figure}[t]\begin{center}
{\includegraphics[width=4.5in]{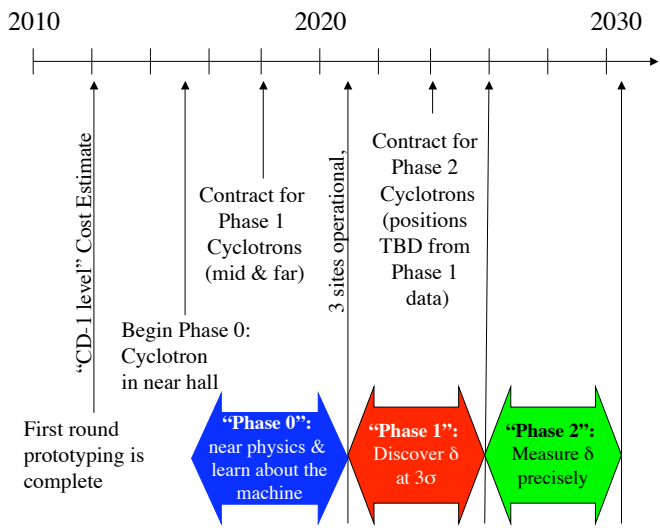}} 
\end{center}
\caption{A possible plan for the staging DAE$\delta$ALUS}
\label{phases}%
\end{figure}

\begin{figure}[t]\begin{center}
{\includegraphics[width=4.5in]{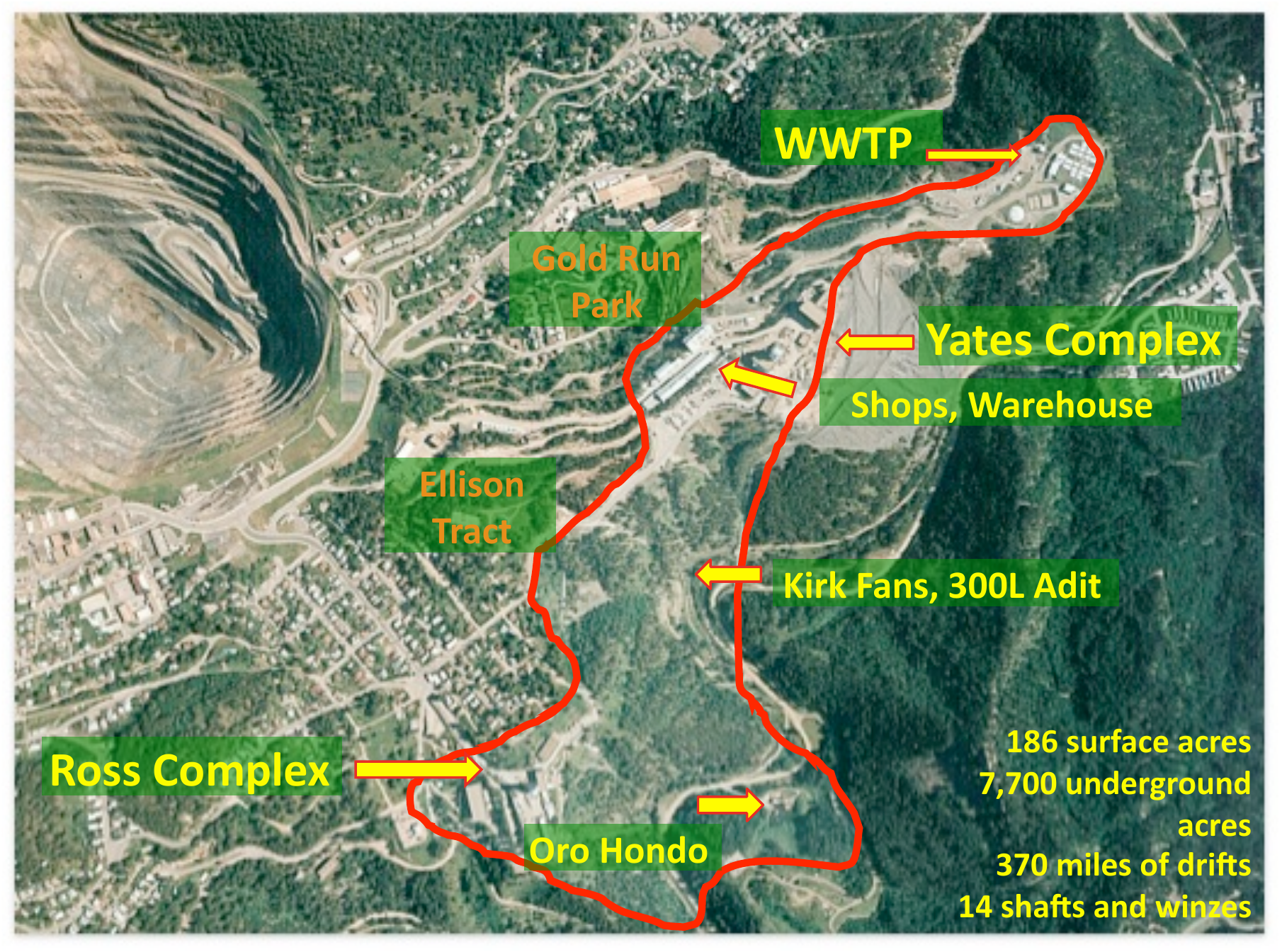}} 
\end{center}
\caption{Sanford Lab footprint, showing the Kirk Adit and Oro Hondo sites.}
\label{footprint}%
\end{figure}

{\begin{figure}[p]
\vspace{-1.in}
{\hspace{-0.25in}\includegraphics[width=6.5in]{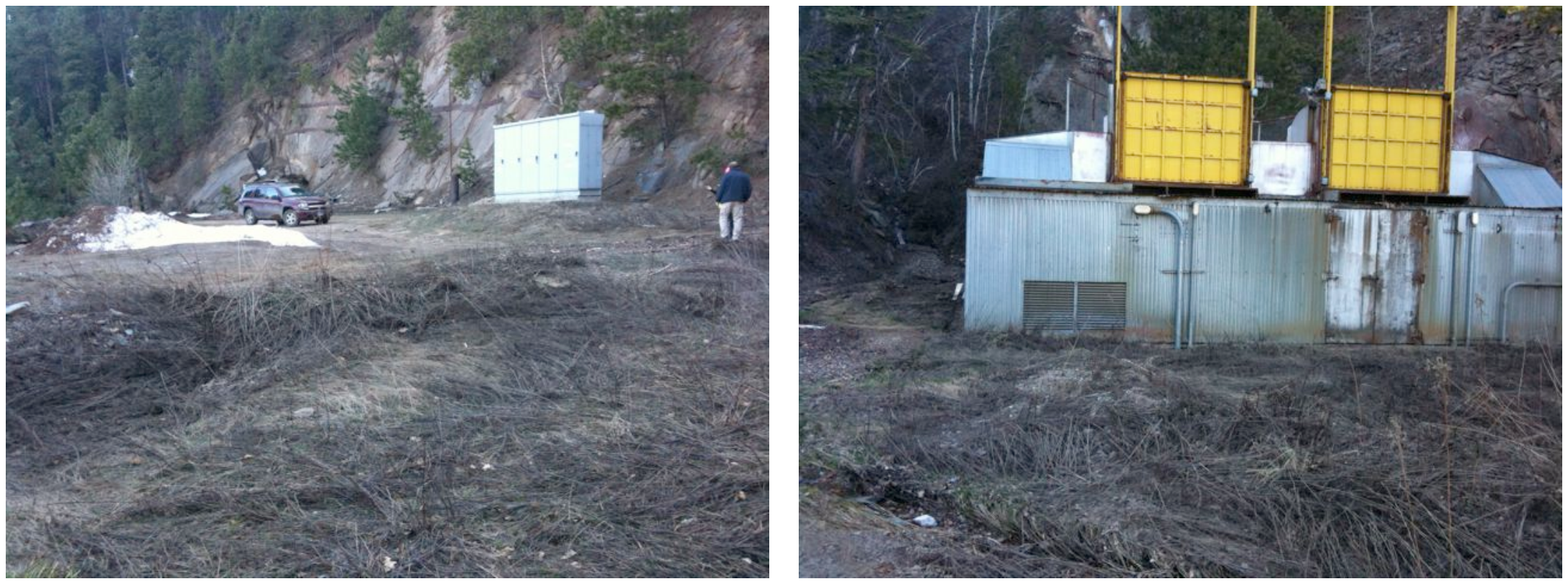}} 
\vspace{-2.25in}
\caption{Region in front of the Kirk Adit.  Left:  a portion of the open space available.   The 300 L entrance is obscured by the car. To the right of the person is 
the fan building, not shown.  Right:  the building housing the unused fans.
\label{kirkadit}}
{\hspace{-0.25in}\includegraphics[width=6.25in]{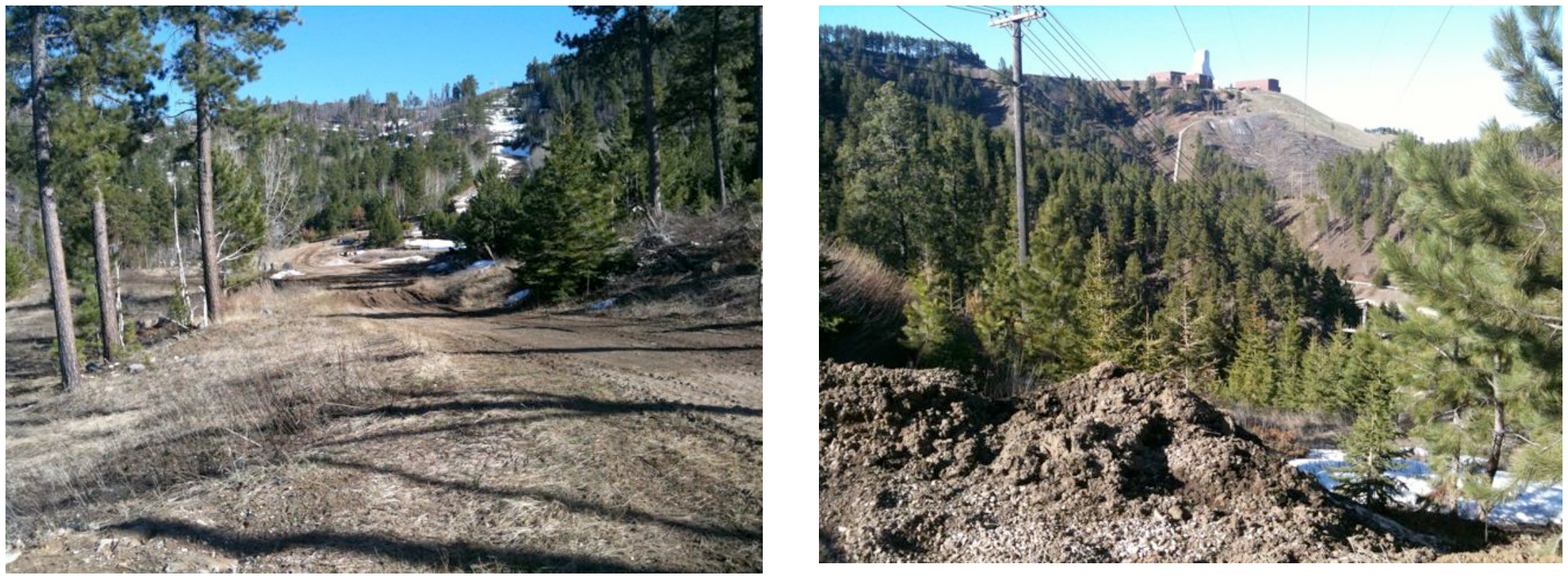}} 
\vspace{-2.25in}
\caption{The Oro Hondo substation site.  Left:  An area which could be cleared
for the accelerator building;  Right: view to the Yates shaft from the site. \label{oro}}
\end{figure}}

We are developing our plans in coordination with the DUSEL Management.
We propose a three-phase plan for staging the experiment, summarized
in Fig.~\ref{phases}.  Before the CP-violation studies begin, we
propose a Phase 0 where we learn to run a single accelerator on-site.
This would allow for near-accelerator physics to begin.   At the same
time, it allows us to understand important aspects of operation.
Phase 1 and 2 represent the CP-violation studies periods.

For the Phase 0 site, we propose the area next to the entrance to the
300 L campus.  This is called the Kirk Adit, and is labeled ``Kirk
Fans, 300 L Adit'' on Fig.~\ref{footprint}.  This site is located just
below the Yates shaft, at the base of a 300 ft rise, and 1.5 km from
the detector. The two photographs in Fig.~\ref{kirkadit} show a
portion of the space available.  There is an open space available
which is large enough for the accelerator (left photograph).  However,
it would be best if the unused fan building (right photograph) could
be removed, as this allows more space.  The entrance to the 300 L is
obscured by the car in the left photograph.  Sufficient power and water are
available at the site.  The entrance road would need improvements in
the first 30 feet.  However, the 300 L entrance may be used as a
construction entrance, in which case the road improvements would be in
place before the accelerator building would be built.

An alternative location for the on-site accelerator is at the Oro
Hondo substation (marked Oro Hondo on Fig.~\ref{footprint}.)  There is
a flat area which could be easily cleared of small trees which would
accommodate the accelerator building (Fig.~\ref{oro}, left).  The area
has power; water availability will need to be investigated.  The road leading to
the area is well maintained, because it leads to the Oro Hondo
substation.  The drawback of this site is that it is far from the
Yates shaft (see Fig.~\ref{oro}, right, which shows the view to the
Yates shaft).  Thus it would not permit a 300 L physics program.

We do not proceed to CD-1 for Phase 1 until 2015 or later.  This
schedule allows time to negotiate the off-site locations.  Several
locations with disturbed land could be considered.  By this
timeframe, information will be available on $\sin^2 2\theta_{13}$ from
Double Chooz, Daya Bay and T2K.  These data can be used to inform our
design.

%% file: coord_v1.tex
In this section, we describe how DAE$\delta$ALUS coordinates 
with other analyses already under consideration for the large
water Cherenkov detector.   We first show the benefits of the
beam for calibration.   We then discuss the impact of the beam 
events on other studies, especially those in 
particle-astrophysics, which are
in the same energy range.  It should be noted that  
there are DAE$\delta$ALUS collaborators participating in each 
of these areas of study, and this allows good coordination 
between our design and those of the other analysis 
groups. 

%% file: calib_v3.tex
Neutrino-electron events from the near accelerator offer a high
statistics ``standard candle'' for calibration of the detector to the
1\% level up to an energy of 55 MeV.  This calibration standard will be valuable to all of the large
detector experiments.

The neutrino-electron scattering process has a well-defined cross-section, as was
discussed in Section \ref{sse:CP}, when we considered these events for
the normalization sample.  There is a net 1\% theoretical error in the
normalization of the cross section. However, there is negligible
theoretical error in both the shape of the visible energy distribution and
the scattering angle as a function of visible and neutrino energy.  As
a result, neutrino-electron scatters provide a very useful sample for
calibration.
\begin{figure}[t]\begin{center}
{\includegraphics[width=4.5in]{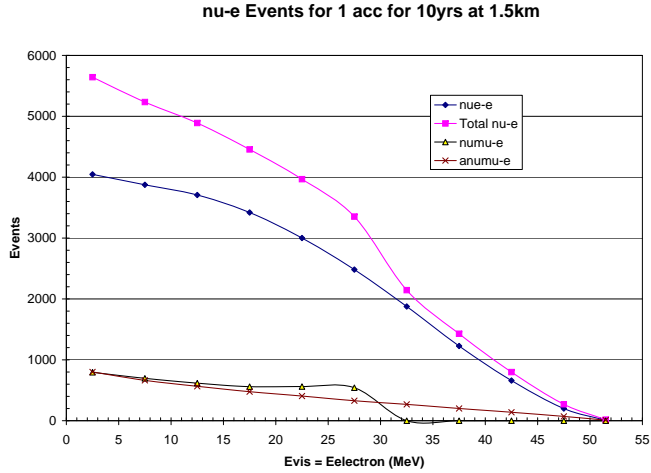} } \end{center} 
\vspace{-0.5in}
\caption{\label{nue-energy} Top curve (magenta): visible energy
 distribution for neutrino-electron scatters from a decay-at-rest
 beam. Lower curves: the $\nu_e$ (blue), $\nu_\mu$ (olive) and $\bar
 \nu_\mu$ (brown) contributions.}
\end{figure}

The visible energy distribution of neutrino-electron scatters from a
decay at rest beam is shown in Fig.~\ref{nue-energy}, top curve
(magenta).  The number of events corresponds to 10 years of running at 1.5 km.
The lower curves show the individual contributions of the three
flavors in the beam: $\nu_e$ (blue), $\bar \nu_\mu$ (brown), and
$\nu_\mu$ (olive).  Neutrino-electron scattering is dominated by the charged
current $\nu_e$-electron interaction.  The scatters of the
monoenergetic $\nu_\mu$ (from $\pi^+$ decay) make an important
contribution below 30 MeV.  The $\bar
\nu_\mu$ scatters are a small contribution.

\begin{figure}[t]\begin{center}
{\includegraphics[width=4.5in]{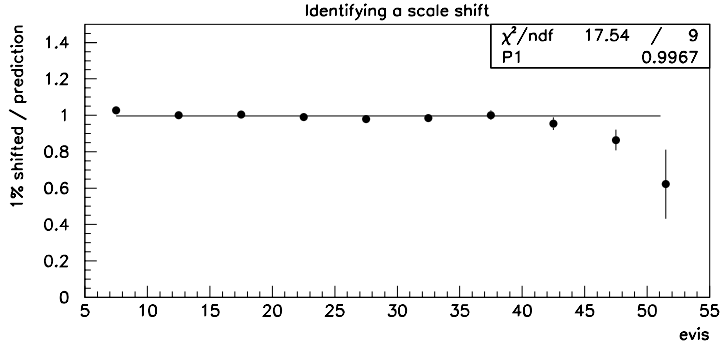} } \end{center} 
\vspace{-3.in}
\caption{\label{nue-calib} An example of an energy miscalibration in
 the detector.  This plot shows the ratio of neutrino-electron events with
 a 1\% energy miscalibration to the predicted shape. Note the poor
 $\chi^2$/DOF and that the region near the endpoint is the most sensitive to miscalibrations.}
\end{figure}

Fig.~\ref{nue-calib} illustrates the strength of the neutrino-electron data for 
energy calibration.   The plot shows the ratio of data with a 1\% energy 
offset to the prediction, with the statistics for the full 10 year run.
The poor $\chi^2/DOF$ indicates that calibration is required. This
shows that this data set is sensitive at the $\sim$1\% level.   Note that
we allow the overall normalization to float since this data set is used
to set the normalization of our analysis and that this calibration relies
only upon shape.

\begin{figure}[t]\begin{center}
{
\includegraphics[width=5.5in]{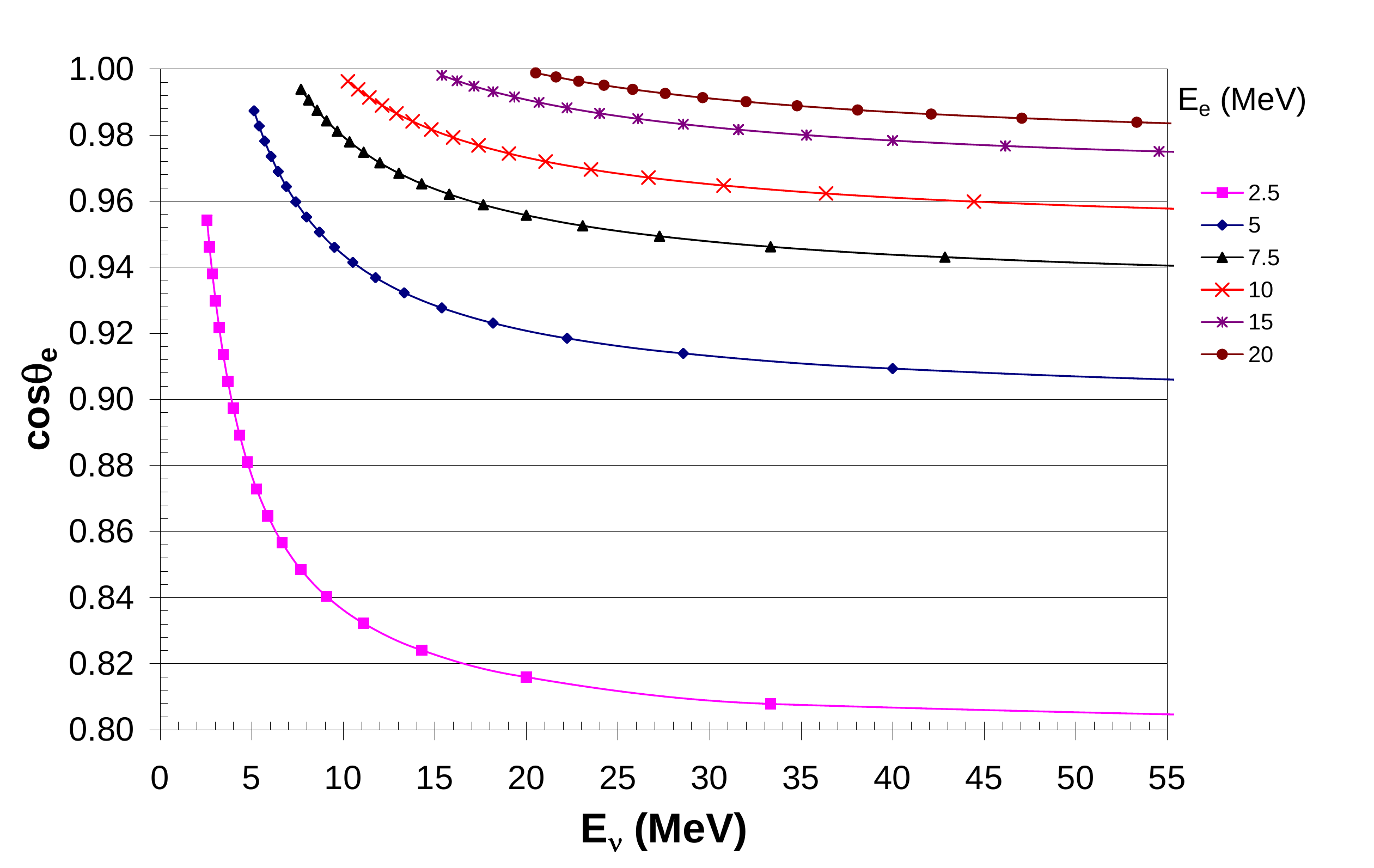}} \end{center} 
\caption{\label{nue-angle} The unique angle vs. visible-energy correlation 
for neutrino-electron scattering at each neutrino energy.}
\end{figure}

There is a strong energy-angle dependence in the kinematics of
neutrino-electron scattering.  This is shown in Fig.~\ref{nue-angle} 
(right).  As discussed in Section~\ref{sse:CP}, this allows the events
from the near accelerator to be identified and separated from other
sources of neutrino-electron scatters.  A cut at $\cos\theta=0.9$ with
respect to the accelerator direction corresponds to an angular cut of
26$^{\circ}$, which should account for angular resolution limitations.  Fitting the
events in this cone as a function of angle and visible energy will
provide a measure of the angular resolution of the detector.

%% file: backgrounds_v4.tex
In this section we consider the impact of DAE$\delta$ALUS
events on the large detector analyses.  We show that DAE$\delta$ALUS has a
negligible effect on all analyses except for the Supernova Relic
Neutrinos (SRN) search.  For the SRN analysis, we estimate that in the worst
case our events degrade the $1\sigma$ limit for zero SRN events by 10\%
to 15\%, while in the best case our events provide a tool to improve
the limit by 20\% to 30\%. We note that all of these analyses benefit
from the neutrino-electron calibration sample provided by DAE$\delta$ALUS.

\subsubsection{High Energy Analyses: 
Long Baseline, Atmospheric and Proton Decay}

The following analyses have signatures which require visible energy well above 55 MeV -- the maximum for DAE$\delta$ALUS -- and/or event signatures which are easily distinguishable from DAE$\delta$ALUS:
\begin{itemize}
\item Long Baseline Neutrino Oscillations, which employs a neutrino
 flux above 300 MeV.
\item Atmospheric Neutrino Studies, which rely on $\nu_\mu$
 charged-current interactions above $\sim 100$
 MeV.
\item Proton Decay, where the signatures are $e^+ \pi^0$ and $K^+
 \nu$.  Both cases have a high visible energy and very different
 signature than DAE$\delta$ALUS-induced interactions.
\end{itemize}

The proton decay analysis has overlapping interest with
DAE$\delta$ALUS because an important proton decay signature includes
observation of a low energy photon.  A 6 MeV photon accompanies proton
decay in the oxygen nucleus, and thus provides a coincident signal
that reduces background.  This is of similar energy to the photons
from the Gd-capture used by DAE$\delta$ALUS. Both experiments will therefore require similar photocathode coverage.

\subsubsection{Solar Analyses}

DAE$\delta$ALUS has a negligible effect on the solar neutrino analyses.
The near accelerator produces a neutrino-electron scattering signal
during the 20\% time period that it is running.  However, these events
are cut by a $\cos\theta<0.9$ requirement, where $\theta$ is the angle
with respect to the near accelerator direction.  This reduces the effective
fiducial volume by 5\%, but only when the accelerator is running.
With only this cut,  the net effect is a 1\% loss of data.    This can 
be cut to 0.5\% if one simply requires that the accelerator run mainly
when the sun is in the opposite hemisphere, which will place the accelerator
events far outside of the angular cut for solar events.

On the other hand, the DAE$\delta$ALUS events provide a very nice,
well-defined, high statistics sample at low electron visible energies
(see Fig.~\ref{nue-energy}).  Along with its value for calibration,
already discussed, this sample allows a measurement of the efficiency
for reconstructing neutrino-electron scatters at low energies.  This
may be valuable to the statistics-limited solar neutrino studies.
\subsubsection{Supernova Burst Analyses}

The Supernova Neutrino Burst Analysis focuses on the same energy range
as DAE$\delta$ALUS.  A supernova neutrino signal will involve mostly
IBD events (the contribution of $\nu$-e scatters is a few percent, and
from other interactions is not more than 10\%) .  For an observed
nearby supernova burst, which should involve at least 24,000 events in
about 10 seconds in 300 kton for a supernova within 20 kpc (edge of
the Galaxy), the expected background of Daedalus events is negligible.
However, DAE$\delta$ALUS events may be a background for extragalactic
SN burst search analyses, which look for multiple events within about
a ten second time window, or look for single events in coincidence
with external triggers such as optical supernovae or gravitational
waves.  In ten years of running there will be $\sim$5000 fake bursts
of two or more DAE$\delta$ALUS events.  By requiring IBD events for
the supernova burst search, this background can be greatly reduced to
the order of 1 in the 10-year period.  For an $\sim$2 hour time window
in coincidence with a visible extragalactic supernova event, there
will be on average about 3 Daedalus events, but the mean number of IBD
events in such a window will be 0.04.  This background can be reduced
to zero by eliminating beam-on time from the supernova analysis,
resulting in a some livetime loss.

\subsubsection{Supernova Relic
 Neutrino Analyses}

The one analysis for which the DAE$\delta$ALUS events have a
substantial overlap is the Supernova Relic Neutrino (SRN) search.  We
have been in contact with the leaders of the SRN analysis, in order to
understand our impact \cite{BeacomPrivate, VaginsPrivate}.  The
conclusion of this interaction is that the two analysis groups can
co-exist and, in fact, help one another.

Both analyses rely on IBD event signatures which will occur as single
events (as opposed to bursts) spread over time.  The SRN analysis
focuses primarily on the 10 to 20 MeV range \cite{VaginsPrivate}.  
Above 20 MeV, invisible
muons, which are below Cherenkov threshold, but stop and decay,
dominate the SRN background.  Below 10 MeV, natural background sources
and reactor neutrinos dominate.  The SRN 10 to 20 MeV window is
adjacent to the DAE$\delta$ALUS 20 to 55 MeV signal window.  The
DAE$\delta$ALUS accelerators do produce a small
number of events below 20 MeV.

\bigskip%
\begin{table}[t] \centering
\begin{tabular}
[c]{|c|c|c|}\hline
Energy bin  & 10-15 MeV & 15-20 MeV  \\ \hline
DAE$\delta$ALUS, 5\% SRN sys &  1.14 & 1.24 \\ \hline
DAE$\delta$ALUS, 10\% SRN sys &  1.09  & 1.14   \\ \hline \hline
Gating out DAE$\delta$ALUS, 5\% SRN sys & 1.12 & 1.19  \\ \hline
Gating out DAE$\delta$ALUS, 10\% SRN sys &  1.08 & 1.12  \\ \hline \hline
Likelihood, gating out DAE$\delta$ALUS, 5\% sys   & 0.5 &  0.59  \\ \hline 
\end{tabular}
\caption{Factor by which the 1$\sigma$ sensitivity will change for the
  SRN analysis, for various scenarios of including DAE$\delta$ALUS, assuming a signal of
  $\delta_{CP}=0$ and $\sin^2 2\theta_{13}=0.04$. See text for explanation of 
  scenarios. \label{tab:SRN}}%
\end{table}%

To assess the impact of DAE$\delta$ALUS events on this analysis,
consider the case of no SRN signal. We consider the case of 5\% and
10\% systematic errors, to bracket a reasonable range for the
analysis.  We consider two energy bins, from 10 to 15 MeV, and 
from 15 to 20 MeV.
We quote the factor by which the 1$\sigma$ sensitivity of the SRN
signal will change, for a $\delta_{CP}=0$ and $\sin^2
2\theta_{13}=0.04$ DAE$\delta$ALUS signal.  If the accelerators are
not gated out, then the well-measured DAE$\delta$ALUS signal can be
subtracted, but that subtraction will have an error which impacts the
SRN analysis.  Rows 1 and 2 of Tab.~\ref{tab:SRN} give the factor by which
the sensitivity will change.  An alternative is to
gate out the 60\% of the time that the three DAE$\delta$ALUS
accelerators are running.  This weakens the SRN 1$\sigma$ sensitivity
by lowering the statistics for a potential SRN signal.  The effect of
this is shown in rows 3 and 4.  In the energy range of the proposed
analysis, in either case, the DAE$\delta$ALUS impact is only a 10\% to
20\% worsening of the sensitivity.

On the other hand, the DAE$\delta$ALUS events provide a substantial
true IBD event sample in the 30 to 55 MeV range.  This can be compared
to the beam-off events in the 30 to 50 MeV range which are dominated
by an order of magnitude by invisible muon events.  These two samples
can be played against one another to develop an algorithm which
accepts true IBD events at high efficiency while rejecting invisible
muon events.  If one imagines developing a likelihood that rejects
50\% of the invisible muon background while maintaining nearly 100\%
efficiency for IBD events, then the $1\sigma$ sensitivity for the SRN
substantially improves.  This is shown by row 5 of Tab.~\ref{tab:SRN}.
Thus, depending on the quality of the likelihood which can be
developed, the DAE$\delta$ALUS events can provide a significant tool
for improvement of the SRN analysis.

From this, one can see that the SRN and DAE$\delta$ALUS analyses work
well together.  DAE$\delta$ALUS provides extra motivation for
Gd-doping and manpower for developing ideas like the likelihood
analysis.  In the other direction, DAE$\delta$ALUS is relying on the
development work on Gd-doping which was developed 
by the SRN community \cite{Beacom:2003nk} and 
is now driven by the SRN and non-proliferation analysis groups.

%% file: conclusion_v1.tex
The DAE$\delta$ALUS experiment offers an opportunity to both expand and enhance the
neutrino physics program at DUSEL. Using a a set of low-cost compact cyclotron
sources, the experiment is able to provide a high-intensity, well-understood
source of neutrinos for various studies from neutrino oscillations to nuclear
spin structure measurements. Each stopped pion produced by incident protons
leads to three different flavor neutrinos allowing the flux to be measured
with little systematic uncertainty. In addition, the energy spectrum of the
neutrinos is also completely determined by the weak decay of the stopped pion
and stopped daughter muon. Thus, a stopped-pion neutrino beam provides an
ideal laboratory for making precision normalized measurements in the region
from 10 MeV to 55 MeV.

For neutrino oscillation measurements of $\theta_{13}$ or $\delta_{CP}$,
DAE$\delta$ALUS would search for $\bar{\nu}_{\mu}\longrightarrow\bar{\nu}_{e}$
oscillations where the $\bar{\nu}_{e}$ is detected through the
inverse-beta-decay process where positron and neutron are detected and
associated through a delayed coincidence. To enhance the neutron capture and
detection, the detector will need to be doped with Gd. For the running
scenario presented in this EOI with multiple MW-cyclotrons at distance of 1.5,
8, and 20 km from a 300 kton water detector, DAE$\delta$ALUS has excellent discovery
and measurement potential for $\theta_{13}$ and $\delta_{CP}$ that is
comparable and complementary to the sensitivity of currently planned LBNE
experiment. Also, since DAE$\delta$ALUS can provide a high-statistics antineutrino
sample, the combination of DAE$\delta$ALUS with $\nu-$only running of LBNE leads to
an enhanced sensitivity (by a factor 2) for observing CP violation and
measuring $\delta_{CP}$.

A broad suite of non-oscillation physics experiments are also possible using
the large LBNE detector or small dedicated detectors placed close to the near
(1.5 km) neutrino source. Coherent neutrino-nucleus scattering measurements
using a 1 ton liquid argon detector placed 30 m from the source can be used to
make measurements of the weak mixing angle or to search for non-standard
interactions. A near source detector could also be used to measure cross
sections relevant to astrophysical processes, supernova explosions and
nucleosynthesis, and to supernova detectors. The high neutrino rates also lead
to the possibility of searching for neutrino magnetic moments or making
measurements of strange quark spin asymmetry, $\Delta s$.

The DAE$\delta$ALUS experiment has two important requirements: the development of
low-cost, high-power cyclotrons and the addition of Gd doping for the large
LBNE detector. Significant R\&D progress is being made in both of these areas
and the prospects are good that attractive technical solutions will come about. If these
requirements can be met, the DAE$\delta$ALUS plus near source experiments will
greatly enlarge the physics program at DUSEL. For these reasons, we request
that the DAE$\delta$ALUS experiment be incorporated as a possible option in the DUSEL
planning with the understanding that the experiment would go forward when
technical issues are addressed.